\documentclass[11pt,a4paper]{article}
\pdfoutput=1
\usepackage{jcappub}
\usepackage[utf8]{inputenc}
\usepackage{amsmath}
\usepackage{graphicx}
\usepackage{caption}
\usepackage{subcaption}
\usepackage{amssymb}
\usepackage{color}
\usepackage{cancel}
\usepackage{framed}
\usepackage{comment}
\usepackage[margin=1in]{geometry}
\usepackage{mathtools}
\usepackage{hyperref}
\usepackage{chngcntr}

\counterwithin{equation}{section}

\newcommand{\je}[1]{{\textcolor{blue}{\tiny (JE)}}\marginpar{\flushleft\scriptsize \textcolor{blue}{\bf JE}: #1}}

\newcommand{\nn}{\nonumber}

\def\a{\alpha} \def\b{\beta}  \def\e{\epsilon}

    \def\r{\rho}
\def\s{\sigma}   \def\l{\lambda}
\def\D{\Delta}

 \newcommand{\Ocal}{{\mathcal O}}

\newcommand{\Bcal}{{\mathcal B}}

\newcommand{\red}[1]{ {\color{red} #1} }
\newcommand{\blue}[1]{ {\color{blue} #1} }

\newcommand\be{\begin{equation}}
\newcommand\ee{\end{equation}}
\newcommand{\bea}{\begin{eqnarray}}
\newcommand{\eea}{\end{eqnarray}}

\def\id{\protect{{1 \kern-.28em {\rm l}}}}

\def\D{\Delta}

\def\a{\alpha}
\def\b{\beta}

\def\e{\epsilon}

\def\s{\sigma}

\def\r{\rho}

\def\1{^{(1)}}
\def\0{^{(0)}}
\def\2{^{(2)}}

\def\id{\protect{{1 \kern-.28em {\rm l}}}}

\newcommand{\lstr}[1]{\textcolor{magenta}{{\bf [LS: #1]}}}

\def\Cincy{\small{Department of Physics, University of Cincinnati, Cincinnati, Ohio 45221, USA}}
\def\Weizmann{\small{Department of Particle Physics and Astrophysics, Weizmann Institute of Science, Rehovot 761001, Israel}}
\def\Harvard{\small{Department of Physics, Harvard University, 17 Oxford St., Cambridge, MA 02138, USA}}

\title{Galactic Condensates composed of Multiple Axion Species}

\author[a]{Joshua Eby,}
\author[b]{Madelyn Leembruggen,}
\author[c]{Lauren Street,}
\author[c]{Peter Suranyi,}
\author[c]{and L.C.R. Wijewardhana}

\affiliation[a]{\Weizmann}
\affiliation[b]{\Harvard}
\affiliation[c]{\Cincy}

\emailAdd{joshaeby@gmail.com}
\emailAdd{madelyn.leembruggen@gmail.com}
\emailAdd{streetlg@mail.uc.edu}
\emailAdd{peter.suranyi@gmail.com}
\emailAdd{rohana.wijewardhana@gmail.com}

\abstract{
Ultralight scalar dark matter has been proposed to constitute a component of dark matter, though the minimal scenarios have increasingly become constrained. In this work, we analyze scenarios where the dark matter consists of more than one ultralight boson, each with different masses. This potentially leads to formation of gravitationally-bound Bose-Einstein condensates with structures that are very different from condensates composed of a single scalar field.
By generalizing from the well-understood single-flavor case, we explore a large range of input parameters, subject to stability criteria, and determine the allowed parameter space for two-flavor condensates as a function of particle physics parameters, paying particular attention to cases where such condensates could compose galactic cores.  We also analyze single-flavor condensates subject to external gravity from massive inner bodies and find that such systems may mimic the size of galactic cores as well.
}


\arxivnumber{2002.03022}

\begin{document}
\maketitle
\flushbottom

\section{Introduction}
Ultralight axions (ULAs) are sub-eV, scalar particles that arise in string compactification~\cite{Arvanitaki:2009fg} and clockwork
theories~\cite{Kaplan:2015fuy}.  These particles could form gravitationally-bound Bose-Einstein condensates (BECs), called axion stars \cite{Kaup:1968zz,Ruffini:1969qy,Colpi:1986ye}; in models of ultralight dark matter (ULDM) \cite{Turner:1983he,Press:1989id,Sin:1992bg,Hu:2000ke,Hui:2016ltb,Lee:2017qve}, these BECs can have astrophysical sizes, as extremely small particle masses of $m \sim 10^{-22} \, \text{eV}$ imply very large Compton wavelengths $1/m \sim \text{pc}$. Simulations \cite{Schive:2014dra,Schive:2014hza,Schwabe:2016rze,Veltmaat:2016rxo,Mocz:2017wlg} suggest that these cored, high-density structures may dominate the mass in the cores of galaxies. For related reasons, ULDM condensates have been proposed to solve the small-scale structure problems of noninteracting cold dark matter (CDM) models \cite{Weinberg:2013aya,Marsh:2015wka}. However, the minimal ULDM model with $m\sim 10^{-22}$ eV has come under increased scrutiny. At present, there are constraints from the observed Lyman-$\a$ forest \cite{Irsic:2017yje,Leong:2018opi}, from stellar streams and lensed quasars \cite{Schutz:2020jox,Benito:2020avv}, as well as kinematic data in low surface-brightness galaxies \cite{Bar:2018acw,Bar:2019bqz,Safarzadeh:2019sre} (among others which are more tentative). It is interesting to ask how the model can be extended to preserve the promising features while avoiding the constraints.

In a recent paper, multi-component condensates composed of two flavors of axions with different masses and no self-interactions were proposed as models of the cores of a few specific astrophysical systems~\cite{Broadhurst:2018fei}. This is in part motivated by existing constraints on single-flavor ULDM, but also by the observation that many theories which gives rise to ULAs typically do not produce a single scalar particle, but rather hundreds of them \cite{Arvanitaki:2009fg,Kaplan:2015fuy}.  As a result, there is little reason to believe that only a single flavor of ULA would be produced in the early universe; more than one scalar may contribute to the total dark matter (DM) relic abundance. Other authors have also begun to investigate the phenomenology of models with more than one axion-like scalar \cite{Amendola:2005ad,Marsh:2013ywa,Fan:2016rda,Berman:2020kko}. 

Here, we analyze two-component condensates as a generalization of the usual one-flavor axion star as a model for galactic cores. These systems are similar to those considered in~\cite{Broadhurst:2018fei}, but in our analysis we include self-interactions, which play an important role in axion star dynamics in situations where the smallness of the interaction coupling $|\lambda|={ {m^2}/{f^2}} \ll 1$ is compensated by a large number of condensate particles $N\gg 1$ (where $f$ is the axion decay constant). For axions with attractive self-interactions, the maximum condensate mass in the single-flavor case is decreased below the Kaup bound ${M_P}^2/m$ \cite{Kaup:1968zz,Ruffini:1969qy} by the ratio $f/M_P$ \cite{Chavanis:2011zi,Chavanis:2011zm,Eby:2014fya}. In this work, self-interactions will limit the parameter space for stable condensates in multi-flavor axion models. (See \cite{Peebles:2000yy,Goodman:2000tg,Li:2013nal,Fan:2016rda,Dev:2016hxv} for previous work analyzing the effect of repulsive interactions in ULDM.)

This work is also general in two important ways. First, our analysis is general in the sense that it can be extended in a straightforward way to the case of $n_F>2$ axion flavors as well, a relevant consideration in the scenario of a true axiverse of hundreds of flavors. In this work we derive the results explicitly for $n_F=2$, leaving theories of higher multiplicity potentially for future work. Second, our results are general because they qualitatively only depend on the ratio of parameters of the two axions, namely, the mass ratio $m_r \equiv m_2/m_1$ and quartic coupling ratio $\l_r \equiv \l_2/\l_1$, where $m_i$ and $\l_i$ are the mass and quartic coupling of the $i$-th axion. While we focus on the ULDM case, where the condensates have astrophysical sizes, our analysis may be useful also for more general axion theories.

Measurements of the rotational velocities inside the central regions of many galaxies have made it possible to perform fits of the density profiles inside these galaxies using two free parameters: the core density ($d_c$) and the core radius ($R_c$) \cite{Rodrigues:2017vto}. The authors of \cite{Deng:2018jjz} analyzed condensate density-radius scaling relationships predicted, assuming that the dark matter (DM) of galaxies is dominated by condensates in various single scalar field models, and found that the observed relationships are not borne out in these models. For axion stars, the mass-radius relation is fixed by the condition of gravitational stability, which uniquely determines the relationship between $d_c$ and $R_c$, as derived by \cite{Chavanis:2011zi,Chavanis:2011zm} and used in the simulations of \cite{Schive:2014dra,Schive:2014hza}. 

The fit parameters $d_c$ and $R_c$ in \cite{Deng:2018jjz}, however, were obtained from density profile fits on entire galaxies, and hence it may not be meaningful to compare the resulting fit to the scaling relation of condensates hypothesized to comprise the \emph{core} of galaxies. As a simple example, when the axion mass is above roughly $m\gtrsim 10^{-21}$ eV, the size of the condensate predicted by simulations \cite{Schive:2014dra,Schive:2014hza} is much smaller than the cores observed in the galaxy samples considered in \cite{Deng:2018jjz}. Though axion condensates would not explain the observation of these cores, the observations are not in tension with the existence of condensates. Still, it calls into question the simplest version of ULDM as a solution to the cusp-core problem, perhaps motivating generalizations of this idea. It is in this spirit that this work was written.

In this paper we point out how the physically-relevant parameter space can be modified by additional scalar states compared to single-flavor theories and show that a wide range of behaviors are possible in theories with multiple axion flavors.  In particular, we show that when modeling galactic cores as two flavors of axion species, the physical parameter space is bounded within a range of scaling exponents between $d_c$ and $R_c$.  However, a given subset of galaxies tends to be widely distributed in the $d_c-R_c$ plane, and cannot be properly described by a simple scaling exponent.


This paper is organized as follows:  we introduce the 
standard relations for axion stars composed of a single species (Section \ref{sec:1flavor}), and generalize it to the case of two flavors 
(Section \ref{sec:mult_flavs}); we discuss the notion of a density-radius scaling relationship in the single-flavor and multi-flavor cases (Section \ref{sec:scaling}); 
we then analyze the extent of the physically-allowed parameter space by determining its boundary in a few tractable cases (Section \ref{sec:calc}), before finally sampling the full range of allowed parameters numerically, subject to stability criteria (Section \ref{sec:random}); 
and finally, we show that modified scaling relations can be obtained analytically by considering single-flavor condensates subject to external gravity by a much smaller, spherically-symmetric inner body (Section \ref{sec:analytic_ex}).  A summary and some concluding thoughts are found in Section \ref{sec:conclusion}.

There is a weak dependence throughout on several $\Ocal(1)$ numbers which depend on a choice of ansatz for the condensate wavefunction shape. For the interested reader, a brief review of the calculation of these constants can be found in Appendix \ref{app:numbers}. Further, while the main text focuses on the case of particular ULDM parameter choices, we illustrate a few generalized (dimensionless) parameters in Appendix \ref{app:MIresults}; this allows our results to be extended to other regions of axion parameter space.


We work in natural units, where $\hbar = c = 1$.

\section{Equations of motion for axion condensates} \label{sec:scaling_intro}



\subsection{Single flavor of axion} \label{sec:1flavor}

We begin with the standard case of a single axion flavor, $n_F = 1$. The simplest way to 
analyze stable solutions for generic scalar field models is to use the non-relativistic energy functional \cite{Chavanis:2011zi,Eby:2016cnq}
\begin{align} \label{Energy}
  E[\psi] = \int d^3r \left[\frac{|\nabla \psi|^2}{2m} 
	+ \frac{m}{2} \Phi_{g}\,|\psi|^2 
	\pm  \frac{|\l|}{16\,m^2}\left|\psi\right|^4 \right],
\end{align}
where $\psi$ is the classical wavefunction, $\Phi_g$ is the Newtonian gravitational potential determined by the field, and $\l$ is some self-interaction coupling. Here, the plus (minus) sign corresponds to repulsive (attractive) self-coupling.  The wavefunction is normalized as $\int d^3r |\psi|^2 = M/m = N$, where $N$ ($M$) is the total particle number (mass) of the condensate. In this work, we analyze only extremely nonrelativstic configurations in which the binding energy is small and the assumption of particle number conservation is appropriate. For discussions of relativistic corrections, see \cite{Braaten:2016kzc,Eby:2017teq,Namjoo:2017nia,Braaten:2018lmj,Eby:2018ufi}.

Given an input profile for the wavefunction, one can compute each term in Eq. (\ref{Energy}) directly; the resulting energy per particle is given by 
\begin{align} \label{energy0}
 \frac{E(\sigma)}{m\,N} &= \frac{D_2}{2\,C_2}\frac{1}{m^2\,\sigma^2} - \frac{B_4}{2\,C_2{}^2}\frac{m^2}{M_P^2}\frac{N}{m\,\sigma} \pm \frac{C_4}{16\,C_2{}^2}\frac{|\l|}{m^3} \frac{N}{\sigma^3} \nn \\
 				&= \frac{a}{m^2\,\sigma^2} - \frac{m^2}{M_P^2}\frac{b\,N}{m\,\sigma} \pm \frac{|\l|}{m^3} \frac{c\,N}{\sigma^3},
\end{align}
where $M_P=1.2\times10^{19}$ GeV is the Planck mass and $\s$ is a variational parameter proportional to the radius of the condensate. In the upper line we use the notation of \cite{Eby:2016cnq,Eby:2018dat}, where
\begin{equation} \label{eq:BCD}
 B_4 \equiv \frac{8\pi}{\s^5\,|\psi(0)|^4} \int_0^\infty dr\,r\,|\psi(r)|^2\,\int_0^r\,d^3s\,|\psi(s)|^2, \qquad 
 C_k \equiv \frac{\int_0^\infty d^3r\,|\psi(r)|^k}{\s^3\,|\psi(0)|^k}, \qquad 
 D_2 \equiv \frac{\int_0^\infty d^3r\,|\psi'(r)|^2}{\s^2\,|\psi(0)|^2},
\end{equation}
but in the lower line we define dimensionless constants
\begin{equation} \label{eq:abc}
 a \equiv \frac{D_2}{2\,C_2}, \qquad b \equiv \frac{B_4}{2\,C_2{}^2}, \qquad c \equiv \frac{C_4}{16\,C_2{}^2},
\end{equation}
for simplicity. We also define 
\begin{equation} \label{eq:A2}
 A_2 = \frac{\int d^3r\,r\,|\psi(r)|^2}{\s^4\,|\psi(0)|^2}
\end{equation}
for future use.
 These constants vary slightly depending on the precise shape of the wavefunction, but do not affect the general scaling behavior; the numerical values for a Gaussian wavefunction are $a = 3/4$, $b = 1/\sqrt{2\pi}$, and $c = 1/(32\pi\sqrt{2\pi})$ (see Appendix \ref{app:numbers}).\footnote{For a comparison of other approximations to the wavefunction shape, see \cite{Eby:2018dat} and references therein.}  
 
 Minimizing the energy of Eq.~\eqref{energy0} with respect to $\sigma$ gives the stable minimum energy solution,
\begin{equation} \label{RofN}
 \sigma_d(N) = \frac{M_P^2}{m^3}\frac{a}{b\,N}\left[1 + \sqrt{1 \pm \left(\frac{N}{\tilde{N}}\right)^2}\right],
\end{equation}
where $\tilde{N} = (M_P / m\,\sqrt{|\l|})(a / \sqrt{3\,b\,c})$ is the critical particle number for condensates with attractive self-interactions; for $N > \tilde{N}$, these condensates are unstable to collapse \cite{Chavanis:2016dab,Eby:2016cnq}. Note that in Eq. (\ref{RofN}), the $\pm$ sign corresponds to $\pm |\lambda|$ in Eq. (\ref{Energy}). It should also be mentioned that $\sigma_d$ is not equal to the core radius $R_c$ of the condensate, though the two quantities are proportional; the constant of proportionality is $\Ocal(1)$, and depends on the shape of the wavefunction $\psi$, as well as the exact definition of $R_c$. For the purpose of this work, we will treat $\s_d$ and $R_c$ as equivalent, which is true up to coefficients of $\Ocal(1)$. For mixed flavor condensates, we will also treat the central density $m\,|\psi(0)|^2$ and the core density $d_c$ as interchangeable, as the results are unchanged given the level of accuracy at which we work.  However, note that the precise definition of the core density will play an important role in Section \ref{sec:analytic_ex}, where we analyze a condensate subject to an external gravitational source.

Following~\cite{Eby:2018dat} we introduce the dimensionless radius, $\rho$, and particle number, $n$, as
\begin{equation} \label{eq:rescale}
 \s = \sqrt{|\lambda|}\frac{M_P}{m^2}\,\rho, \qquad N = \frac{M_P}{m\,\sqrt{|\lambda|}}\,n.
\end{equation}
Then Eq. (\ref{RofN}) in dimensionless units reduces to
\begin{equation} \label{single}
\rho_d(n) = \frac{a}{b \, n} \left( 1 + \sqrt{1 \pm \left(\frac{n}{\tilde n}\right)^2} \right),
\end{equation}
where
\begin{equation} \label{eq:tildes}
 \tilde{n} = \frac{a}{\sqrt{3\,b\,c}}, \qquad \tilde{\r} \equiv \r_d(n=\tilde{n}) = \sqrt{\frac{3\,c}{b}}
\end{equation}
are the reduced critical particle number and corresponding radius. For the remainder of this work, the subscript $d$ will be omitted for notational simplicity, in which case $\s$ ($\r$) will represent the radius (rescaled radius) at which the energy is minimized.

\subsection{Multiple axion flavors} \label{sec:mult_flavs}

We now consider condensates formed from $n_F > 1$ species of axion.
We will assume that each flavor of axion can interact with each other only gravitationally, and that each has an attractive self-interaction; the $i$-th axion flavor will have mass $m_i$ and interaction coupling $\l_i$, which are all independent parameters in principle. It is possible to include some point-like interaction coupling between different flavors as well; in such a scenario, the equations of motion would be modified but also there could be scattering processes whereby 2 heavier axions annihilate to produce 2 lighter ones. This is an interesting scenario, worthy of further exploration; in this work, we neglect this possibility for simplicity.

We approximate each axion component as spherically symmetric with similar wavefunction shapes, implying that they differ only in radius and normalization. As in \cite{Eby:2016cnq,Eby:2018dat}, we parameterize the profiles by
\be \label{eq:wavefn}
\psi_i(r)=\sqrt{\frac{N_i}{C_2\,\sigma_i^3}}\,F(r/\s_i) \equiv \psi_i(0)\,F(r/\s_i),
\ee
where we have defined a dimensionless function $F(\xi)$ determined by the wavefunction shape with ${F(0) = 1}$.
The normalization of the wavefunction determines the number of axions of each flavor
\be
N_i=\int d^3r {|\psi_i(r)|}^2,
\ee
which in turn determines the total condensate mass
\be
M=\sum_{i=1}^{n_F} m_i\, N_i.
\ee
The gravitational potential is similarly just the sum of contributions from each condensate,
\begin{align} \label{eq:PhiG}
\Phi_g &= \frac{1}{M_P^2}\int \frac{{d^3}r'}{|\vec{r}-\vec{r}\,'|}\sum_{j=1}^{n_F} m_j {|\psi_j(r')|}^2.
\end{align}
The energy functional for such a multi-axion condensate is found by generalizing Eq. \eqref{Energy} as 
\begin{equation}\label{E}
E[\{\psi_1,...,\psi_{n_F}\}]=\int d^3r\,\sum_{i=1}^{n_F}\left[\frac{{|\nabla\psi_i|}^2}{2\,m_i} + \frac{m_i}{2}\Phi_g{|\psi_i|}^2 - \frac{\lambda_i}{16\,m_i^2}|\psi_i|^4\right],
\end{equation}
where we used the minus sign and set $\l_i>0$ in the self-interaction term, so that the self-interactions are all attractive.

Let us consider now the simplest nontrivial case of $n_F = 2$ axion flavors. It is straightforward to directly compute each term in the energy functional of Eq. \eqref{E}: 
\begin{equation} \label{E2}
 E(\s_1,\s_2) = \sum_{i=1}^2\left(\frac{a\,N_i}{m_i\,\s_i^2} - \frac{b\,m_i^2\,N_i^2}{M_P^2\,\s_i}   - \frac{c\,\l_i\,N_i^2}{m_i^2\,\s_i^3}\right) 
 		- \frac{2\sqrt{2}\,b\,m_1\,m_2\,N_1\,N_2}{M_P^2\,\sqrt{\s_1^2+\s_2^2}},
\end{equation}
which is just a sum of separate energy functionals for the two flavors of axion, plus a cross term from the gravitational potential which couples them.
Again rescaling the radius and particle number using Eq. (\ref{eq:rescale}), as
\begin{equation} \label{eq:rescale2}
 \s_i = \sqrt{\lambda_i}\frac{M_P}{m_i^2}\,\rho_i, \qquad N_i = \frac{M_P}{m_i\,\sqrt{\lambda_i}}\,n_i,
\end{equation} 
we find
\begin{equation} \label{Eres}
 \frac{\l_2^{3/2}\,M_P}{m_1^2} E = \l_r^{3/2}\left(\frac{a\,n_1}{\r_1^2} - \frac{b\,n_1^2}{\r_1} - \frac{c\,n_1^2}{\r_1^3}\right) + m_r^2\left(\frac{a\,n_2}{\r_2^2} - \frac{b\,n_2^2}{\r_2} - \frac{c\,n_2^2}{\r_2^3}\right)
 				- \frac{2\sqrt{2}\,b\,n_1\,n_2\,m_r^2\,\l_r}{\sqrt{m_r^4\,\r_1^2 + \l_r\,\r_2^2}},
\end{equation}
where we define the ratios $m_r \equiv m_2/m_1$ and $\lambda_r \equiv \lambda_2/\lambda_1$. Without loss of generality, we will set $m_r \geq 1$.

Variation of the energy functional in Eq. (\ref{Eres}) with respect to the two variational parameters $\r_1$ and $\r_2$ leads to a set of equations of motion (EOM) for the condensate, of the form
\begin{align} 
 0 &= 3\,c\,n_1 - 2\,a\,\r_1 + b\,n_1\,\r_1^2 + \frac{2\sqrt{2}\,b\,n_2\,\r_1^5}{\left(m_r^4\,\r_1^2 + \l_r\,\r_2^2\right)^{3/2}}\frac{m_r^6}{\sqrt{\l_r}}, \label{eq:eom1} \\
 0 &= 3\,c\,n_2 - 2\,a\,\r_2 + b\,n_2\,\r_2^2 + \frac{2\sqrt{2}\,b\,n_1\,\r_2^5}{\left(m_r^4\,\r_1^2 + \l_r\,\r_2^2\right)^{3/2}} \l_r^2. \label{eq:eom2}
\end{align}
In contrast to the single-flavor case, where the radius $\rho$ can be determined algebraically, in this case solutions for $\r_1$ and $\r_2$ form a two-dimensional space, whose extent will depend on
$n_1$, $n_2$, and the ratios of particle physics inputs $m_r$ and $\l_r$. Note also that  the generalization of Eqs. (\ref{eq:eom1}-\ref{eq:eom2}) to more than two flavors is straightforward.

In the numerical results 
of later sections, we will use a Gaussian profile for $F(r/\s_i)$ (see Appendix \ref{app:numbers}), though other ans\"atze do not change the results appreciably. The numerical coefficients change by $\Ocal(1)$ for different profile shapes, but this does not affect the 
general behavior. Exact solutions can be obtained, in principle, by solving the set of Schr\"odinger equations for $n_F$ wavefunctions $\psi_1,\psi_2,...,\psi_{n_F},$ coupled through the gravitational potential of Eq.~\eqref{eq:PhiG}. This procedure gives more precise results, but should not change our conclusions qualitatively.

\section{Density-radius scaling relationship} \label{sec:scaling}

\subsection{Single flavor of axion}

The authors of ~\cite{Deng:2018jjz} pointed out that the core density and core radius in a large observational sample of galaxies \cite{Rodrigues:2017vto} can be fit to a power law of the form
\begin{equation} \label{eq:scaling}
d_c {{R_c}^{\beta}}= constant
\end{equation}
with $\beta \sim 1$. Using Eq. (\ref{single}) we can derive the predicted scaling relationship, if these cores consist of massive axion stars; in that case we have
\be\label{core_density}
d_c=m\,|\psi(0)|^2\sim \frac{m\,N}{\s^3}, 
	\qquad R_c\sim \sigma. 
\ee
For example, if $N \ll \tilde{N}$ then the self-interactions essentially decouple, and one finds $M \propto R_c^{-1}$; this implies $d_c \propto R_c^{-4}$, or $\beta=4$. For attractive interactions and $\tilde{N}/2 \lesssim N \lesssim \tilde{N}$, Eq. (\ref{single}) also implies $\beta\simeq 3$,\footnote{One can obtain this relation by taking $N \rightarrow \tilde{N} - \epsilon$ in Eq. (\ref{single}) and expanding in $|\e| \ll \tilde{N}$, 
where $\epsilon$ is 
some small number that is always positive for attractive self-interactions, and can be positive or negative for repulsive self-interactions.} a scaling relation which is less steep but still far from $\b=1$.
On the other hand, if $\l>0$ then we can have $N \gg \tilde{N}$ (known as the Thomas-Fermi regime \cite{Boehmer:2007um,Eby:2018dat}), in which case the relation is $R_c(M) = constant$; then $M(R_c)$ is a vertical line, which also implies that $d_c(R_c)$ is a vertical line (which can be represented as $\beta \to \infty$). One can also investigate the scaling relations for configurations consisting of complex scalar particles. Using similar arguments, the authors of \cite{Deng:2018jjz} found $\beta < 0$ for Q-balls, $\beta = 2$ in the strong gravity regime, and $\beta = 2p/(p-1)$ for a general polytrope of index $p$. None of these cases appear to reproduce $\beta \simeq 1$.

In this discussion, we have assumed something about the nature of the galaxy core, namely (1) that it is DM dominated (any baryonic effect is neglected), and (2) the DM in the core is a single condensate of minimum energy.  Each of these assumptions could be relaxed: the core may consist of some admixture of DM and baryons, or may be instead a
condensate composed of multiple species. In the former case, the condensate may be affected by the gravitational potential of the baryons, or (as is the case for $m\gtrsim 10^{-21}$ eV) there may be a small condensate contained inside of a larger core whose size is determined by some other dynamical process. In the latter case, we show below that with the addition of multiple axionic species, one is not restricted to the simple single flavor scaling relation above.



\subsection{Two Flavors of Axion} \label{sec:toy}

To determine the scaling relationship for the case of two flavors, we define the core density and core radius by
\begin{align}\label{rhocR}
d_c&=m_1|\psi_1(0)|^2+m_2|\psi_2(0)|^2 = \frac{1}{C_2}\left[\frac{m_1\,N_1}{\s_1^3} + \frac{m_2\,N_2}{\s_2^3}\right] = 
	\frac{1}{C_2}\frac{m_1^6}{M_P^2 \lambda_1^2}\left[\frac{n_1}{\rho_1^3}+\frac{m_r^6}{\lambda_r^2}\frac{n_2}{\rho_2^3}\right],\nonumber\\
R_c&= \frac{\int d^3r\,r\,\left(|\psi_1|^2+|\psi_2|^2\right)}{\int d^3r\,\left(|\psi_1|^2+|\psi_2|^2\right)} 
	= A_2 \frac{N_1\,\s_1 + N_2\, \s_2}{N_1 + N_2}
	= A_2 \frac{M_P \sqrt{\lambda_2}}{m_2^2} \left[ \frac{m_r^3\,n_1\,\rho_1+n_2\,\rho_2}{m_r\,\sqrt{\lambda_r}\,n_1+n_2}\right].
\end{align}
where $C_2$ and $A_2$ are given in Eq.~\eqref{eq:BCD} and \eqref{eq:A2}, respectively.
In what follows, we will use Eq.~\eqref{rhocR} to determine the range of allowed scaling exponents $\beta$ in the two-flavor case using simple analytical models. This has implications in the context of diversity in rotation curves across the wide range of observed galaxies \cite{Kaplinghat:2019dhn}.

It will be instructive to consider a simple toy model.
We simplify the previous discussion by neglecting the interaction between the two axions through gravity. In that case each condensate is independent of the other, and one may use the standard scaling relations $n_i \propto \rho_i^{-1}$ for $i\in\{1,2\}$, as shown in Eq.~\eqref{single}. Then Eq. (\ref{rhocR}) simplifies to
\begin{align}\label{rhocRsimple}
d_c&\varpropto\frac{1}{\rho_1^4}+\frac{m_r^6}{\lambda_r^2}\frac{1}{\rho_2^4},\nonumber\\
R_c&\varpropto \left(\frac{m_r\,\sqrt{\lambda_r}}{\rho_1}+\frac{1}{\rho_2}\right)^{-1}.
\end{align}
Interestingly, the contribution of each condensate is equal in determining $d_c$ and $R_c$ at two particular values of the rescaled radii: $\r_2/\r_1 = m_r^{3/2}/\sqrt{\l_r}$ and $\r_2/\r_1 = (m_r\,\sqrt{\l_r})^{-1}$.
Given our definition that $m_r\geq 1$, this defines three distinct regions in the range of allowed ratios $\rho_2\,/\rho_1$:
 \begin{align}
I:\,\,&\frac{ \rho_2}{\rho_1}>\frac{m_r^{3/2}}{\sqrt{\lambda_r}},\nonumber\\
II: \,\,&\frac{m_r^{3/2}}{\sqrt{\lambda_r}}>\frac{ \rho_2}{\rho_1}>\frac{1}{m_r\sqrt{\lambda_r}},\nonumber\\
III:\,\,&\frac{ \rho_2}{\rho_1} < \frac{1}{m_r\sqrt{\lambda_r}}.
 \end{align}
 In a rough approximation, 
 we omit subdominant terms in Eq.~\eqref{rhocRsimple}, and find the following behaviors in the three regions:
 \begin{align} 
 I:\,\,& d_c \varpropto\frac{1}{\rho_1^4},\,\, R_c\varpropto \rho_1,\nonumber\\
 II:\,\,& d_c \varpropto\frac{1}{\rho_2^4},\,\,R_c\varpropto \rho_1,\nonumber\\
 III:\,\,&d_c\varpropto\frac{1}{\rho_2^4},\,\,R_c\varpropto \rho_2.
 \end{align}
 Now  the scaling exponent $\beta=4$ in regions I and III, reproducing the single-flavor result. However, if $m_r \gg 1$, then in region II $d_c$ is independent of $R_c$, implying that the $\beta \to 0$. This can be understood by the fact that, in this approximation, the two condensates are completely independent of one another. As a result, it is possible for one condensate to dominate the core density (determined at small radial distances $r$), and the other to dominate the total radius (determined at large $r$).
 
Of course, in reality, the gravitational interaction between flavors will affect the boundaries between the regions.  Near the boundaries, the simple scaling relations do not hold, and so we expect that the average scaling parameter lies somewhere between $0\lesssim\beta\lesssim4$.  Further proof of that fact is given below when we discuss analytic estimates of the scaling parameter.

This very simple example also illustrates that the mass ratio $m_r$ is potentially more important in determining the scaling behavior than the coupling ratio $\l_r$, as the latter appears with a small exponent. This is consistent with the standard claim that self-interactions become important only near the boundary of stability (the maximum stable mass discussed in Section \ref{sec:1flavor}).
 
\section{Boundaries of the physical region} \label{sec:calc}

It is necessary to emphasize the generic problem with predictions based on a theory containing more than one boson flavor. Consider the scaling exponent described in the previous section. For single-axion theories, this exponent for a condensate core can be unambiguously determined; this owes to the fact that $\rho$ and $n$, appearing in Eq. (\ref{core_density}), have a one-to-one relationship determined by the EOM, so the expressions for $d_c$ and $R_c$ can both be expressed by a single variable, say $n$. Then one can eliminate the variable $n$ and express $d_c$ directly in terms of $R_c$, and one arrives at the scaling relation. In the case of two bosons, we can use the EOM of Eqs. (\ref{eq:eom1}-\ref{eq:eom2}) to eliminate two variables, say $n_1$ and $n_2$.  However, the relationship between $\rho_1$ and $\rho_2$ remains indeterminate. 


To analyze the behavior of two-flavor condensates, we need to determine the physically admissible region $\Bcal$ in the space of points $(\r_1,\r_2$), which reduces to the problem of finding a corresponding boundary function $B(\r_1,\r_2)=0$. Now, in general the boundary function $B(\rho_1,\rho_2)$ cannot be determined analytically.  Therefore, we will follow two alternative routes: first, we find the boundary function $B(\r_1,\r_2)$ for simple limits of $m_r$ (this section); 
then, in the full theory, we generate a large ensemble of random points in the first quadrant of the $(\rho_1,\rho_2)$ plane and impose physical requirements on each point, thereby defining the allowed region $\Bcal$ (Section \ref{sec:random}). 
 
 At each input point of $\rho_1,\rho_2 >0$, we impose the following requirements:
 \begin{itemize}
  \item[1.] $n_1>0$ and $n_2>0$ as determined by Eqs. (\ref{eq:eom1}-\ref{eq:eom2});
  \item[2.] The trace $\text{TR} \equiv {\rm Tr}\left[\frac{\partial^2 E}{\partial \r_i\partial \r_j}\right] > 0$;
  \item[3.] The determinant $\text{DET} \equiv {\rm Det}\left[\frac{\partial^2 E}{\partial \r_i\partial \r_j}\right] > 0$.
 \end{itemize}
The first requirement is straightforward, as a randomly-selected pair $\r_1,\r_2>0$ will not necessarily satisfy the physical condition that $n_1,n_2>0$ in the EOM. The second and third are requirements on the matrix of second derivatives of the energy functional; the positivity of the trace and determinant ensure the stability of a configuration under small perturbations along $\r_1$ and $\r_2$. We keep points only if they satisfy all of the above requirements; they define the physical region $\Bcal$ and the corresponding boundary function $B(\rho_1,\,\rho_2)$.  

Before moving to detailed calculations, in this section we analyze a number of simple limits in which the boundary function $B(\r_1,\r_2)$ can be determined analytically.

 \subsection{Boundary region for $m_r=1$, $\lambda_r = 1$}
 
The simplest limit to analyze is $m_r=1$, $\lambda_r = 1$, which is a two-flavor theory with identical masses and couplings (we assume the existence of some additional quantum number that distinguishes the two species). To find the boundary function $B(\rho_1,\,\rho_2)$ in this case, consider the further simplifying assumption $\rho_1=\rho_2=\rho$.  We obtain the following expressions, required to analyze the requirements above:
 \begin{align}\label{eq:constraints1}
  n_1 = &\,n_2=\frac{2\,\tilde{n}\,(\r/\tilde\r)}{1+2\,(\rho/\tilde{\r})^2},\nonumber\\
    \text{TR}&= \frac{8\,a^2}{3\,c}\frac{\frac{7}{2}(\r/\tilde\r)^2 - 1}{\r^3\left(1 + 2(\r/\tilde{\r})^2\right)^2},\nonumber \\
    \text{DET}&=\frac{16\,a^4}{9\,c^2}\frac{\left(2(\r/\tilde\r)^2 - 1\right)\left(5(\r/\tilde\r)^2 - 1\right)}{\r^6\left(1 + 2(\r/\tilde\r)^2\right)^4},
  \end{align}
  where we have used the definitions of Eq. (\ref{eq:tildes}). The quantities TR and DET are both positive only if $\r > \tilde{\r}/\sqrt{2} = \sqrt{3\,c / 2\,b}$.
  
 Consider now a small difference in radii $\delta \rho=\rho_1-\rho=\rho-\rho_2 \ll 1$. Two boundaries of $\mathcal B$ are determined by the zeros of the product $n_1\,n_2$ which violate the requirement that $n_1,n_2>0$.  In this case, we can identify $B(\r_1,\r_2) = n_1\,n_2\Big|_{\rm boundary} = 0$.   Expanding $n_1\,n_2$  to $\Ocal(\delta \rho^2)$, we obtain 
\begin{equation}
 n_1\,n_2 \simeq \frac{4\,a^2}{9\,c^2\left(1+2(\r/\tilde\r)^2\right)^2} \left[\r^2 - \delta\r^2\,\frac{1 + 5(\r/\tilde\r)^2 + 35(\r/\tilde\r)^4 + 50(\r/\tilde\r)^6}{\left(1+2(\r/\tilde\r)^2\right)}\right].
\end{equation}
Then it follows that $\mathcal B$ is bounded by the lines
\begin{equation}
 \delta\r \simeq \pm \frac{\r \sqrt{1 + 2(\r/\tilde\r)^2}}{\sqrt{1 + 5(\r/\tilde\r)^2 + 35(\r/\tilde\r)^4 + 50(\r/\tilde\r)^6}}.
\end{equation}
At the threshold of stability $\r = \tilde{\r}$, we find $\delta\rho=0.18\tilde{\r}$; however, the range of $\delta\rho$ decreases rapidly at larger values of $\rho>\tilde{\r}$, implying that the physical domain for $\rho_1$ and $\rho_2$ rapidly shrinks to a very narrow strip around $\rho_1=\rho_2$.  


\subsection{Boundary region for $m_r\gg1$} \label{sec:mtoinf}

A second simplifying limit is $m_r \to \infty$, as Eqs. (\ref{eq:eom1}-\ref{eq:eom2}) simplify considerably,
so much so that they can be directly solved for  $\rho_1$ and $\rho_2$. We obtain on the stable branches
\begin{align}
\rho_1&= \frac{a}{b}\left(n_1 + \frac{2\sqrt{2}\,n_2}{\sqrt{\lambda_r}}\right)^{-1}\left[1 + \sqrt{1 - \frac{n_1}{\tilde n^2}\left(n_1 + \frac{2\sqrt{2}\,n_2}{\sqrt{\lambda_r}} \right)}\right],\label{eq:infeq1} \\
\rho_2&= \frac{a}{b\,n_2} \left[1 + \sqrt{1 - \left(\frac{n_2}{\tilde n}\right)^2}\right] \label{eq:infeq2}.
\end{align}
For $\rho_2$, one reproduces the single-condensate case, given in Eq. (\ref{single}). On the other hand, $\r_1$ interpolates roughly between its single condensate result $\r_1 \simeq a/(b\,n_1)$ (when $n_2 \ll n_1 \ll \tilde{n}$) and what might be called the ``gravitational atom" limit $\r_1 \sim a\,\sqrt{\l_r}/(b\,n_2)$ (when $n_1 \ll n_2 \ll \tilde{n}$); the latter occurs roughly when condensate 2 is supported by the gravitational potential of condensate 1.

We can use Eq. (\ref{eq:infeq2}) to eliminate $n_2$ in the expression for $\r_1$ in Eq. (\ref{eq:infeq1}):
\begin{equation}
 \r_1 = \frac{a}{b} \frac{n_1}{\tilde{n}^2} \left[1 - \sqrt{1 - \left(\frac{n_1^2}{\tilde{n}^2} + \frac{4\,\sqrt{2}\,(\r_2/\tilde\r)}{\sqrt{\lambda_r}\left(1 + (\r_2/\tilde\r)^2\right)}\,\frac{n_1}{\tilde n}\right)}\right]^{-1}. \label{eq:r1_largemr}
\end{equation}
As a function of $n_1$, $\r_1$ is maximized at $n_1 \to 0$ and minimized when the expression under the above square root vanishes; the latter gives a value for $n_1$ of
\begin{equation} \label{eq:n1max}
n_1 = \tilde{n}\left[-\frac{2\sqrt{2}\,(\r_2/\tilde{\r})}{\sqrt{\lambda_r}(1 + (\r_2/\tilde \r)^2)} + \sqrt{1 + \frac{8(\r_2/\tilde \r)^2}{\lambda_r(1 + (\r_2/\tilde \r)^2)^2}}\right].
\end{equation}
The values $n_1=0$ and $n_1$ from Eq.~(\ref{eq:n1max}) define the following 
limits on $\rho_1$ as  functions of $\rho_2$:
\be\label{eq:rho1limits}
-\frac{2\sqrt{2}\,(\r_2/\tilde\r)}{1 + (\r_2/\tilde \r)^2} + \frac{\sqrt{1 + 2(1 + 4\l_r)(\r_2/\tilde \r)^2 + (\r_2/\tilde\r)^4}}{1 + (\r_2/\tilde \r)^2} < \frac{\rho_1}{\tilde{\r}} < \frac{\sqrt{\lambda_r}}{2\sqrt{2}}\frac{(1 + (\r_2/\tilde \r)^2)}{(\r_2/\tilde \r)}.
\ee
In the next section, we will confirm numerically by explicit sampling of physical parameters that for large $m_r$, these boundaries approximately agree with the physically admissible region $\mathcal{B}$.

Next, we analyze the relationship between the central density and the core radius. 
In the large $m_r\gg 1$ limit, the scaling relations of Eq. \eqref{rhocR} simplify; solving Eqs. \eqref{eq:infeq1} and \eqref{eq:infeq2} for $n_1$ and $n_2$ and substituting into Eq. (\ref{rhocR}) we obtain,
\begin{align}\label{rhocinf2}
d_c& \approx \frac{1}{C_2} \frac{m_1^6}{M_P^2 \lambda_1^2} \frac{m_r^6}{\lambda_r^2}\frac{n_2}{\rho_2^3}
		=  \frac{1}{C_2} \frac{m_1^6}{M_P^2 \lambda_1^2} \frac{m_r^6}{\lambda_r^2} \frac{2\,a}{3\,c\,\r_2^2}\frac{1}{(1+(\r_2/\tilde\r)^2)},\nonumber\\
R_c& = A_2 \frac{M_P \sqrt{\lambda_2}}{m_2^2} \frac{m_r^3\,n_1\,\rho_1}{m_r \sqrt{\lambda_r} \,n_1 + n_2}
		\approx A_2 \frac{M_P \sqrt{\lambda_2}}{m_2^2}\frac{m_r^2\, \r_1}{\sqrt{\lambda_r}} \left(1 - \frac{n_2}{m_r \sqrt{\lambda_r}\,n_1}\right) 
\nn \\
		&= A_2 \frac{M_P \sqrt{\lambda_2}}{m_2^2} \frac{m_r^2}{\sqrt{\lambda_r}} \left[\r_1 - \frac{(1 + (\r_1/\tilde\r)^2)}{(1 + (\r_2/\tilde\r)^2)\sqrt{\lambda_r} - 2\sqrt{2}(\r_1\,\r_2/\tilde\r^2)}\frac{\r_2}{m_r}\right],
\end{align}
where we computed $R_c$ to leading order in $1/m_r$. This is consistent with the rest of derivation above, because the input equations of motion (\ref{eq:eom1}-\ref{eq:eom2}) contain terms of $\Ocal(1/m_r^6)$, which are much more suppressed in the $m_r \gg 1$ limit. However, in the true $m_r \to \infty$ limit, the scaling exponent $\beta \to 0$, since $d_c$ depends only on $\r_2$ whereas $R_c$ depends only on $\r_1$.

We can obtain a range of scaling exponents in the large, but finite, limit of $m_r \gg 1$. Since  in Eq.~(\ref{rhocinf2}) $d_c$ depends only on $\rho_2$, it is convenient to look at the dependence of $R_c$ as a function of $\rho_2$.  We obtain two limiting expressions for $R_c$ by replacing $\rho_1$ by its values at the lower and upper bounds in Eq.~(\ref{eq:rho1limits}).  At the upper bound of $\rho_1$, 
\be
R_{\rm upper}\propto \rho_2,
\ee
and we obtain exactly the same scaling relation as in the single flavor case, with scaling exponent $\beta\simeq 4$.\footnote{Strictly speaking, the upper bound of $\r_1$ is obtained at $n_1\to 0$, but at this point the second term in Eq.~(\ref{rhocinf2}) diverges. It is sufficient to approach this this value of $\r_1$ until the second term dominates over the first, and the scaling exponent $\beta \to 4$. This is also easily obtained directly from Eq.~(\ref{rhocR}) when $n_1\to 0$.}  However, at the lower bound of $\rho_1$ we obtain a complicated expression, which has a form
\be
R_{\rm lower} \propto \tilde\r + \Delta \r(\r_2),
\ee
where $\Delta \r$ vanishes at $\rho_2=0$ and approaches $\tilde\r$ at large $\rho_2$ (independently of $m_r$).  Thus to a good approximation, the small-$\r_1$ boundary of  $\mathcal B$  is independent of $\rho_2$. This implies the independence of $d_c$ from $R_c$, or in other words a scaling exponent of $\beta\to0$. Thus, depending on the allowed (or observed) values of $\r_1$, the sampling region will be bounded within the range $0\lesssim\beta\lesssim4$. 


Now we must stress again that without further information concerning the  distribution of $\rho_1$ and $\rho_2$, one cannot have unambiguous information about the scaling exponent $\beta$.  Assuming $\Bcal$ is populated indiscriminately over the range of physically allowed parameters, the extent of the physical parameter space can be very wide along $d_c$ and $R_c$ depending on the value of $m_r$ (and to a lesser degree, $\l_r$).  This implies that in the two-flavor case, these two quantities are weakly correlated, at best.  In this case, when modeling observed galactic cores as two species of axions, the analytical estimates of the scaling exponents obtained previously can be an important tool in determining the boundaries of the physical parameter space.  However, due to the wide scatter in the $d_c - R_c$ plane obtained from numerically sampling the physical region, scaling exponents become poor descriptors of the sampling regions.  Therefore, we do not fit a scaling exponent to the numerical results of the next section.


\section{Results from sampling of the physical region} \label{sec:random}

As explained previously, the relationship between $d_c$ and $R_c$ for two axion flavors depends on the two radial parameters $(\rho_1,\rho_2)$, and so points  representing individual galaxies do not form a line but rather a two dimensional set.  
Taking an agnostic view, we weight points inside $\mathcal B$ equally and analyze the resulting parameter space. This approach has the added benefit of being more readily generalizable to other (sub-galactic scale) physical systems.

Of course, the assumption that galaxies can be sampled randomly from the physical region of $\Bcal$ is not in general valid. The actual distribution of $\rho_1$ and $\rho_2$ inside $\mathcal B$ will depend on many factors, including the mass and radius distribution of axion condensates at the time of their formation, as well as complex galactic dynamics. The situation is further complicated by limitations in experimental observations of galaxies, which can surely introduce bias in the sample.
In light of this lack of information, we will use  a homogeneous distribution of the radial parameters $\rho_i$ in $\mathcal B$. 
This work can thus be thought of as a proof of concept determination of the range of physically admissible two-flavor condensates, rather than a prediction of any particular set of model parameters. Because we focus on galactic-scale condensates, in many of the figures below we have included a rectangle which roughly corresponds to the density and radius scales of galaxies currently observed; this may help to guide the eye of the reader to the most physically-relevant region of parameter space.

In addition to assumptions about sampling, the physical region will depend on three important inputs: (1) the value of $m_r$, which we will see determines the ``width" of the physical region in the plane of $\r_1$ and $\r_2$; (2) the value of $\l_r$, setting the stability boundary for the two condensates; and (3) the large-radius cutoff of the physical region, which we call $\rho^{\rm max}$. The first two have already been discussed at length, and in any case only depend on the parameter choices of the multi-axion theory. 
On the other hand, it is not so clear how to choose the large-radius cutoff in the analysis. 
Here we extend $\r^{\rm max}$ to very large values to show a large range in the stable parameter space, even if the final results in some regions are highly unphysical; in reality physical cutoffs should also be provided by the formation history of such axion condensates, as well as observational limitations.

Our sampling procedure is as follows: as a first step we generate random pairs, $\rho_1>0$ and $ \rho_2>0$
on a grid of $3\times 10^4$ points in the $(\r_1,\r_2)$ plane defined by $\r_1,\r_2 \in \left[\r^{\rm min},\r^{\rm max}\right]$; we set the limit $\r^{\rm max}\simeq 10^5-10^6$, an arbitrary value which is large enough to extend into the physically relevant range, and $\r^{\rm min} \simeq 0.005$, a value small enough to reach the boundary of stability at low $\r_1,\r_2$. We then test each point against the physical requirements described in Section \ref{sec:calc}: $n_1>0$, $n_2>0$ (as obtained from Eqs. (\ref{eq:eom1}-\ref{eq:eom2})), $\text{TR}>0$ and $\text{DET}>0$, discarding as unphysical any points that do not satisfy all four constraints.
In practice, the resulting range of rescaled particle numbers spans more than 7 orders of magnitude, where the upper limit for both $n_1$ and $n_2$ is roughly $\tilde{n}\simeq\Ocal(10)$. Finally, we determine both $d_c$ and $R_c$ for each pair using Eq. \eqref{rhocR}. 

In this section, we report the physical results using two sets of benchmark inputs: we focus mostly on $m_1 = 10^{-22}\, \text{eV}$ and $\l_1=10^{-94}$ (Benchmark 1); but also give results for $m_1 = 10^{-19}$ eV and $\l_1 = 10^{-88}$ (Benchmark 2). Benchmark 1 corresponds roughly to the standard ULDM parameters, where $\l = (m/f)^2$ with $f=10^{16}$ GeV, whereas Benchmark 2 represents a set of ULDM parameters that is experimentally mostly unconstrained. For the purpose of greater generality we also present the corresponding dimensionless quantities in Appendix \ref{app:MIresults}.

\subsection{Core radii}

\begin{figure}[t]
\centering 
 \includegraphics[scale=0.245]{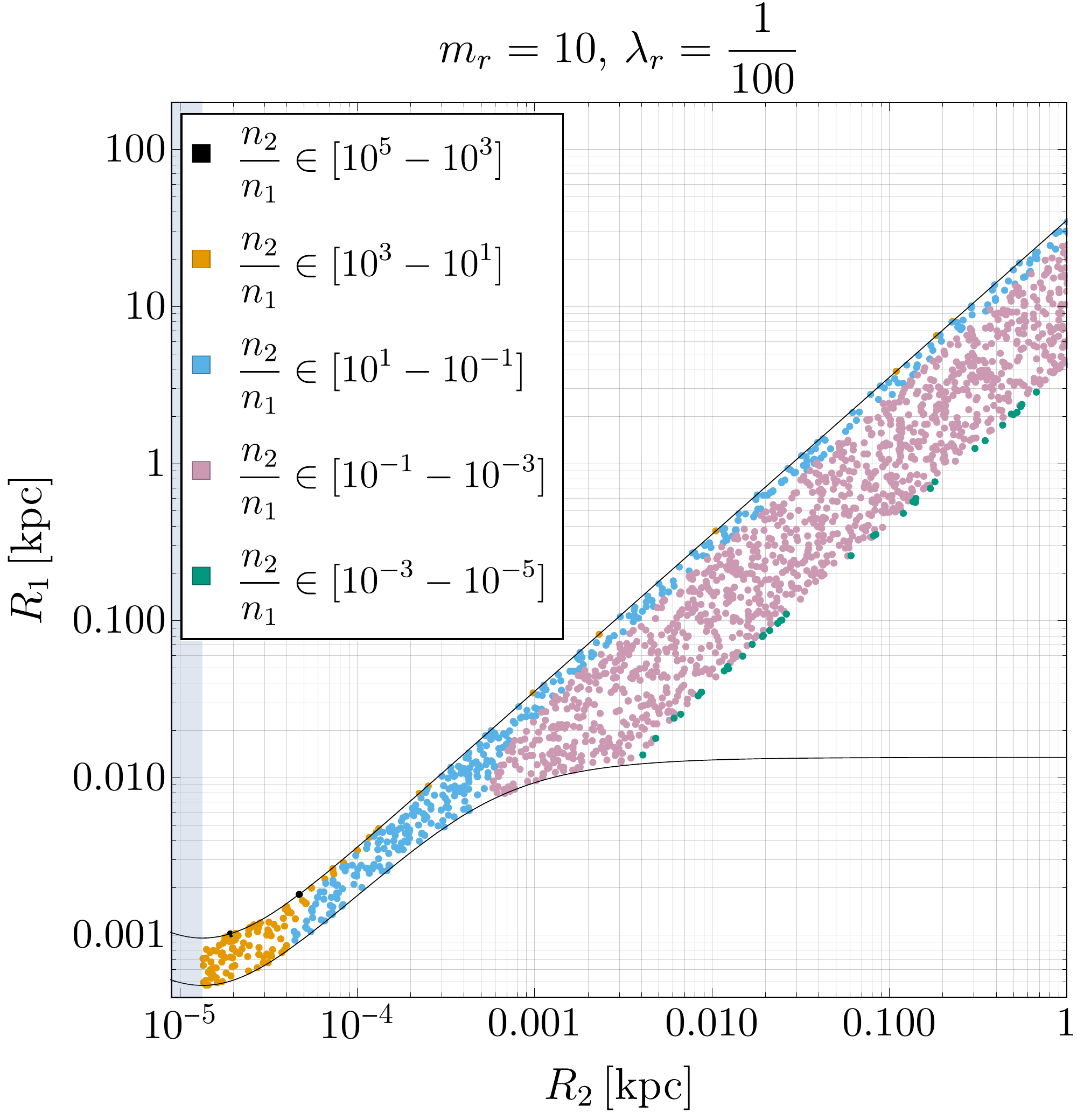} \,
 \includegraphics[scale=0.24]{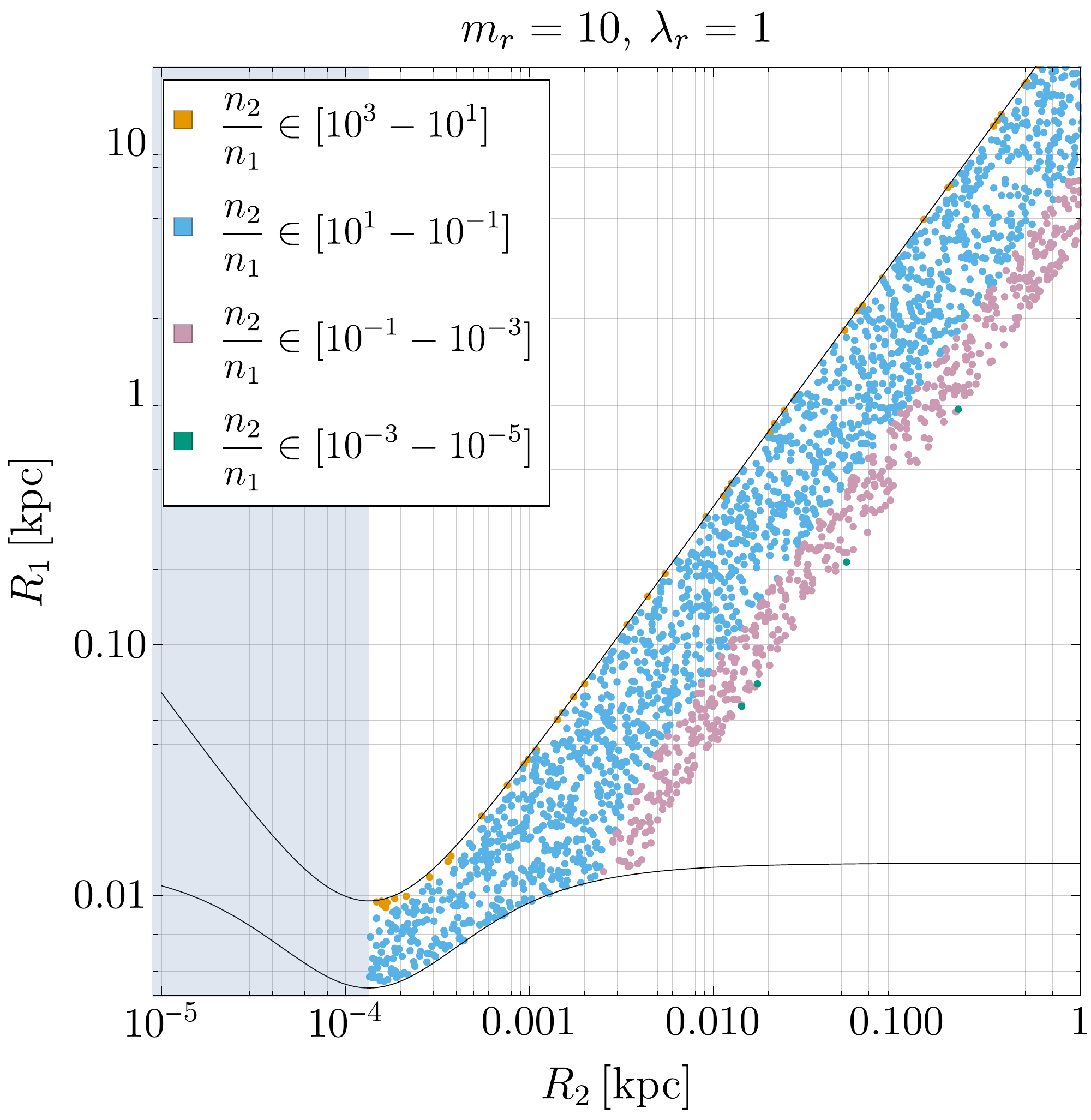} \,
 \includegraphics[scale=0.24]{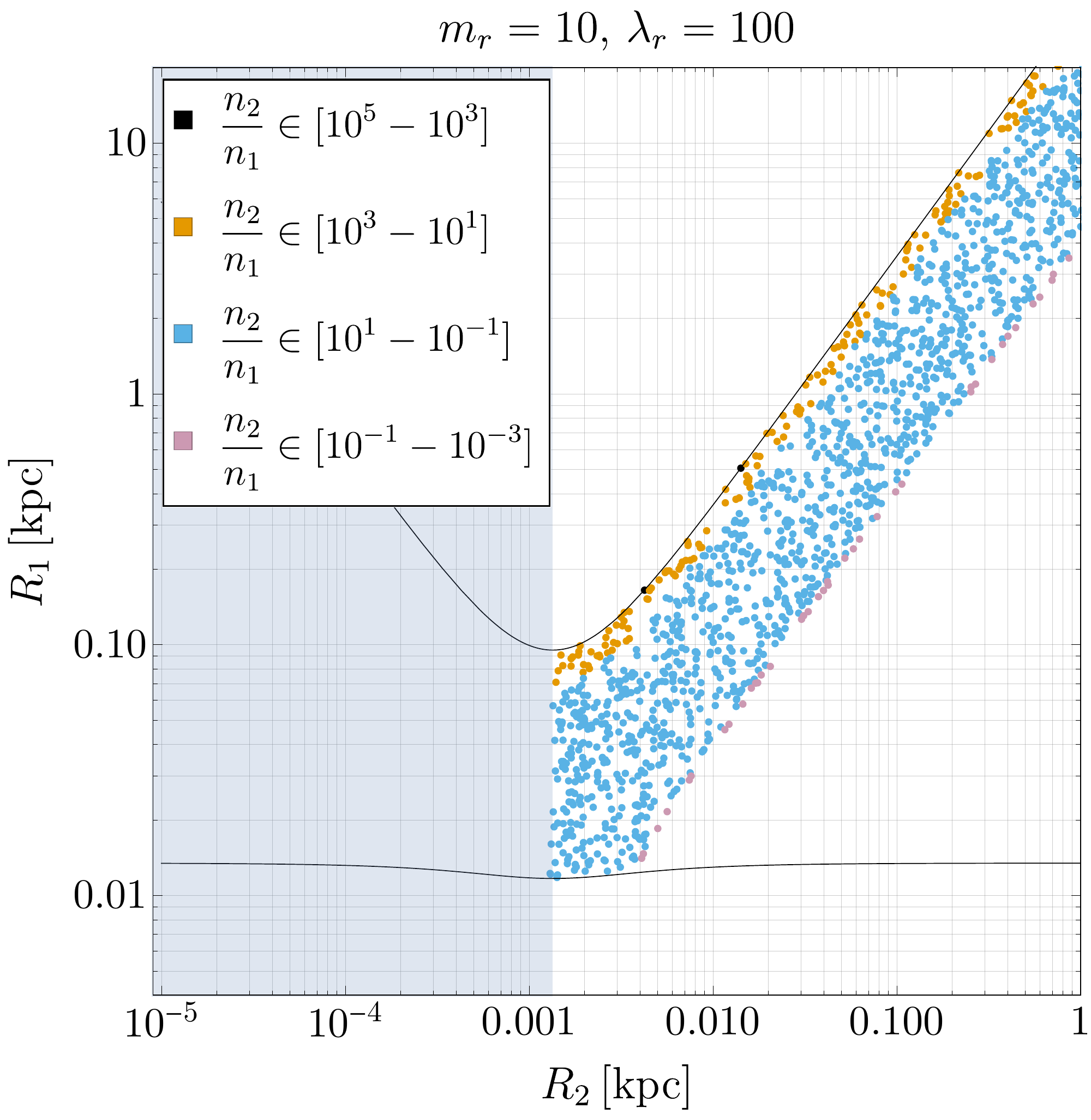}
 \\
  \includegraphics[scale=0.256]{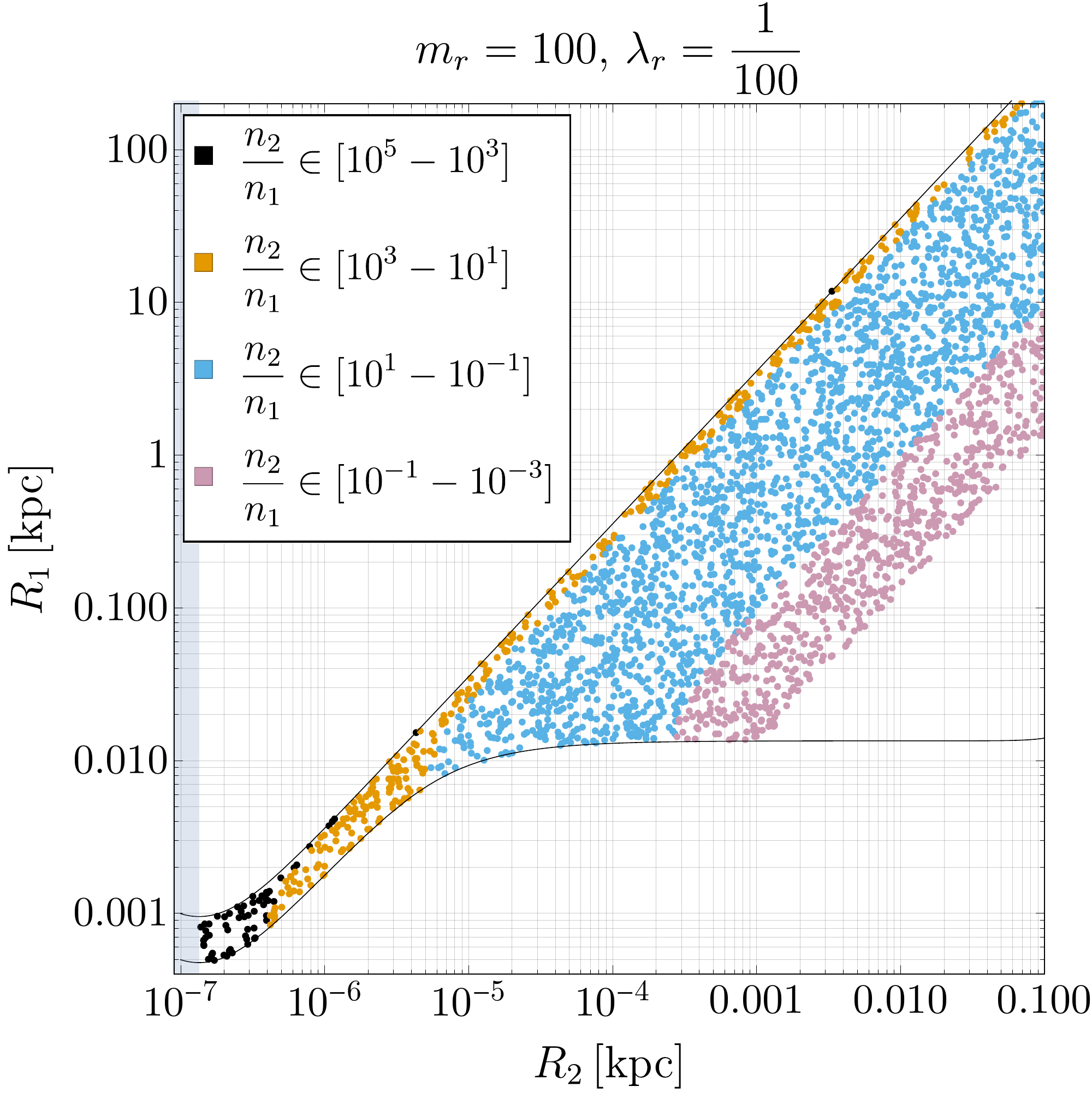} \,
 \includegraphics[scale=0.24]{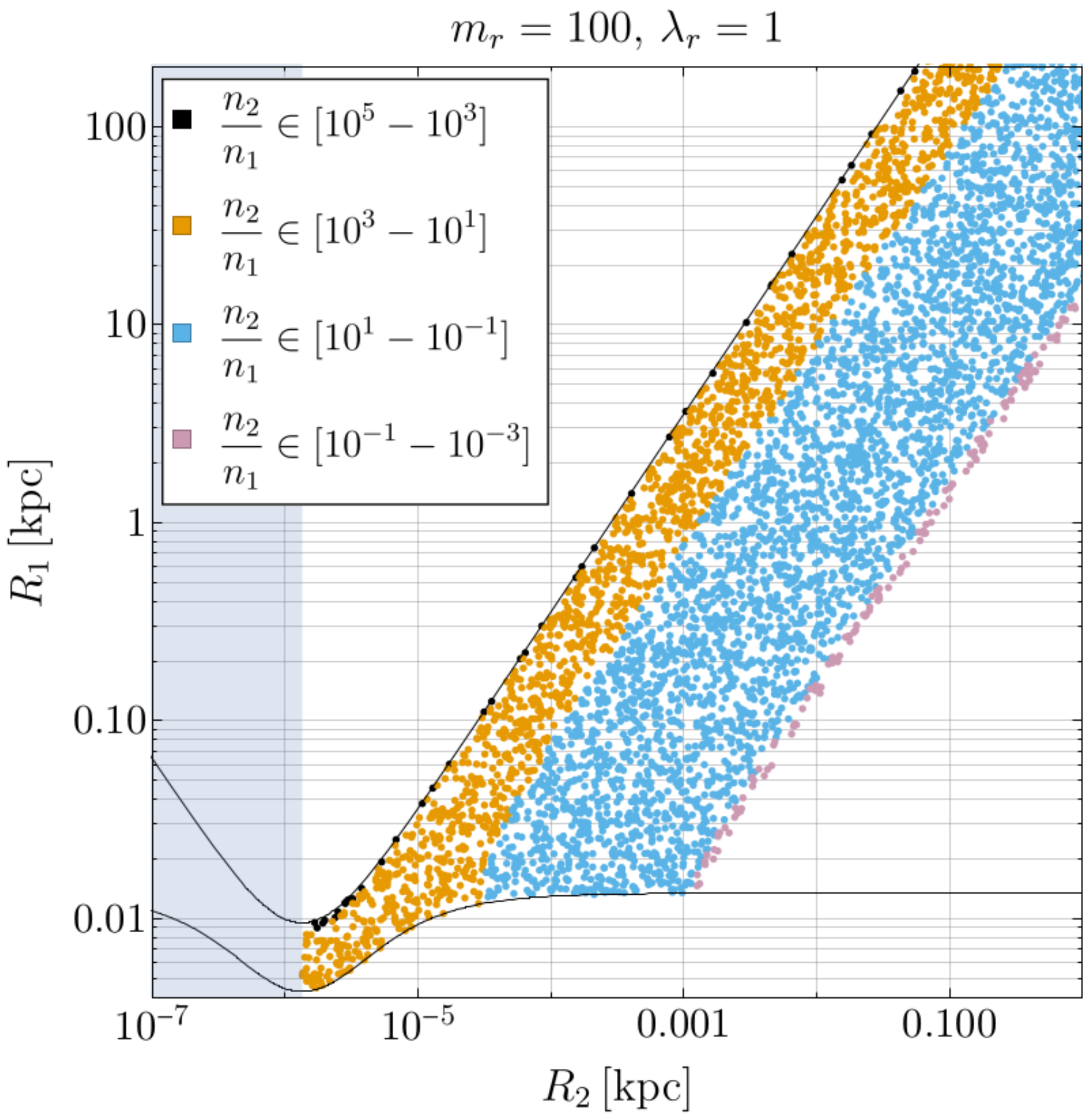} \,
 \includegraphics[scale=0.24]{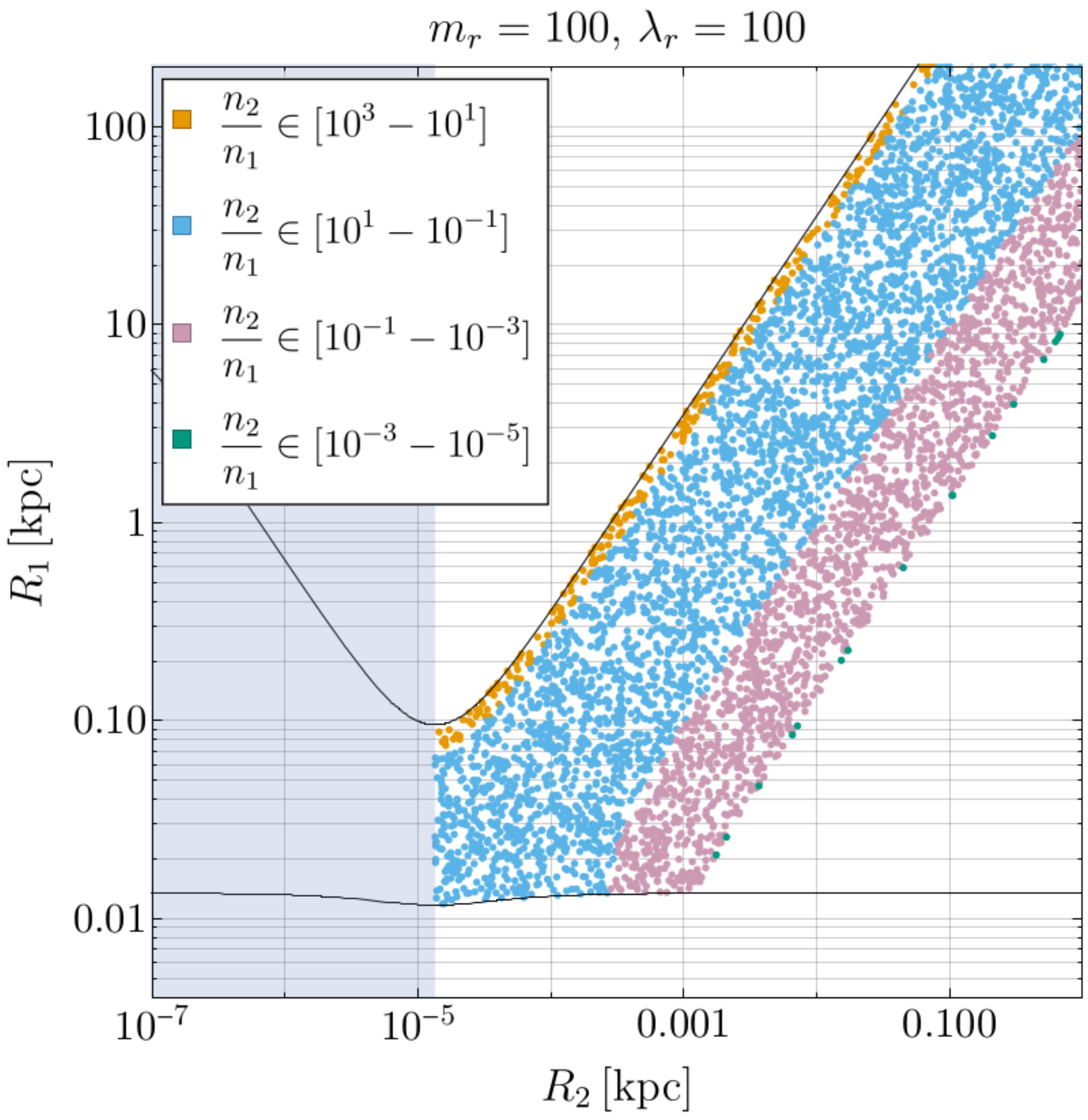}
\caption{The physical region $\mathcal B$, as defined by a random sampling of radial parameters $R_1$ and $R_2$, for the Benchmark 1 values $m_1 = 10^{-22}\, \text{eV}$ and 
$\l_1=10^{-94}$. The black solid lines are the boundaries of $\Bcal$ in the $m_r\to\infty$ limit, as discussed in Section \ref{sec:mtoinf}, and the shaded region corresponds to $\r_2<\tilde\r$. 
The color of the points are determined by the rescaled number ratios $n_2/n_1$, as described in the text.
The top (bottom) row corresponds to the choice $m_r = 10$ ($m_r = 100$), whereas the left, center, and right panels correpond to $\l_r=1/100,1,100$ (respectively).}
\label{Binf}
\end{figure}

First, we analyze the range of allowed condensate radii $R_1$ and $R_2$ consistent with the above conditions; the results are illustrated in Figure \ref{Binf} using the Benchmark 1 values of $m_1=10^{-22}$ eV and $\l_1=10^{-94}$.
We analyzed several choices of the ratios $m_r = 10,100$ (top and bottom rows) and $\l_r= 1/100,1,100$ (left, center, and right columns). The shaded region along the left side of each panel represents $\tilde{R}_2 = M_P\,\sqrt{\l_2}\,\tilde{\r}_2/m_2^2$, which remains a boundary of stability even in the two-flavor case. Eq. (\ref{eq:rho1limits}) 
determines the boundary of the physical domain $\mathcal B$ in the large $m_r$ limit, and these limits are shown as solid lines in Figure \ref{Binf}; we observe good agreement with the analytic results.
The resulting physical parameters $d_c$ and $R_c$ will be considered in the next subsections; for now, we simply note that at large $m_r$, the boundary of the physical region $\Bcal$ is (up to rescaling) only weakly dependent on other input parameters, and spans a large range of possible values of $\r_1$ and $\r_2$.

The color of the points in Figure \ref{Binf} represent the range of allowed rescaled particle number ratios, $n_2 / n_1$. The black, yellow, blue, pink, and green points each span two orders of magnitude and are centered around $n_2/n_1 = 10^4,10^2,1,10^{-2},10^{-4}$ (respectively); the colors also appear in this order roughly from the left side to the right side of the plot, though note that not every range is accessible for particular choices of parameters. Very asymmetric two-component condensates appear to be especially difficult to stabilize, as illustrated by the lack (or complete absence) of black and green points across the parameter space we consider.

In a full theory with two flavors of axion DM, it is plausible that the mass fraction 
\begin{equation}
\frac{M_2}{M_1} = \frac{\sqrt{\l_1}}{\sqrt{\l_2}}\left(\frac{n_2}{n_1}\right) = \frac{n_2/n_1}{\sqrt{\l_r}}
\end{equation}
will, in a typical galaxy, track roughly the ratio of relic densities of the two axion flavors, $\Omega_2/\Omega_1$; the latter could be computed from the full theory in the early universe. We observe then that given a choice of $m_r$ and $\l_r$, some mass fractions will produce few, or no, stable two-component condensates. An example of this is the top-right panel of Figure \ref{Binf}, where for $m_r=10$ and $\l_r=100$, there are almost no stable condensates in the range $n_2/n_1 < 0.1$, corresponding to $M_2/M_1 < 0.01$ (observe the lack of pink points and absence of green ones). This implies $M_2 \gtrsim 0.01\,M_1$ in the vast majority of the stable parameter space for this choice of parameters.

\subsection{Core density profile shape}

\begin{figure}[t]
\centering 
 \includegraphics[scale=0.35]{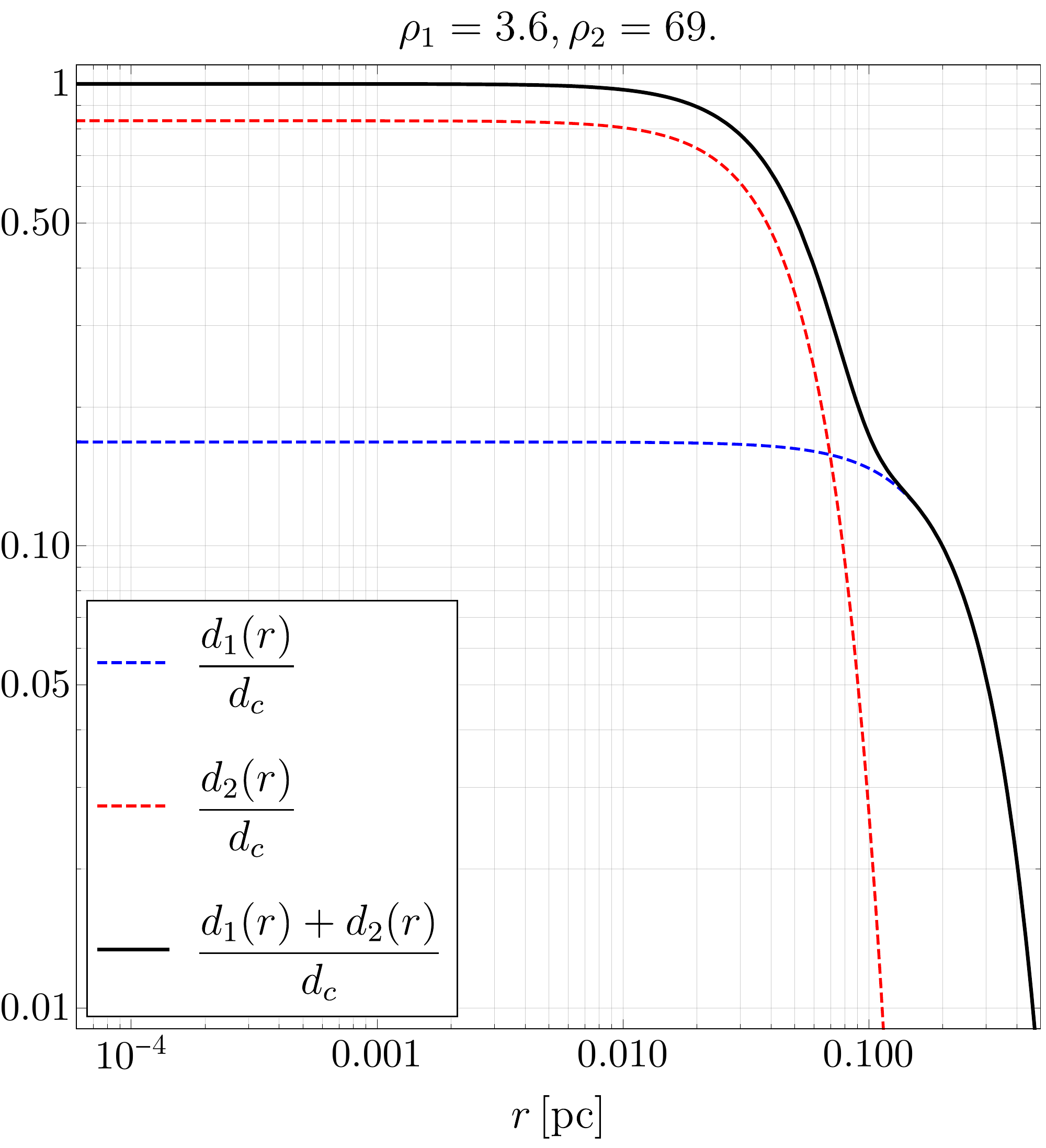} \,
 \includegraphics[scale=0.35]{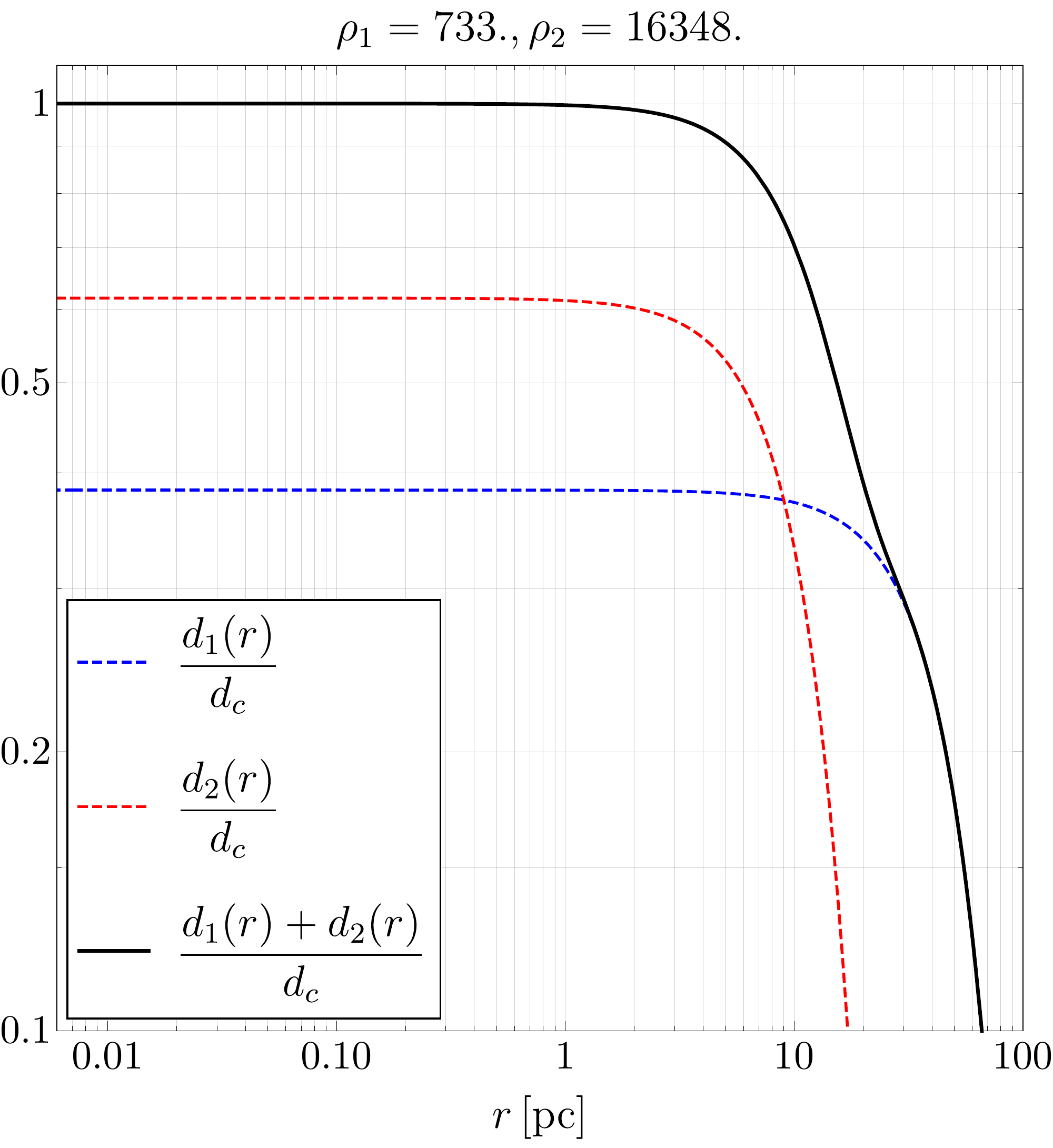}
\caption{Individual normalized condensate densities $d_1/d_c$ (blue dashed line) and $d_2/d_c$ (red dashed), as well as the sum $(d_1+d_2) / d_c$ (black thick), for the stable parameter choices $(\r_1,\r_2) = (3.6,69)$ (left) and $(\r_1,\r_2) = (733,16348)$ (right). Physical parameters are determined by the benchmark values $m_1=10^{-22}$ eV and $\l_1=10^{-94}$, using the ratios $m_r = 10$ and $\l_r = 1$.}
\label{fig:wavefunctions}
\end{figure}

Throughout this work, we have assumed Gaussian shapes for the wavefunctions of both condensates. This choice affects only the $\Ocal(1)$ numbers described in Section \ref{sec:scaling_intro}, like $a$, $b$, $c$, etc. (see Appendix \ref{app:numbers}). Given this selection for the shapes, then a choice of $\r_1$ and $\r_2$ in the numerical sample will uniquely determine $n_1$ and $n_2$, and thus fixes the wavefunctions $\psi_1(r)$ and $\psi_2(r)$ defined by Eq.~\eqref{eq:wavefn}. 

In Figure \ref{fig:wavefunctions}, we illustrate the resulting densities $d_1(r) = m_1|\psi_1|^2$ (blue dashed line) and $d_2(r) = m_2 |\psi_2|^2$ (red dashed), as well as the sum $d_1(r)+d_2(r)$ (black thick), at two stable points: one at relatively small $\r_1$ and $\r_2$ (left panel) and one at larger $\r_1$ and $\r_2$ (right panel). All curves are normalized to the central value $d_c=d_1(0)+d_2(0)$ for ease of comparison. As described in Section \ref{sec:toy}, we see in the figure the typical result that one component of the condensate (in this case, condensate 2) may dominate the core density, whereas the other (condensate 1) may dictate the total radius. Although the two condensates are not independent, this approximate decoupling of density from radius explains the wide scatter of points in the plane of $d_c$ from $R_c$ that we will describe in the next section.

\subsection{Core density vs. core radius}

\begin{figure}[t]
\centering 
 \includegraphics[scale=0.242]{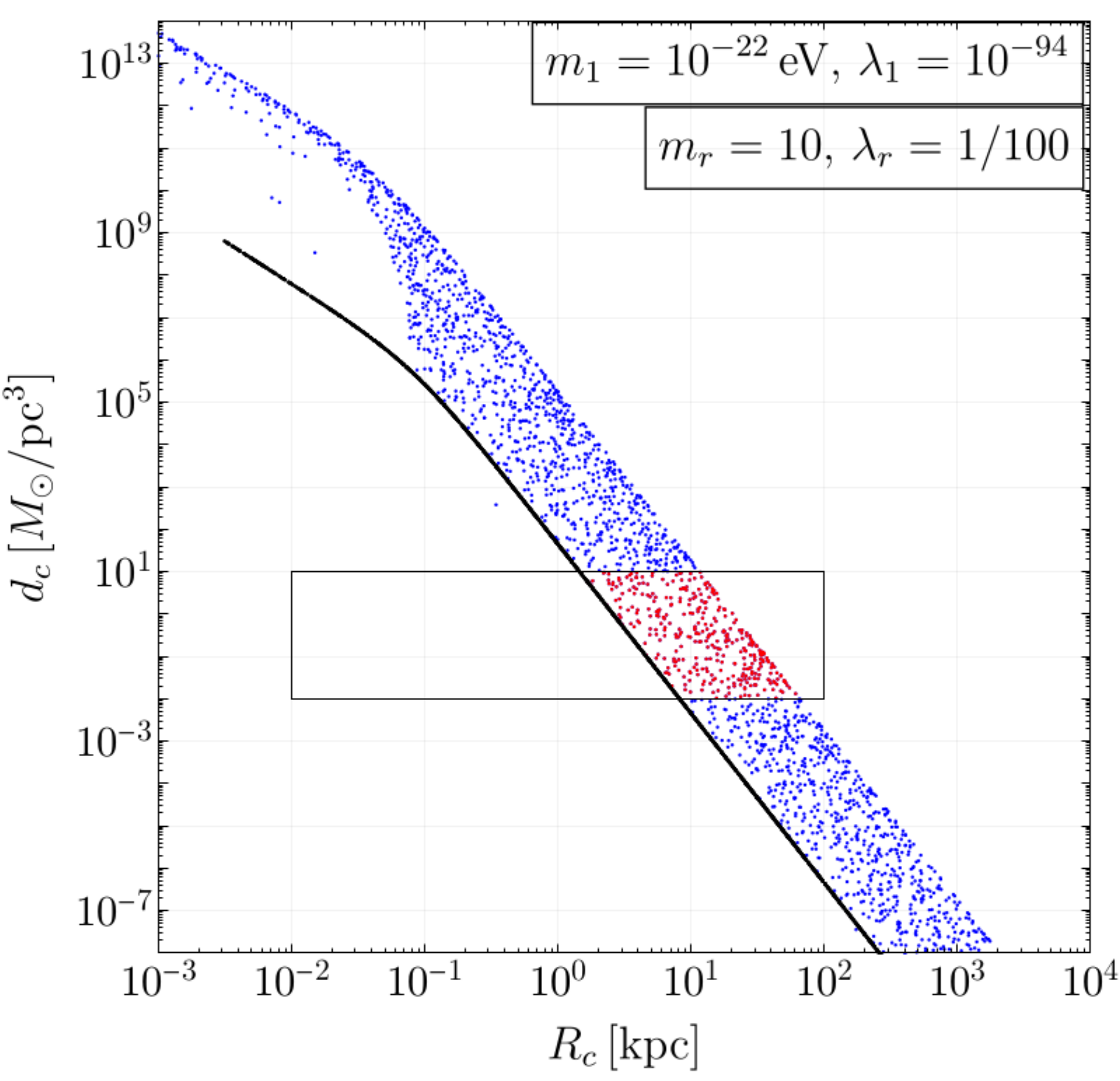} \,
 \includegraphics[scale=0.242]{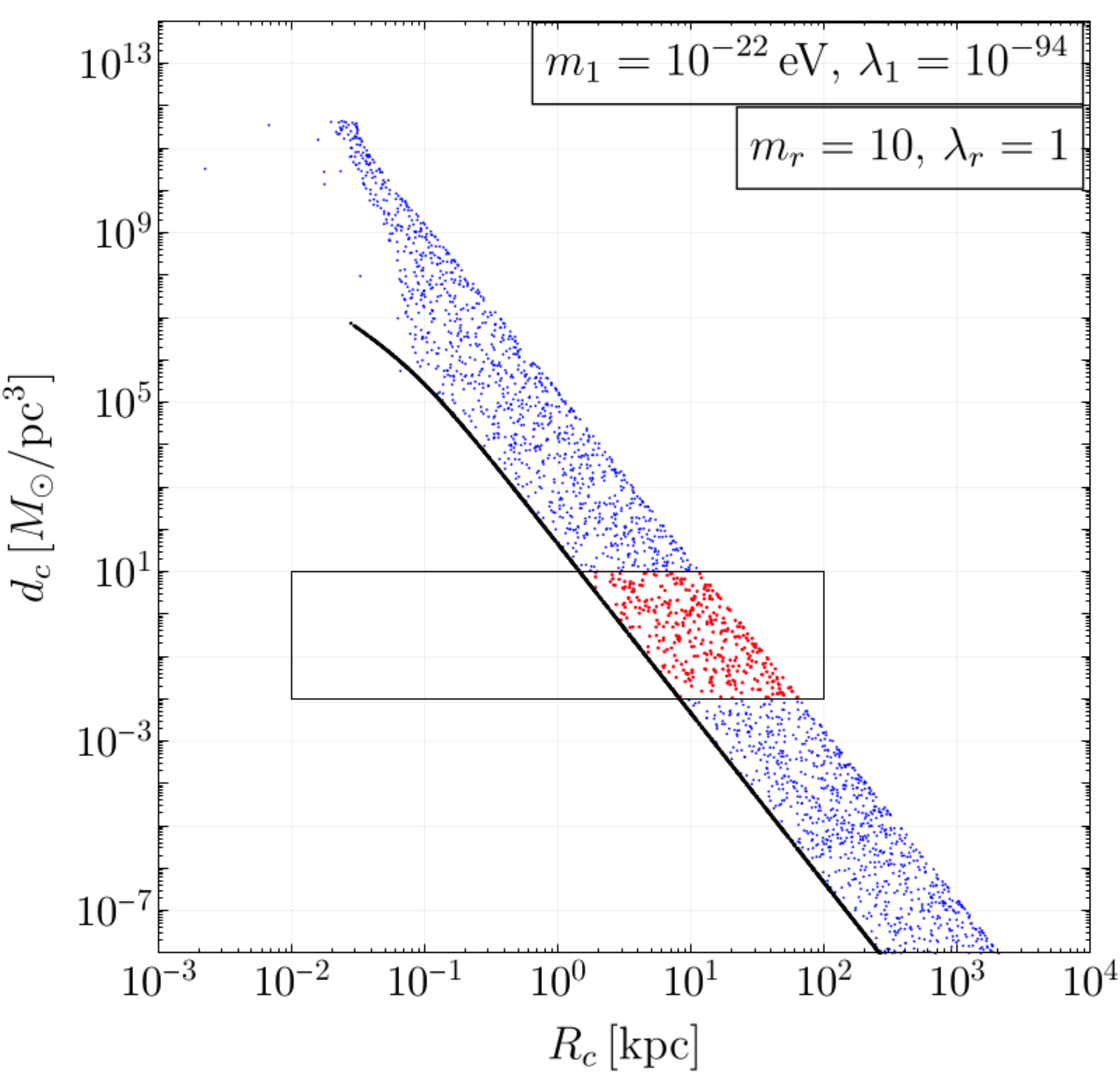} \,
 \includegraphics[scale=0.242]{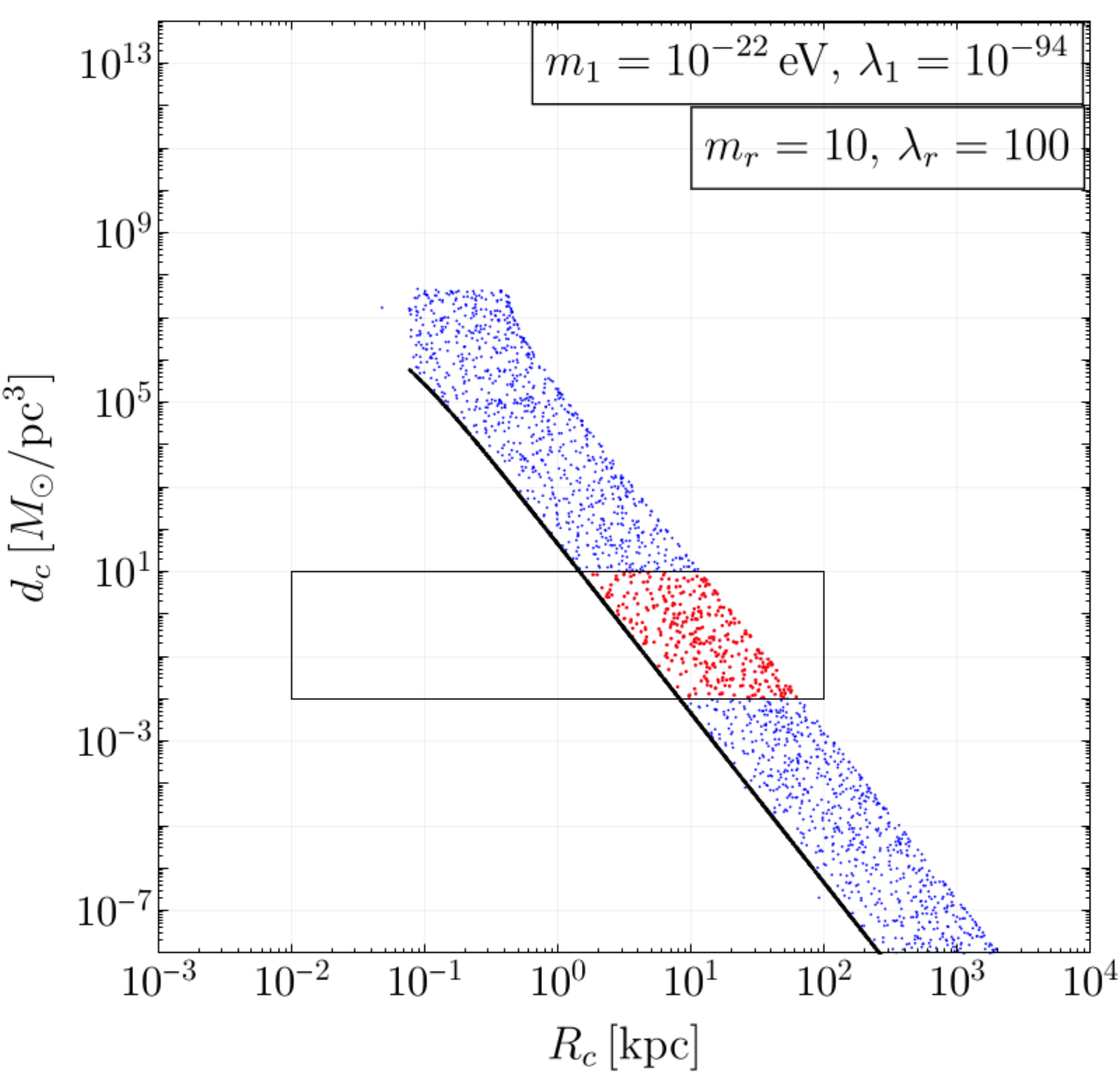}
 \\
  \includegraphics[scale=0.242]{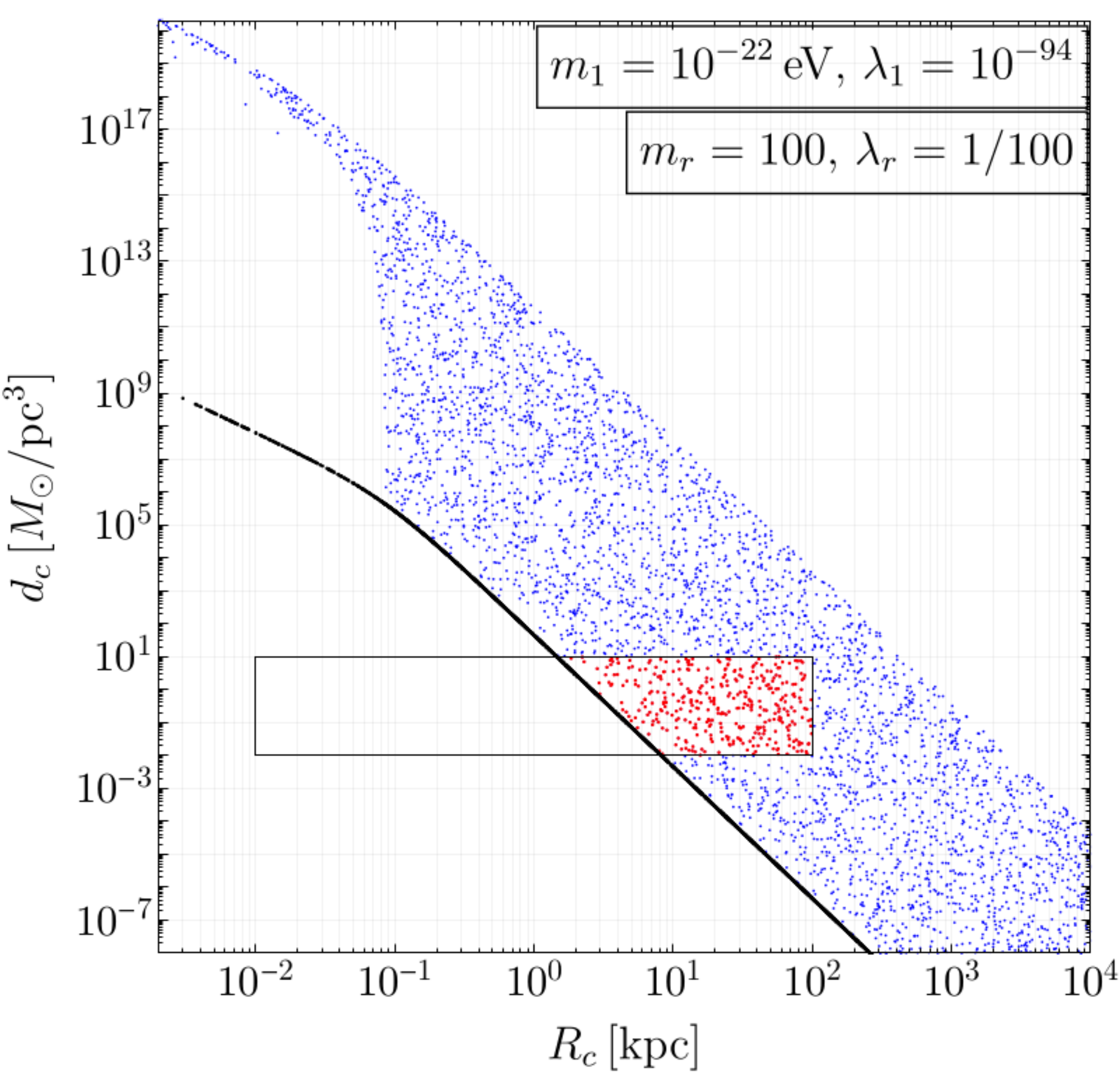} \,
 \includegraphics[scale=0.242]{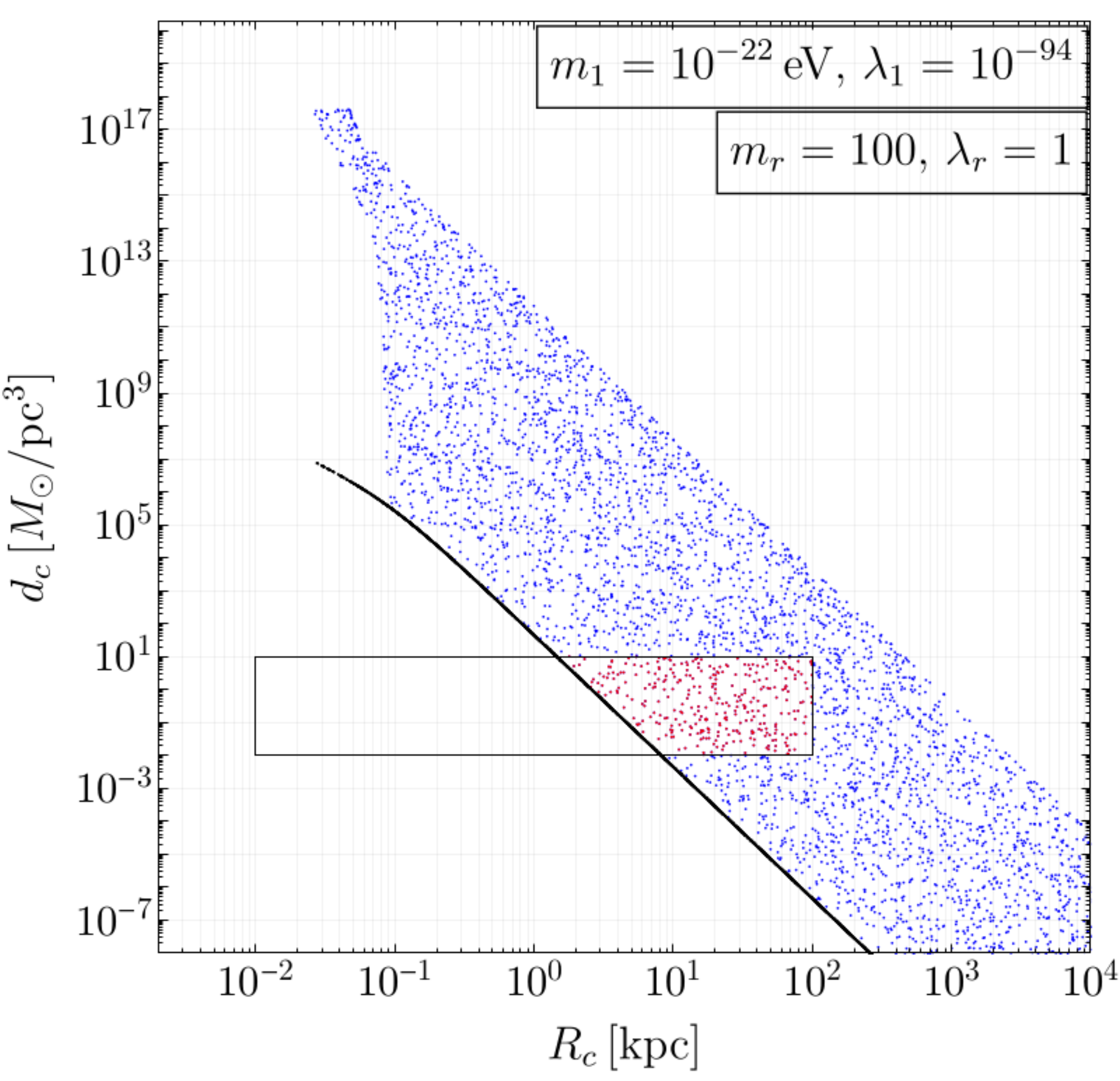} \,
 \includegraphics[scale=0.242]{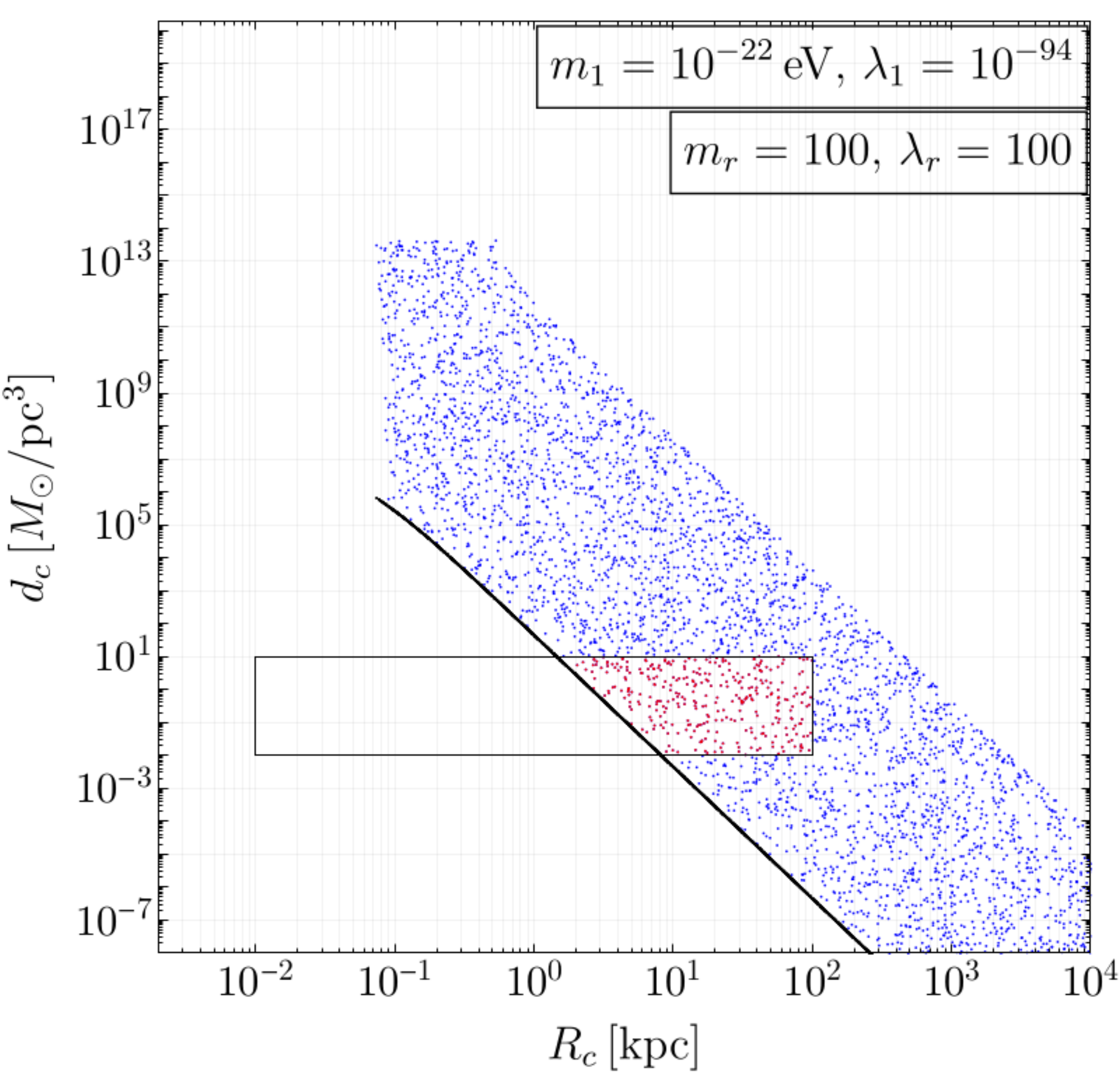}
\caption{Density $d_c$ and radius $R_c$ for stable two-component condensates, determined by the same sampling as Figure \ref{Binf} for Benchmark 1 values $m_1 = 10^{-22}\, \text{eV}$ and 
$\l_1=10^{-94}$. The blue points represent stable configurations, and the red points (bounded by the thin rectangle) is the subset in the range of observed galactic parameters. The black lines bound the physical range on the left and represents the relationship for a single condensate composed of only the first flavor ($n_2 \to 0$).
The top (bottom) row corresponds to the choice $m_r = 10$ ($m_r = 100$), whereas the left, center, and right panels correpond to $\l_r=1/100,1,100$ (respectively).}
\label{fig:dcRc}
\end{figure}

Now we turn to the observable parameters, the core density $d_c$ and radius $R_c$, illustrated in Figure \ref{fig:dcRc}; the blue points represent the stable two-component condensates and span a very large range. 
 The points are bounded to the left by the black line, representing the single-condensate result with flavor 1 only (i.e. $n_2 \ll n_1$). The opposite limit $n_2 \gg n_1$ bounds the points on the right, though this does not reduce to the flavor 2-only result; this is due to the asymmetry of $m_r \neq 1$, and can be understood by taking appropriate limits of Eqs. (\ref{eq:infeq1}-\ref{eq:infeq2}).
 
The general effect of the parameter choices $m_r$ and $\l_r$ are clearly visible in Figure \ref{fig:dcRc}. The coupling ratio $\l_r$ affects only the boundary of stability, where larger $\l_r$ implies a higher value of $\l_2$ and a higher likelihood that condensate 2 becomes unstable for larger particle numbers; therefore, at large $\l_r$ there are fewer points at the top-left of the parameter space in the $d_c-R_c$ plane. See, for example, in Figure \ref{fig:dcRc} that as $\l_r$ increases (moving from the left to right panels) there are fewer high mass/small radius points. The mass ratio $m_r$, on the other hand, determines the `width' of the stable region, where larger $m_r$ gives rise to a wider space of stable configurations.

If we fit a simple function $d_c \propto R_c^{-\b}$ to the full data set, the resulting scaling exponent remains roughly $\b \simeq 4$, as in the single-condensate case, albeit with much more scatter.  
Of course, if the true catalogue of galaxies in nature resembled Figure \ref{fig:dcRc}, then a simple linear fit over the entire range would be increasingly poor as $m_r$ increases, as the space of physical parameters becomes increasingly two-dimensional in the plane of $d_c$ and $R_c$. 
As mentioned previously, we, therefore, do not fit a scaling exponent to our numerical results.

\begin{figure}[t]
\centering 
 \includegraphics[scale=0.242]{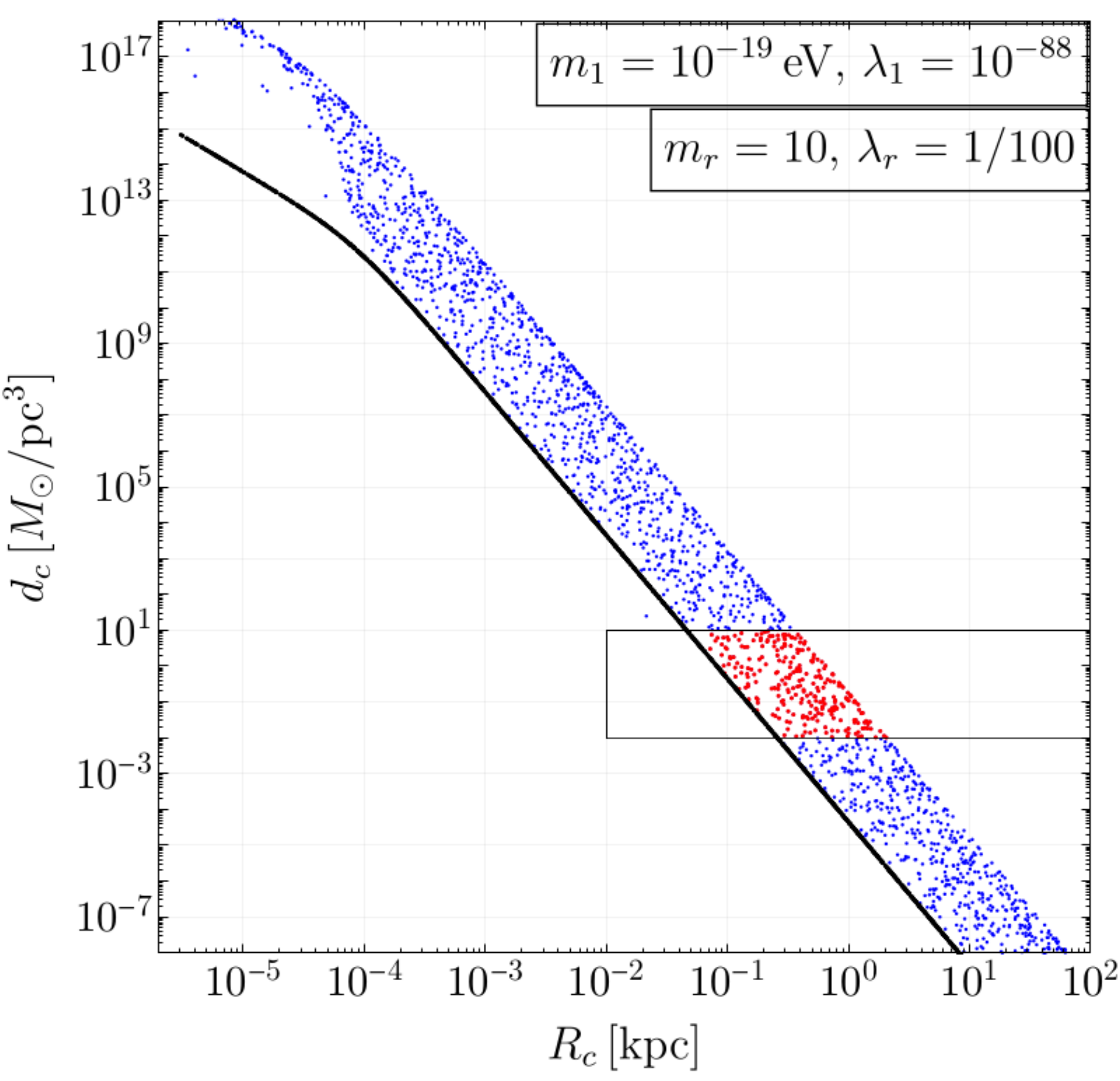} \,
 \includegraphics[scale=0.242]{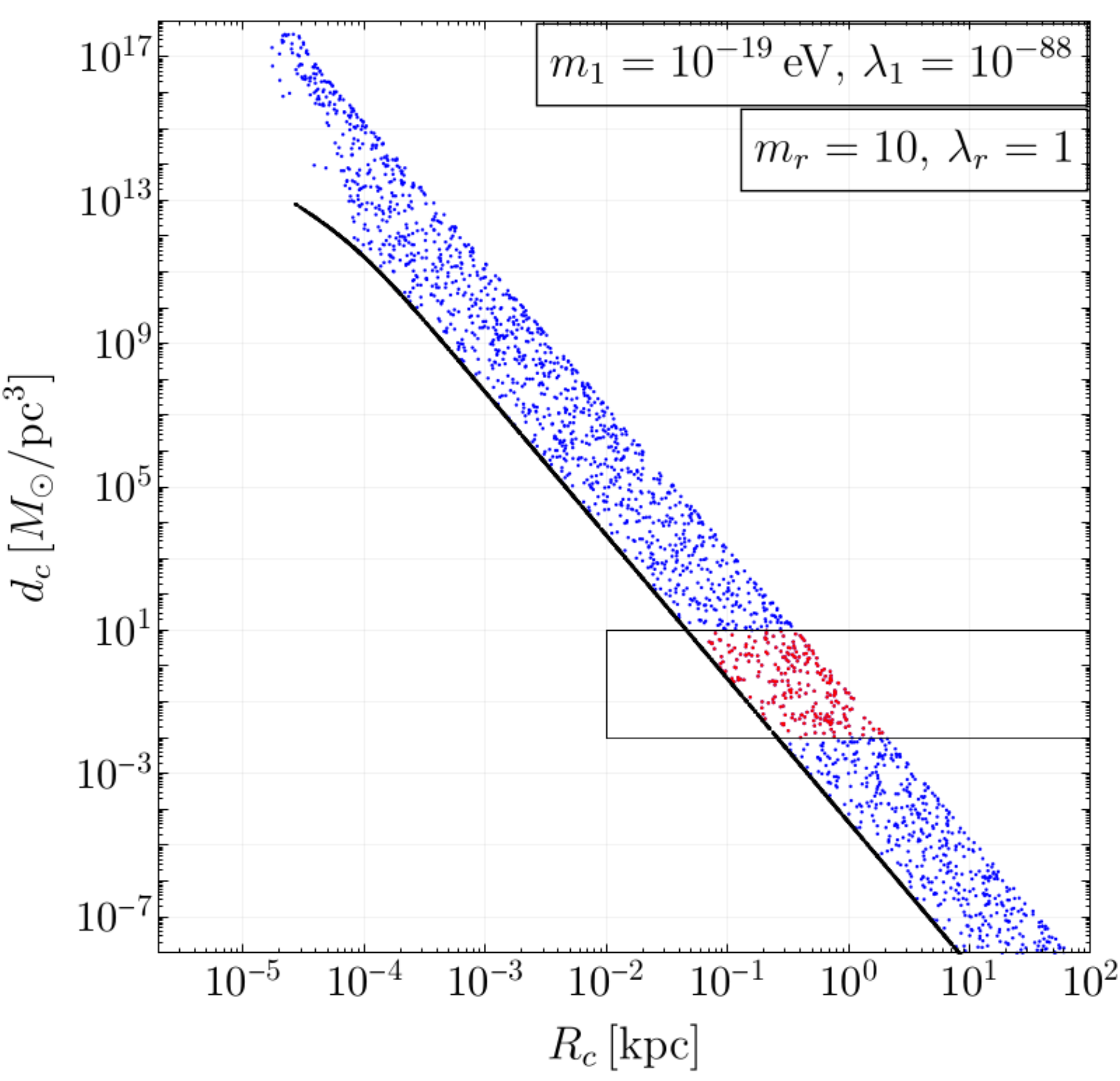} \,
 \includegraphics[scale=0.242]{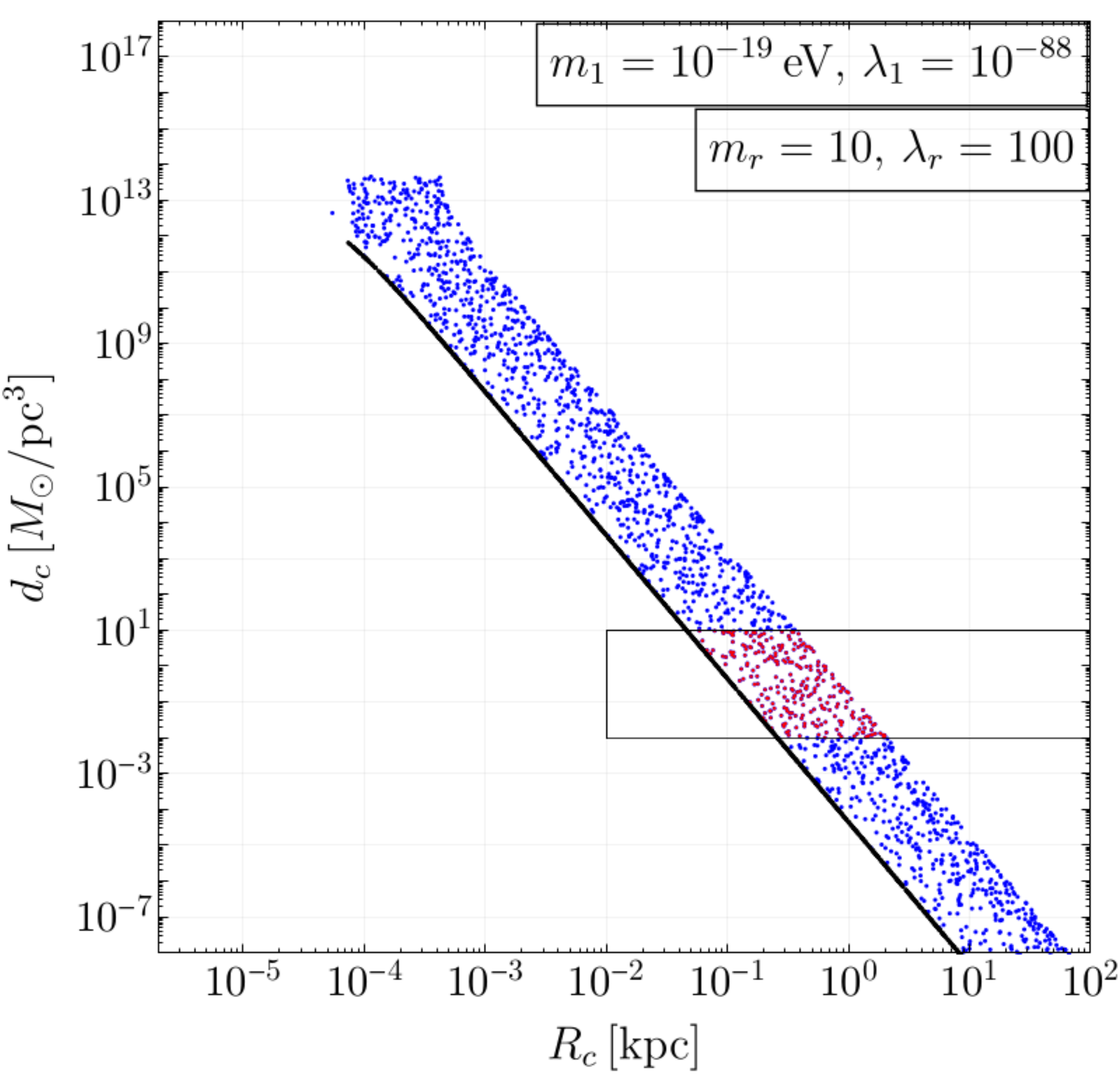}
 \\
  \includegraphics[scale=0.242]{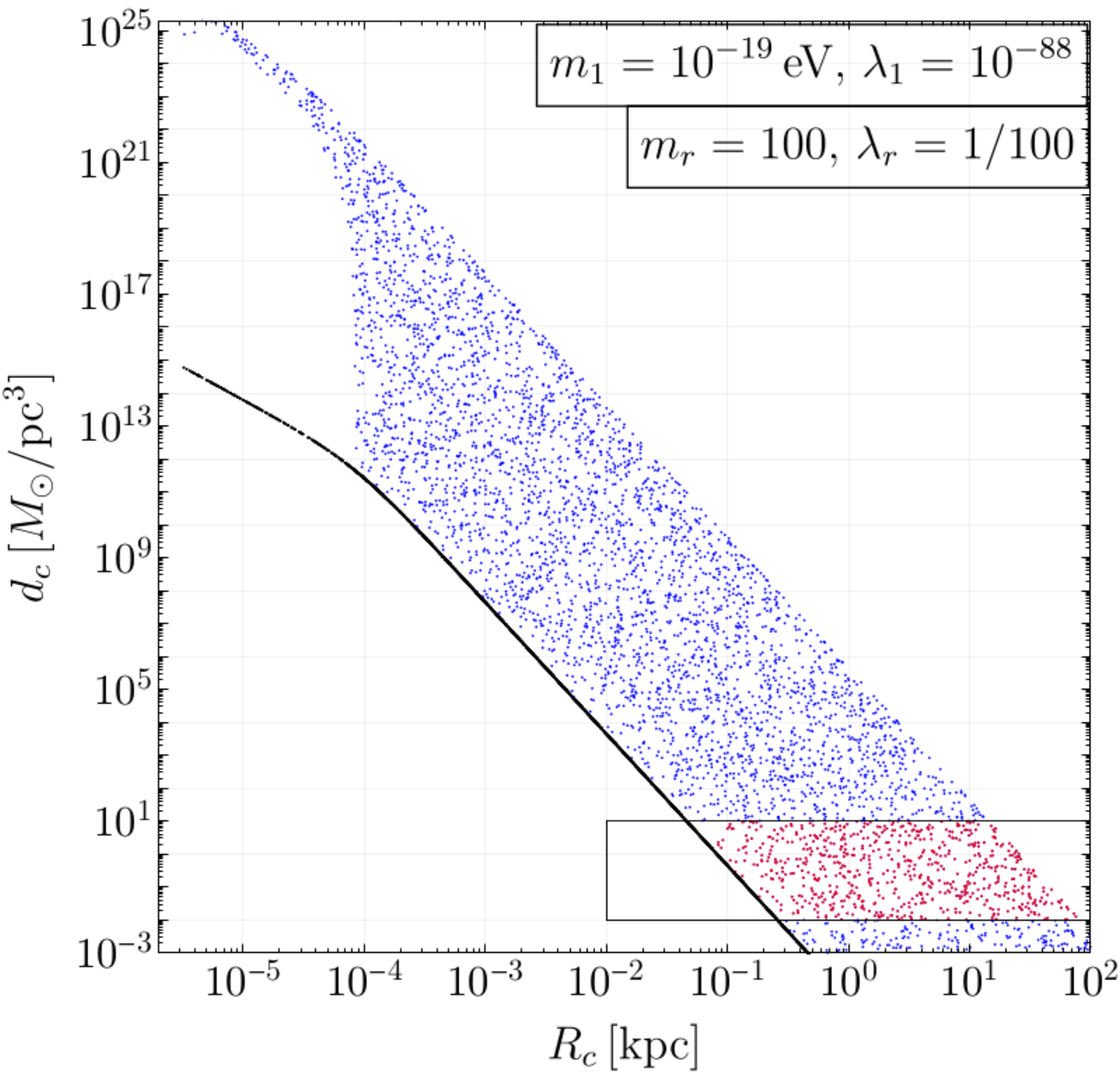} \,
 \includegraphics[scale=0.242]{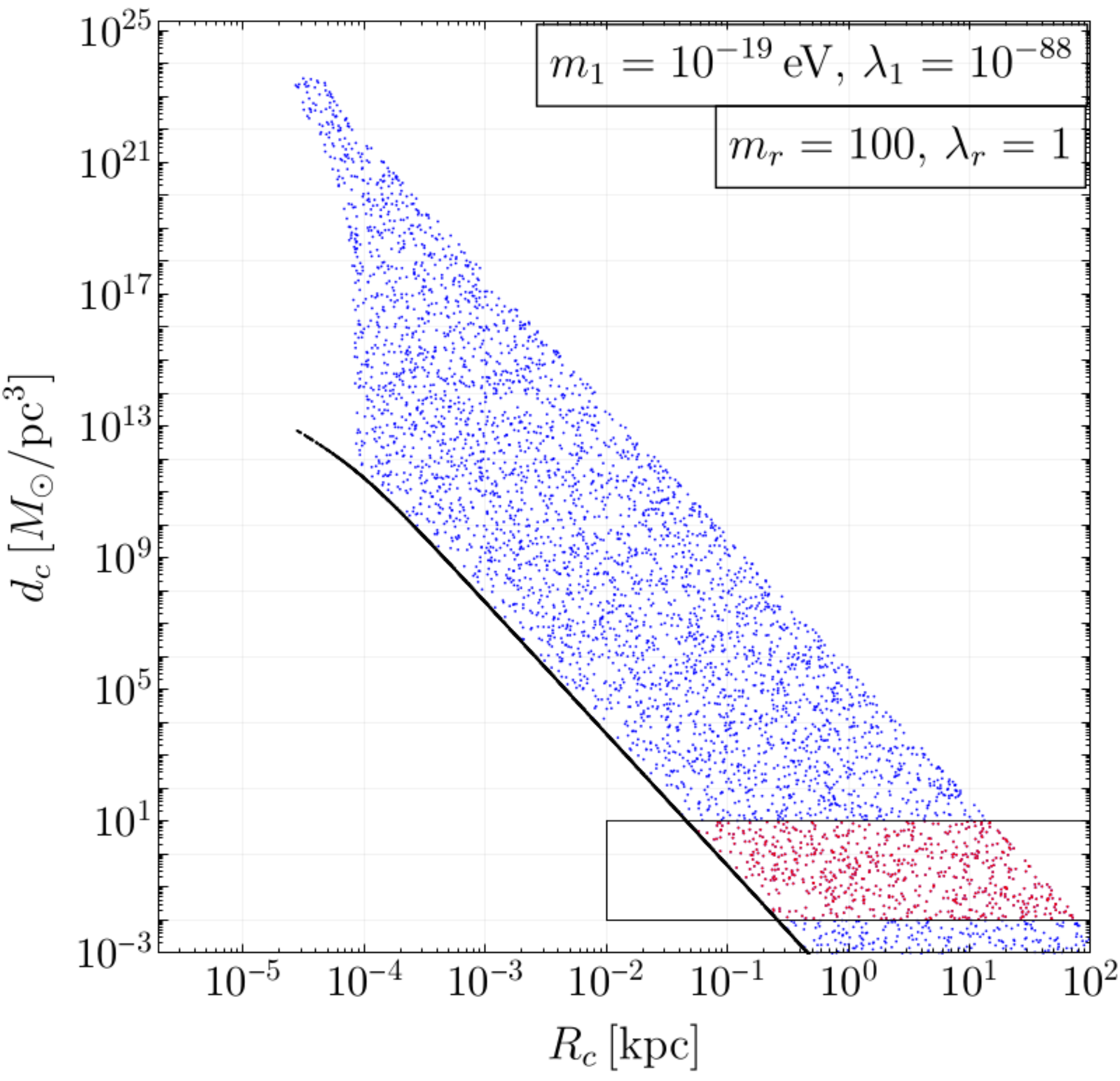} \,
 \includegraphics[scale=0.242]{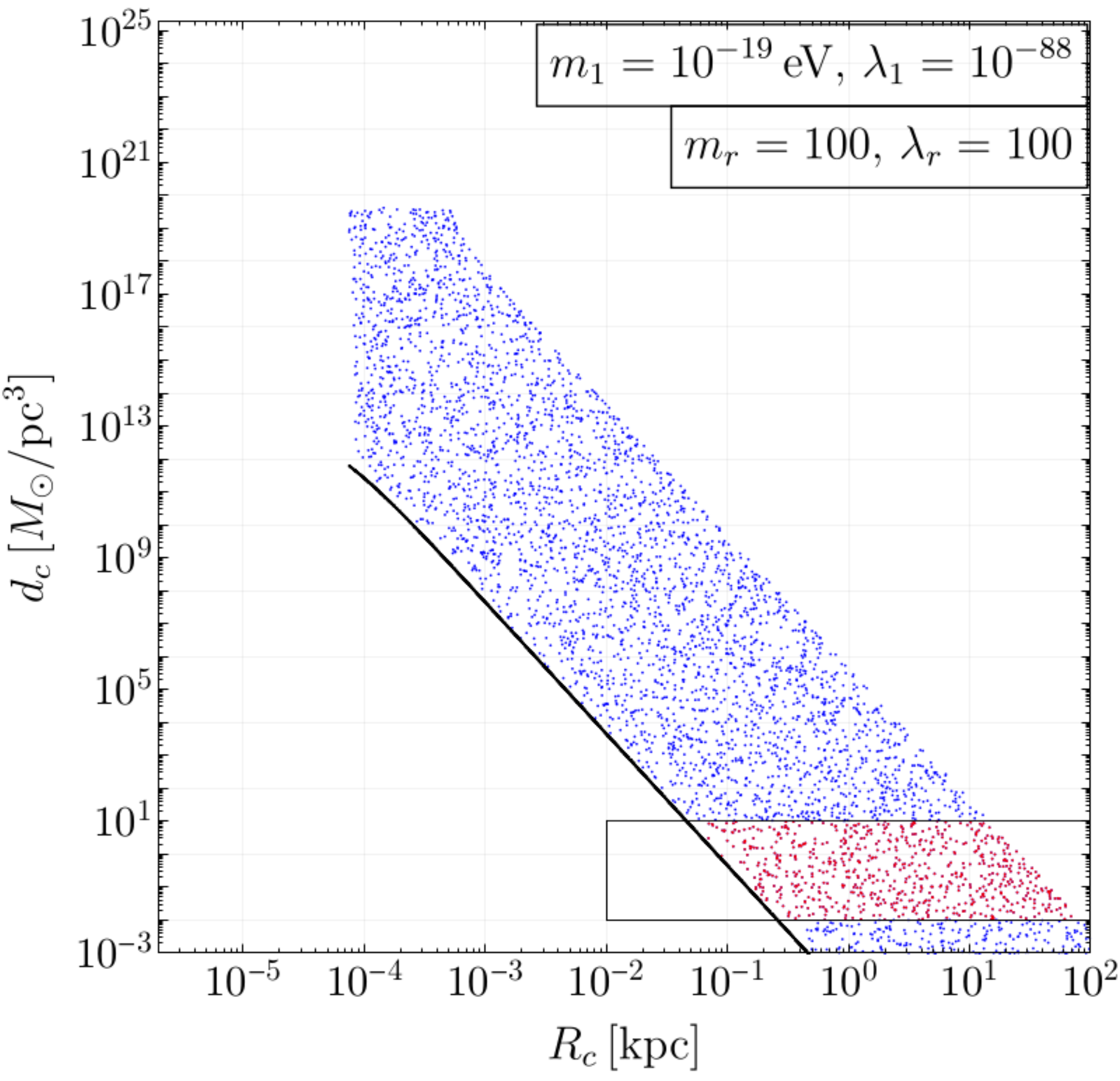}
\caption{Density $d_c$ and radius $R_c$ for stable two-component condensates, determined by the same sampling procedure as Figure \ref{fig:dcRc}, but rescaled using Benchmark 2 values $m_1 = 10^{-19}\, \text{eV}$ and 
$\l_1=10^{-88}$. The color and nature of the points and lines are the same as in Figure \ref{fig:dcRc}.}
\label{fig:dcRc19}
\end{figure}

Moreover, the blue points, while stable, extend very far to the top-left (large $d_c$, small $R_c$) and arbitrarily far to the bottom-right (small $d_c$, large $R_c$). In both directions, galaxies are either not expected to form or, if they do, they would be extremely difficult to detect. For this reason we highlight the subset of points that lie in the range where galaxies are known to exist, given in the figure by the red points bounded by the gray rectangular box. We define this range by examining fits to various galaxy samples, for example SPARC \cite{Lelli:2016zqa}, MASSIVE \cite{Veale:2017eqa}, as well as that of \cite{Rodrigues:2017vto}, on which the work of \cite{Deng:2018jjz} was based; this spans a wide range of roughly $0.01\,M_\odot/$pc$^3\lesssim d_c\lesssim10\,M_\odot/$pc$^3$ and $10$ pc $ \lesssim R_c \lesssim 100$ kpc. This range is only approximate, highlighted mostly to guide the reader's attention to the most physically significant range, however imprecise the boundary may be. It is possible that new galaxies with very large or very small densities or radii will be discovered and the relevant range will grow.

The ULDM model defined by Benchmark 1 has come under increasing scrutiny; for models with $m \lesssim 10^{-21}$ eV particles making up $100\%$ of DM, there now exist strong constraints from Lyman-$\a$  \cite{Irsic:2017yje,Leong:2018opi} as well as kinematic data in large galaxy samples \cite{Bar:2018acw,Bar:2019bqz}. In a two-flavor theory, the lighter particle $m_1$ would make up less than $100\%$ of DM, weakening existing constraints; nonetheless, it is interesting to consider alternative parameter choices where neither flavor obtains $m\lesssim 10^{-21}$ eV. In Figure \ref{fig:dcRc19}, we illustrate the resulting stable plane of $d_c$ and $R_c$ for a second benchmark, $m_1 = 10^{-19}$ eV and $\l_1 = 10^{-88}$. The two-flavor condensates tend to be denser than for Benchmark 1, but it is interesting that the physical region (red points) remains extremely relevant, even for such `heavy' ULDM particles. This may suggest two-component condensates with `large' ULDM masses $m_1 \sim 10^{-19}$ eV and $m_2 \gtrsim 10^{-18}$ eV as a model for galactic cores. This topic deserves a more dedicated analysis.\footnote{Note also that we neglect complications related to long relaxation times for axion stars with $m \gtrsim 10^{-19}$ eV; see e.g. \cite{Bar:2019pnz}.}

We emphasize that our analysis should not be interpreted as a prediction of any particular scaling exponent in a multi-flavor theory of axion condensates. As stated previously, because of the wide range of scatter in the $d_c-R_c$ space for a large range of parameters, a fit of our numerical results to a scaling exponent is a poor method of analysis.  Moreover, the observation of a scaling exponent will depend on the sampling assumptions and experimental sensitivities discussed throughout this work. It is also possible that different collections of galaxies, which may have relatively different formation histories, would display differing scaling exponents; this is, in fact, true of the data set \cite{Rodrigues:2017vto} used by \cite{Deng:2018jjz}, which had values $\beta$ at several different values of $\Ocal(1)$. 

\section{Condensate subject to an external gravity source} \label{sec:analytic_ex}

In the previous section, we have shown how the parameter space for condensates in a two-axion theory is significantly modified compared to the one-flavor case.
We now show how the parameter space is modified for a singly flavored condensate, if it forms in the presence of a background gravitational source. This background source might, in the context of this work, be thought of as the baryonic component or supermassive black hole of a large galaxy, where the ULDM forms only a fraction of the mass of the core; for heavier axions $10^{-18}$ eV $\lesssim m \lesssim 10^{-8}$ eV, a large background potential from the sun or a planet gives rise to a so-called axion solar halo or earth halo, as in the model described in \cite{Banerjee:2019epw}.  This analysis will be similar to that of \cite{Chavanis:2019bnu}, who considered a condensate surrounding a central black hole.

Consider a single-flavor axion condensate composed of some species of bosons subject to attractive $\lambda\,\phi^4$ self-interactions and self-gravity, exactly as described in Section \ref{sec:1flavor}. However, suppose at the center of this condensate there is a smaller, spherically-symmetric body with total mass $M_*$ and radius $R_*$.  Assuming the two systems interact only through gravity, the total energy of the condensate is given by Eq. (\ref{Energy}) (where we take the negative sign for the self-interaction term corresponding to attractive self-interactions) with the additional term,
\begin{align} \label{eq: Eg_ext}
E_{g,\text{ext}} = m\, \int d^3 r \, \Phi_{g,\text{ext}}\,\left| \psi \right|^2
\end{align}
where the external gravitational potential due to the inner body with a constant density is
\begin{align} \label{eq:V_ext}
\Phi_{g,\text{ext}} = - \frac{M_*}{M_P^2} \times
\begin{cases}
\displaystyle{\frac{3}{2 R_*} - \frac{r^2}{2R_*^3}} & \text{for}\quad r \leq R_*
\\ \\
\displaystyle{\frac{1}{r}} & \text{for} \quad r > R_*
\end{cases}.
\end{align}
We scale the physical input parameters in analogy to Eq. (\ref{eq:rescale}), using
\begin{align} \label{eq:rescale1}
R_* = \sqrt{|\lambda|}\frac{M_P}{m^2}\r_*,	\qquad
M_* = \frac{M_P}{\sqrt{|\lambda|}} n_*.
\end{align}

Assuming the external gravitational source has a small radius, we can expand in $\rho_*/\rho \ll 1$ and obtain the total energy per particle, which is of the form,
\begin{align} \label{eq: dil_eng}
\frac{E}{N} = \frac{m^2}{M_P}\frac{1}{\sqrt{|\lambda|^3}} \left( \frac{a + a' \,n_* \, \rho_*}{\rho^2} - \frac{b\, n + b'\, n_*}{\rho}  - \frac{ c \, n - c' n_* \, \rho_*^2}{\rho^3} \right)
\end{align}
where the constants $a^{(\prime)}$, $b^{(\prime)}$, and $c^{(\prime)}$ depend on the shape of the wavefunction (see Appendix \ref{app:numbers}). Note that we have kept terms in the expansion which modify the standard energy functional up to $\Ocal(1/\rho^3)$, though the $a'$ and $c'$ terms are suppressed by powers of the small quantity $\rho_*/\rho$. The variational parameter $\r$ for which the energy per particle is minimized is given by
\begin{align} \label{eq: sigma_sol}
\r &=  \frac{a +  a' n_* \, \rho_* + \sqrt{\left( a +  a' n_* \, \rho_*\right)^2 + 3 \left( b \, n + b' n_* \right) \left( -c \, n + c' n_* \, \rho_*^2 \right)}}{b \, n + b' n_*} \nn \\
	&= \frac{a +  a' n_* \, \rho_*}{b \, n + b' n_*} \left[ 1 +
		\sqrt{1 - \frac{3 \left( b \, n + b' n_* \right) \left( c \, n - c' n_* \, \rho_*^2 \right)}
					{\left( a +  a' n_* \, \rho_*\right)^2}}\right].
\end{align}
Of course, we also recover Eq. \eqref{single} if we set $a'=b'=c'=0$.
From Eq. (\ref{eq: sigma_sol}), one can see that the critical particle number $\tilde{n}'$ beyond which no stable minimum energy solutions exist is given by
\begin{align} \label{eq: Ncrit}
\tilde{n}' = \frac{3 \left( b \, c'n_* \, \rho_*^2 - b' \, c \, n_*\right)}{6\,b\,c}\left[1 + \sqrt{1 
			+ \frac{4 \, b \, c \left[ \left( a +  a' n_* \, \rho_* \right)^2 + 3 \, b' \, c' n_*^2 \, \rho_*^2\right]}{3 \left( b \, c'n_* \, \rho_*^2 - b' \, c \, n_*\right)^{2}}}\right].
\end{align}

Just as in Sections \ref{sec:scaling} and \ref{sec:calc}, we find that analytical estimates of the scaling relation between $d_c$ and $R_c$ fall within some range.  However, in contrast to Sections \ref{sec:scaling} and \ref{sec:calc}, the scaling relation also significantly depends on the definition of the core density.  We find for a fixed $n_*$ and $\rho_*$, the range of possible scaling exponents obtained by varying the condensate particle number $n$ differs for two possible definitions of the core density, as explained below.  Also, just as in Section \ref{sec:random}, we find that taking a large range of inner and outer body masses as a sampling region results in a wide distribution in the $d_c - R_c$ plane.  Because of this, the scaling exponent becomes a poor method to analyze this space.

As an example to demonstrate the different scaling exponent ranges through analytical estimates using two possible definitions of the core density, we consider a large range of possible condensate sizes by varying both the density of the inner body and the number of particles of the outer condensate.  One can take the core density of this system to be the analogue of Eq. (\ref{core_density}) (where the core density is defined as the central density $m|\psi(0)|^2$).  We note, however, that this definition of the core density is unphysical because the density at the center of a galactic core is not measured in practice.  We discuss the results for this definition to highlight the fact that the resulting scaling relations differ depending on the definition of the core density.  However, we emphasize that this definition should not be used unless the core density can be defined this way without loss of generality as in Section \ref{sec:random}.  For the model of this section, the central density includes both the condensate and the contribution from the external source with a constant density,
\begin{align} \label{eq: rho_c_ex}
d_{c,\text{cent}} = \rho_\text{tot}(0) = \frac{3\pi \,M_*}{4\,R_*^3} + m\,|\psi(0)|^2 \propto \frac{n_*}{\rho_*^3} + \frac{n}{\r^3},
\end{align}
whereas by assumption $\r \gg \r_*$ dominates the core radius, giving $R_c \simeq \rho$.  At fixed $n_*$ and $\r_*$, we can determine $\r(n,n_*,\rho_*)$ and $d_c(n,n_*,\rho_*)$ from Eqs. (\ref{eq: sigma_sol}) and (\ref{eq: rho_c_ex}), which then vary only with $n$. After rescaling parameters according to Eqs. (\ref{eq:rescale}) and (\ref{eq:rescale1}), we obtain the physical sampling region.  In this case, the maximum steepness obtainable for $d_{c,\text{cent}} \sim R_c^{-\beta}$ is $\beta = 4$ which corresponds to a very low-density for the inner body of $n_*/\rho_*^3 \lesssim n/\r^3$; on the other hand, the minimum steepness $\beta = 0$ corresponds to high-density $n_*/ \rho_*^3 \gtrsim n/\r^3$.  In the latter case, the core density is independent of the condensate radius since the condensate density is negligible.  If we then find the scaling exponent for each set of points corresponding to a given $n_*$ and $\rho_*$, we obtain a range $0 \leq \beta \leq 4$.  However, because this definition of the core density is unphysical as described above, we choose to give a more physical definition of core density as described below.

We take the core density as the the density just outside the external source (more specifically, the density at a scaled radius of $\bar{r}_\text{out} = (1 + 10^{-6}) \rho_*$).  In this case, the core density is,
\begin{align}\label{eq:rho_c_ex_out}
d_{c,\text{out}} = m |\psi(r_\text{out})|^2 \propto \frac{n}{\rho^3} \exp{\left[-\left(\frac{\bar{r}_\text{out}}{\rho}\right)^2\right]}.
\end{align}
For this definition, the scaling exponent actually takes a range from $\beta \geq 4$.  One can see from Eq. (\ref{eq: sigma_sol}), that for small enough inner body masses, the core radius is independent of the inner body mass, and the resulting scaling exponent is the same as that for an single flavor condensate (i.e. $\beta \sim 4$).  However, for large enough inner body masses and small enough condensate particle number, the condensate radius becomes independent of the condensate particle number.  In this case, as the condensate particle number decreases, the core density becomes smaller, yet the condensate radius remains the same resulting in a scaling exponent $\beta >4$.  Therefore, for this more physical definition of the core density, we find a range $\beta \geq 4$.

\begin{figure}[t]
\centering
\includegraphics[width=0.45\textwidth]{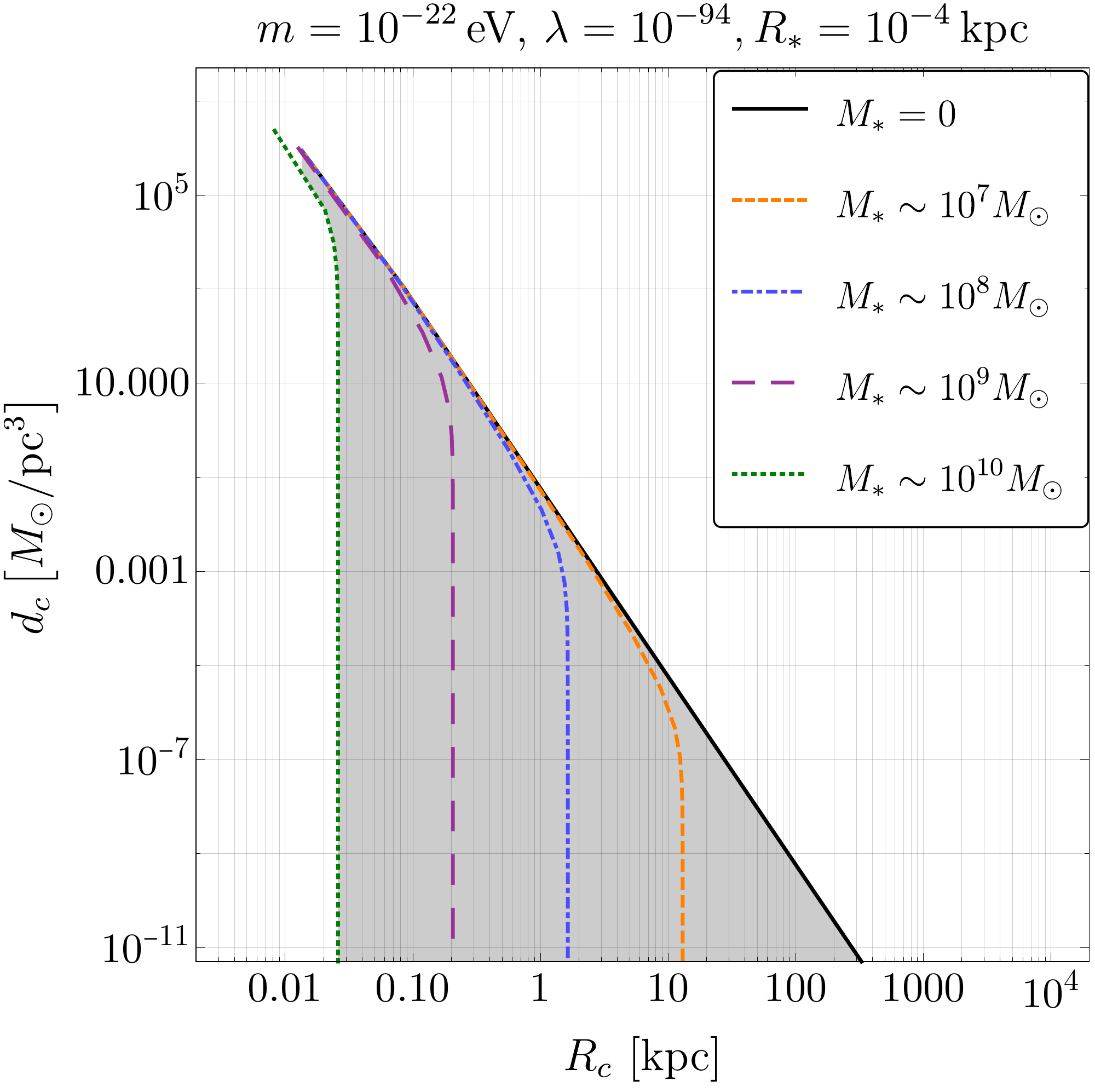}
\,
\includegraphics[width=0.45\textwidth]{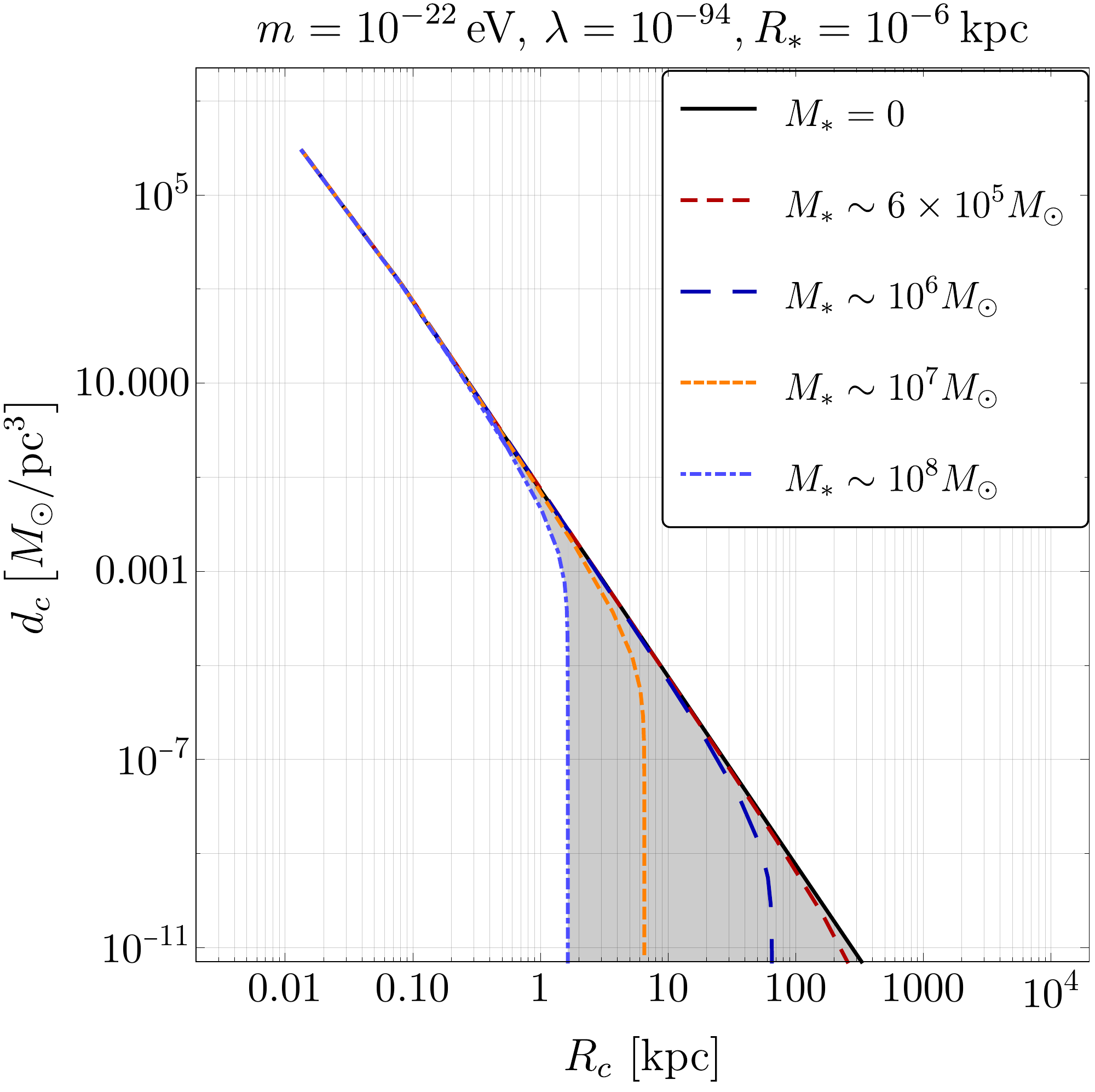}
\caption{Core density vs. core radius for condensate particle parameters $m = 10^{-22} \, \text{eV}$, $\lambda = 10^{-94}$, and for the core density defined by Eq. (\ref{eq:rho_c_ex_out}).  Left:  The possible parameter space for an inner body radius of $10^{-4}\,\text{kpc}$ and a range of inner body masses $0 \leq M_* \leq 10^{10}\,M_\odot$.  Right:  The possible parameter space for an inner body radius of $10^{-6}\,\text{kpc}$ and a range of inner body masses $0 \leq M_* \leq 10^{8}\,M_\odot$.  For each possible point in the parameter spaces, the constraints that $R_* \geq 10^2 R_S$ and $\rho_* \leq 10^{-2} \rho$ are satisfied as explained in the text.}
\label{fig:analytic_scaling}
\end{figure}

If physical galactic cores sampled are modeled as condensates subject to gravity from a spherically symmetric inner body, then there will be a range of possible inner body and condensate densities that can describe each galactic core.  For this reason, we take a sampling region that consists of a range of inner body and condensate densities subject to the constraints that the inner body radius is much greater than the Schwarzschild radius, $R_* \geq 10^2 R_S$ where $R_S$ is the Schwarzschild radius and the scaled inner body radius is much less than the scaled condensate radius, $\rho_* \leq 10^{-2} \rho$.  Figure \ref{fig:analytic_scaling}~ shows the possible parameter space for a given value of the inner body radius $R_*$ and a range of inner body masses $M_*$ where the core density is defined by Eq. (\ref{eq:rho_c_ex_out}).

We show Figure \ref{fig:analytic_scaling}~ as a guide to see how, for a fixed value of the inner body radius $R_*$, the possible parameter space for $d_c$ vs. $R_c$ depends on the range of inner body masses $M_*$.  Taking arbitrarily low condensate particle numbers, the possible parameter space technically extends to arbitrarily low core densities.  Hence, stable configurations are bounded by the density-radius line corresponding to a very low inner body mass which results in $\beta \sim 4$ (see the black diagonal lines of both panels) and the density-radius line corresponding to the maximum inner body mass analyzed resulting in $\beta > 4$ (see the green dotted line of the left panel and the blue dot-dashed line of the right panel).  One can see that the possible parameter space is largely degenerate, meaning that for a given inner body mass range, a different value of inner body radius results in many of the same $d_c$ vs. $R_c$ points.  This degeneracy can be seen especially along the line corresponding to $\beta \sim 4$ and for low density-large radius points.  For the left panel, we take the largest inner body mass to be of the order of the largest supermassive black holes, and the inner body radius to be such that both constraints $R_* \geq 10^2 R_S$ and $\rho_* \leq 10^{-2} \rho$ are comfortably satisfied.  Notice that for a smaller $R_*$ (right panel), the largest $M_*$ in the left panel are no longer physical because of the constraint $R_* \geq 10^2 R_S$.

We have shown that, in theory, the stable configurations of condensates subject to gravity from spherically symmetric inner body densities will have a parameter $\beta$ that is bounded from below as $\beta \geq 0$, depending on the definition of the core density used.  However, as the core density defined by Eq. (\ref{eq: rho_c_ex}) is unphysical, the theoretically possible boundary on the physical region is actually $\beta \geq 4$.  This highlights the importance of a strict definition of galactic core densities when using this model.

It is interesting to compare the results of Sections \ref{sec:scaling} and \ref{sec:calc} for the mixed flavor condensates to the results shown above. In contrast to Sections \ref{sec:scaling} and \ref{sec:calc}, we find that the choice of the definition of the core density results in significantly different results.  For the case of the mixed flavor condensates, we find no change in the analytical estimates of the scaling exponents for two different definitions of the core density, and hence we can take the definition of the core density as the central density (Eq. (\ref{core_density})) without any loss of generality.  This is because the particle numbers $n_1$, $n_2$ and radii $\rho_1$, $\rho_2$ of the two condensates are related in a way that a change in any one of these parameters results in a change in the other three.  Conversely, for the model described in this section, the radius $\rho$ and particle number $n'$ of the condensate changes depending on the choice of the inner body mass $M_*$ and radius $R_*$, however, the inner body is chosen to have a constant density independent of the parameters of the condensate.  In this case, the analytical estimates of the density-radius scaling relations depend on if the core density is defined inside or outside of the inner body.


\section{Conclusion} \label{sec:conclusion}

In this work, we generalized previous analyses of gravitationally-bound scalar field condensates, known as axion stars, to the case of more than one flavor interacting gravitationally. We determined in detail the conditions for stability of two-component condensates using a variational procedure to approximate the solution to the coupled equations of motion. Then, focusing on the scenario of ultralight dark matter (ULDM), we scanned a large range of input parameters which could be physically-motivated by the observation of cored density profiles in galaxies.

Though we focused on the case of ULDM, we attempted throughout to maintain as general an analysis method as possible, so that one can use these results to determine the properties of two-component axion stars in other models. The majority of our results depend only on mass and coupling ratios $m_r = m_2/m_1$ and $\l_r = \l_2 / \l_1$, and only in the final numerical results did we assume some physical scale for the input values $m_1$ and $\l_1$. Still, for both ULDM benchmarks $m_1 = 10^{-22}$ eV and $m_1 = 10^{-19}$ eV considered, we showed that two-component condensates with mass ratios $m_r = 10$ and $m_r = 100$ can mimic the density and radius of galaxy cores in a large physical range. This is true as long as the attractive self-interaction couplings $\l_i$ ($i=1,2$) were sufficiently small to prevent the onset of instability.

It has been long known that single-flavor condensates have a predictable relationship between their central density $d_c$ and their core radius $R_c$ \cite{Chavanis:2011zi,Chavanis:2011zm}, which has the form $d_c\propto R_c^{-\b}$ with $\b=4$. Moreover, it was recently pointed out \cite{Deng:2018jjz} that an empirical relationship has been observed in large galaxy samples which suggests $\b \simeq 1$, in tension with the condensate prediction. In this work, we point out that such tension only exists if the single-flavor condensate dominates the density in the core; given that this assumption can break down in several physically-relevant systems, one should consider the corresponding conclusions with great care. This type of analysis may represent a direct constraint on the parameters in other models, for example for ULDM condensates with repulsive interactions and sizes of the order of galaxies. It could also be a useful analysis for ultra compact dwarf (UCD) galaxies, whose sizes are of the order of galactic cores \cite{Eby:2018zlv,Broadhurst:2019fsl}.  We leave such analyses for further work.

If, as predicted in many models of ULDM, two (or more) flavors of axions exist, then the density-radius relationship is no longer predictive, as the space of stable configurations is two-dimensional and can be very wide. We also pointed out that observational constraints will further limit the relevant space on which these scaling relations are determined; the range of galaxy parameters currently observed remains narrow, relative to the full space of stable two-component condensates. Therefore it is possible that the observation of a relationship like $\b\simeq 1$ is simultaneously affected by (a) the physics of condensate stability, (b) the formation history of galaxies, and (c) experimental limitations.

We also presented a more simple model in which the scaling exponent can be modified, namely, that of a single-flavor condensate subject to external gravity from massive inner bodies; the central body might physically be identified with a central black hole or a baryonic bulge inside of a galaxy. We showed that for this model, the range of possible scaling exponents depends both on the definition of the core density and the sampling region assumed.  We analyzed two different definitions of the core density and found that for both definitions, if the inner body is very light one recovers the standard single-flavor result $\beta \sim 4$.  If the inner body is much more massive than the condensate, then it will dominate the core density, defined as the central density, and the scaling exponent $\beta \sim 0$.  However, for the more physical definition of the core density defined outside the inner body radius, the inner body mass dominates the condensate radius, and the scaling exponent $\beta > 4$.  However, in analyzing the full space of configurations no scaling exponent emerges, because like the two-flavor case, this space is two-dimensional and therefore cannot be summarized by a single exponent.

There is another interesting scenario which could modify the scaling relationship, which is the case of a single axion but a mixed condensate consisting of multiple bound states, e.g. an admixture of the ground state and excited states. Then the radius of the object will be enhanced by the increased size of the excited state wave functions, and the ratio of particles in each state would modify the scaling relation. A full discussion of this scenario is postponed to a future work.


\appendix

\section{Evaluation of numerical constants} \label{app:numbers}

In this section we illustrate the computations of the numerical coefficients used in the main text. The first subsection is based primarily on previous work \cite{Eby:2016cnq,Eby:2018dat} and is reproduced here for completeness; the second subsection is a new derivation for the effect of an external gravitational source on the axion star energy functional.

\subsection{Axion Star(s) Only}

First, note that given the notation of Eq. \eqref{eq:wavefn}, we can rewrite Eqs. \eqref{eq:BCD} and \eqref{eq:A2} much more compactly as
\begin{align}
 A_2 &= \int_0^\infty d^3\xi\,\xi\,F(\xi)^2 \\
 B_4 &= 32\pi^2\int_0^\infty d\xi\,\xi\,F(\xi)^2\int_0^\xi d\eta\,\eta^2\,F(\eta)^2 \\
 C_k &= \int_0^\infty d^3\xi  \,F(\xi)^k \\
 D_2 &= \int d^3\xi \,F'(\xi)^2.
\end{align}
Throughout this work, we have used the Gaussian approximation for each axion wavefunction
\begin{equation} \label{wavefn}
\psi_i(r)=\sqrt{\frac{N_i}{\sigma_i^3 \, \pi^{3/2}}}\,\exp\left(-\frac{r^2}{2\,\s_i^2}\right),
\end{equation}
which implies $F(\xi) = \exp\left(-\xi^2/2\right)$. Therefore the integrals above can be evaluated directly, giving
\begin{equation}
 A_2 = 2\pi, \quad 
 B_4 = \sqrt{2\pi}\,\pi^2, \quad
 C_2 = \pi^{3/2}, \quad
 C_4 = (\pi/2)^{3/2}, \quad
 D_2 = 3\,\pi^{3/2}/2,
\end{equation}
which we used in this work. Then the coefficients of the energy functional of Section \ref{sec:scaling_intro}, defined in Eq. \eqref{eq:abc}, are
\begin{equation}
 a = 3/4, \quad
 b = 1/\sqrt{2\pi}, \quad 
 c = 1/(32\pi\sqrt{2\pi}).
\end{equation}
This in turn determines the values
\begin{equation}
 \tilde n = 2\pi\sqrt{3}, \qquad \tilde\r = \sqrt{\frac{3}{32\pi}}
\end{equation}
for the critical rescaled particle number and corresponding radius of the single-axion condensate. 
Note that using other (non-Gaussian) approximate wavefunctions would not change the results appreciably; see \cite{Eby:2018dat} for details.

In the case of multiple axion flavors, the gravitational potential is determined as the sum of contributions from each condensate separately, which for the Gaussian ansatz is
\begin{align}
\Phi_g &= \frac{1}{M_P^2}\int \frac{{d^3}r'}{|\vec{r}-\vec{r}\,'|}\sum_{j=1}^{n_F} m_j {|\psi_j(r')|}^2 = \sum_{j=1}^{n_F} \frac{m_j}{M_P^2}\frac{{\rm erf}\left(r/\s_j\right)N_j}{r}.
\end{align}
Then, for the two-flavor model, the equations of motion (\ref{eq:eom1}-\ref{eq:eom2}) are
\begin{equation} \label{eq11}
0=-96\, \pi^{3/2}\rho_1+\sqrt{2}\,n_1(3+32\,\pi\,\rho_1^2)+\frac{m_r^6}{\sqrt{\lambda_r}}\,\frac{128\,\pi\,n_2\,\rho_1^5}{\left(m_r^4\rho_1^2+\lambda_r\,\rho_2^2\right)^{3/2}},
\end{equation}
\begin{equation} \label{eq21}
0=-96\, \pi^{3/2}\rho_2+\sqrt{2}\,n_2(3+32\,\pi\,\rho_2^2)+\lambda_r^2\,\frac{128\,\pi\,n_1\rho_2^5}{\left(m_r^4\rho_1^2+\lambda_r\,\rho_2^2\right)^{3/2}}.
\end{equation}

\subsection{Axion Star in Background Potential}

 In Section \ref{sec:analytic_ex}, we analyzed an axion star in the presence of a background potential, characterized by a mass $M_*$ and a radius $R_*$. In analogy to the above, we can determine the contribution to the axion star energy functional by using a generic ansatz for the external density profile,
\begin{equation}
 d_*(r) = d_0\,F_*(r/R_*)^2.
\end{equation}
The normalization is fixed as 
\begin{equation}
 M_* = \int d^3r\,d_*(r),
\end{equation}
and we rescale parameters as given by Eq. \eqref{eq:rescale1}
\begin{align} 
R_* = \sqrt{|\lambda|}\frac{M_P}{m^2}\r_*,	\qquad
M_* = \frac{M_P}{\sqrt{|\lambda|}} n_*.
\end{align}
This implies that
\begin{equation}
 d_0 = \frac{1}{C_2'}\frac{M_*}{R_*^3} = \frac{1}{C_2'} \frac{m^6}{\l^2\,M_P^2} \frac{n_*}{\r_*^3},
\end{equation}
where we define the constant
\begin{equation}
 C_2' \equiv \int d^3\eta\,F_*(\eta)^2.
\end{equation}

The resulting gravitational potential is
\begin{align}
 \Phi_*(r) &= -G\,\int d^3r'\frac{d_*(r')}{\left| \vec{r}\,' - \vec{r}\right|} \nn \\
 		&= -\frac{4\pi G}{r}\,\int_0^r dr'\,r'^2 d_*(r') - 4\pi\,G\,\int_r^\infty dr'\,r'\,d_*(r') \nn \\
		&= - \frac{4\pi G}{C_2'}\,M_* \left[\frac{1}{r}\int_0^{r/R_*}\,d\eta\,\eta^2\,F_*(\eta)^2 + \frac{1}{R_*}\int_{r/R_*}^\infty\,d\eta\,\eta\,F_*(\eta)^2 \right].
\end{align}
This potential, coupled to the axion star, induces an additional term in the energy functional of the form
\begin{align} \label{eq: Eg_ext}
E_{g,\text{ext}} &= m\, \int d^3 r \, \Phi_{*}(r)\,\left| \psi(r) \right|^2 \nn \\
			&= -\frac{16\pi^2\,m\,G\,M_*}{C_2'\,M_P^2}\frac{N}{C_2\,\s} \int_0^\infty d\xi\,\xi^2\,F(\xi)^2
						\left( \frac{1}{\xi}\int_0^{\xi(\s/R_*)}\,d\eta\,\eta^2\,F_*(\eta)^2 
							+ \frac{1}{R_*/\s}\int_{\xi(\s/R_*)}^\infty\,d\eta\,\eta\,F_*(\eta)^2\right) \nn \\
			&= -\frac{B_4'(\r)}{2\,C_2\,C_2'}\frac{m^2}{M_P\,\l^{3/2}}\frac{n\,n_*}{\r},
\end{align}
where we defined
\begin{equation} \label{B4p}
 B_4'(\r) \equiv 32\pi^2 \int_0^\infty d\xi\,\xi\,F(\xi)^2
						\left(\int_0^{\xi(\r/\r_*)}\,d\eta\,\eta^2\,F_*(\eta)^2 + \frac{\xi\,\r}{\r_*}\int_{\xi(\r/\r_*)}^\infty\,d\eta\,\eta\,F_*(\eta)^2\right)
\end{equation}
and used the fact that $\s/R_* = \r/\r_*$.
Importantly, $B_4'$ depends on $\r$, unlike the constant $B_4$.

In the numerical calculations of Section \ref{sec:analytic_ex}, we assumed a constant-density profile for the external source; in our notation this means $F_*(\xi) = 1$ for $r\leq R_*$ and $0$ elsewhere. In that case, $C_2' = 4\pi/3$, and the gravitational potential is given by Eq. \eqref{eq:V_ext}. Using Eqs. \eqref{eq: Eg_ext} and \eqref{B4p}, we can now compute the contribution to the axion star energy directly. Due to the discontinuity at $r=R_*$, it is more straightforward to compute $B_4'$ in two separate regions and add the results. 

First, for $r\leq R_*$, we have $\xi = r/\s \leq R_*/\s = \r_*/\r$; in this region we find
\begin{align} \label{B41}
 B_4'(\r)\Big|_{r\leq R_*} &= 32\pi^2\,\int_0^{\r_*/\r} d\xi\,\xi\,F(\xi)^2\left(\int_0^{\xi(\r/\r_*)}d\eta\,\eta^2\,F_*(\eta)^2 + \frac{\xi\,\r}{\r_*}\int_{\xi(\r/\r_*)}^1\,d\eta\,\eta\,F_*(\eta)^2\right) \nn \\
		&= 32\pi^2\left[-\frac{\r^3}{6\,\r_*^3}\int_0^{\r_*/\r} d\xi\,\xi^4\,F(\xi)^2 + \frac{\r}{2\,\r_*}\int_0^{\r_*/\r} d\xi\,\xi^2\,F(\xi)^2 \right] \nn \\
		&=  \frac{2\pi^2}{3}\left[\left(\frac{6\,\r^2}{\r_*^2} - 8\right)\exp\left(-\frac{\r_*^2}{\r^2}\right) - 3\sqrt{\pi}\,\left(\frac{\r^3}{\r_*^3} - \frac{2\,\r}{\r_*}\right)\,{\rm erf}\left(\frac{\r_*}{\r}\right)\right],
\end{align}
where in the last step we used the Gaussian profile of Eq.~\eqref{wavefn} for the axion wavefunction. In the other region $r>R_*$, we have $\xi = r/\s > R_*/\s = \r_*/\r$; in that case only the first integral in Eq. \eqref{B4p} contributes, and we obtain
\begin{align} \label{B42}
 B_4'(\r)\Big|_{r> R_*} &= 32\pi^2\,\int_{\r_*/\r}^\infty d\xi\,\xi\,F(\xi)^2\left(\int_0^{1}d\eta\,\eta^2\,F_*(\eta)^2\right) \nn \\
		&= \frac{32\pi^2}{3}\int_{\r_*/\r}^\infty d\xi\,\xi\,F(\xi)^2 \nn \\
		&= \frac{16\pi^2}{3}\,\exp\left(-\frac{\r_*^2}{\r^2}\right),
\end{align}
where in the last step we again used the Gaussian profile for $F(\xi)$. Combining Eqs. (\ref{eq: Eg_ext}), \eqref{B41}, and \eqref{B42}, we obtain
\begin{equation}
 \frac{\l^{3/2}\,M_P}{m^2\,n} E_{\rm g,ext} = \frac{3}{4} \left[\left(\frac{\r^3}{\r_*^3} - \frac{2\,\r}{\r_*}\right){\rm erf}\left(\frac{\r_*}{\r}\right) - \frac{2}{\sqrt{\pi}}\left(\frac{\r^2}{\r_*^2}\right)\exp\left(-\frac{\r_*^2}{\r^2}\right)\right] 
 					\frac{n_*}{\r}.
\end{equation}
Finally, expanding in $\r_* \ll \r$, we obtain
\begin{equation} \label{Egexp}
 \frac{\l^{3/2}\,M_P}{m^2\,n} E_{\rm g,ext} \approx \frac{a'\,n_*\,\r_*}{\r^2} - \frac{b'\,n_*}{\r} + \frac{c'\,n_*\,\r_*^2}{\r^3} + \Ocal\left(\r_*/\r\right)^4
\end{equation}
with
\begin{equation}
 a' = 0, \qquad b' = \frac{2}{\sqrt{\pi}}, \qquad c' = \frac{2}{5\sqrt{\pi}}.
\end{equation}

\section{Model-independent results} \label{app:MIresults}

\begin{figure}[t]
\centering 
 \includegraphics[scale=0.24]{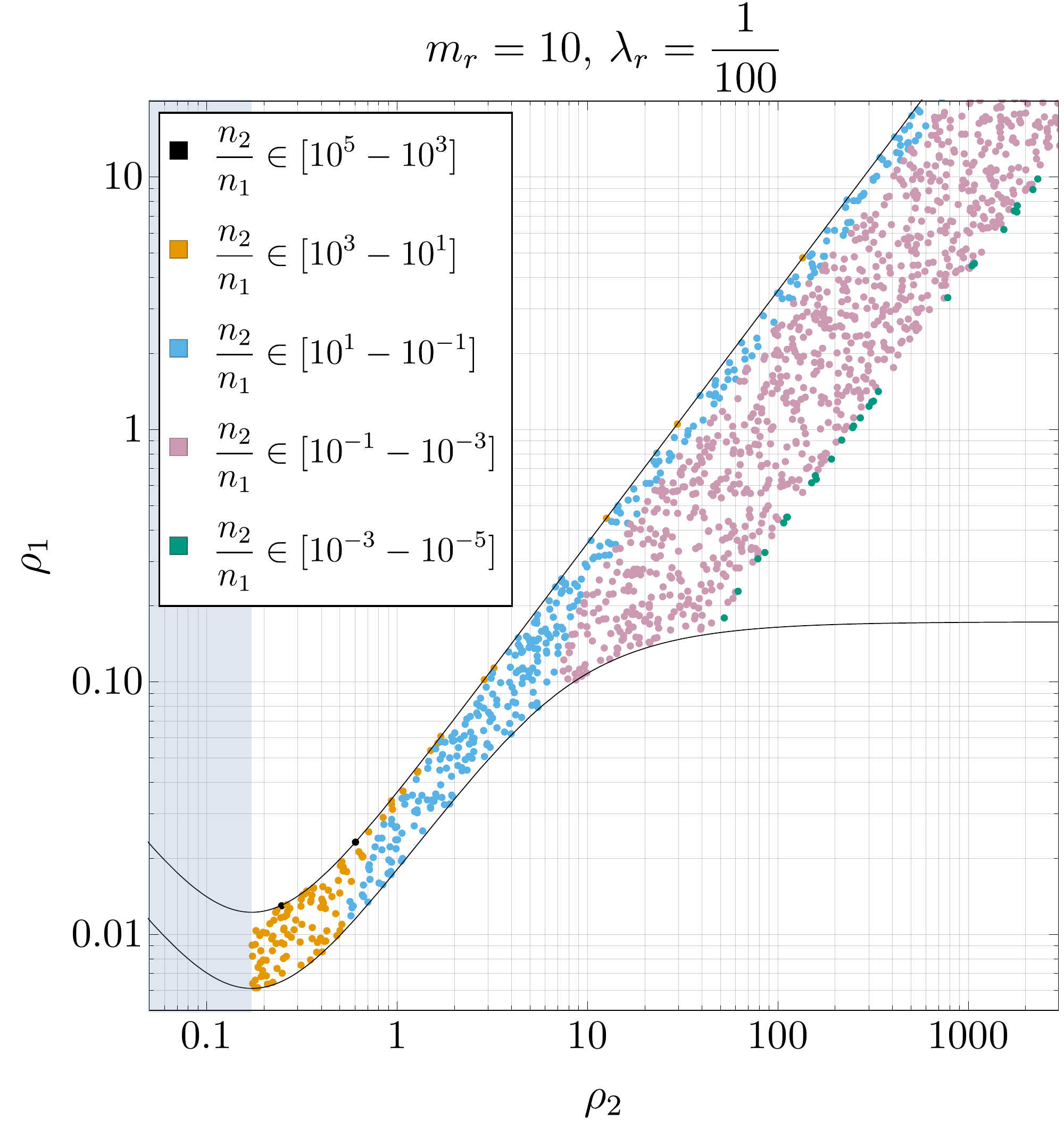} \,
 \includegraphics[scale=0.24]{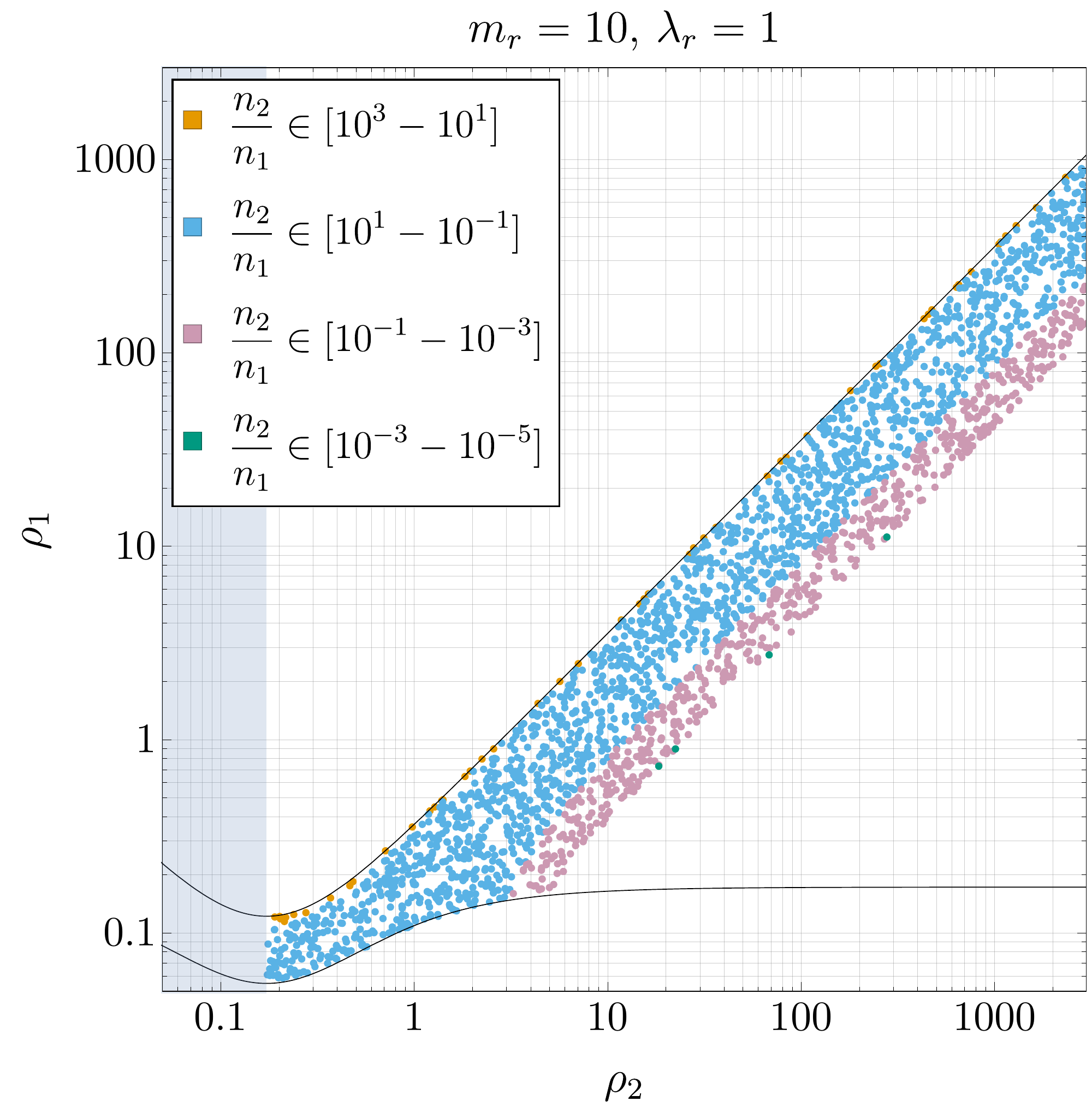} \,
 \includegraphics[scale=0.24]{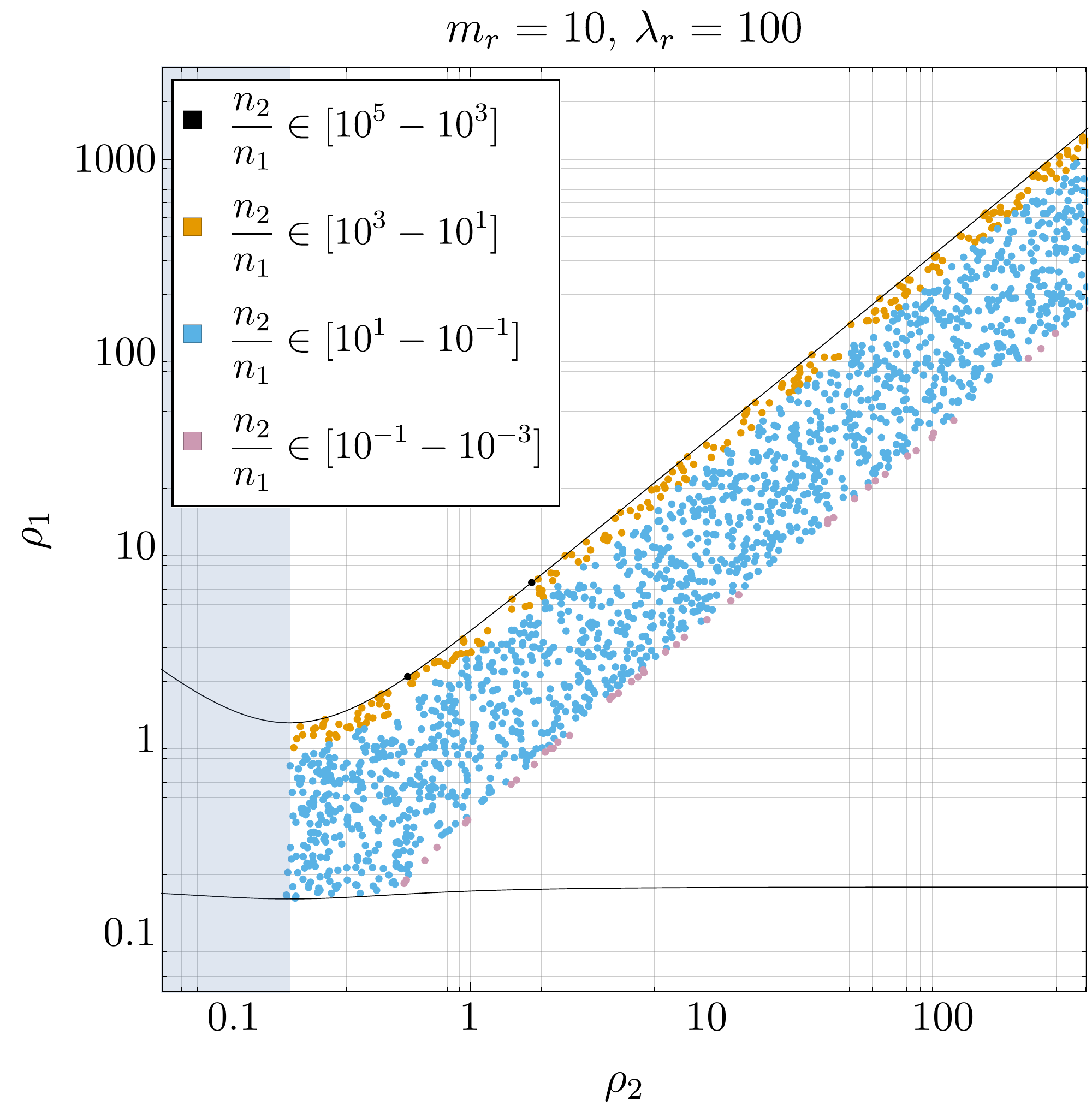}
 \\
  \includegraphics[scale=0.24]{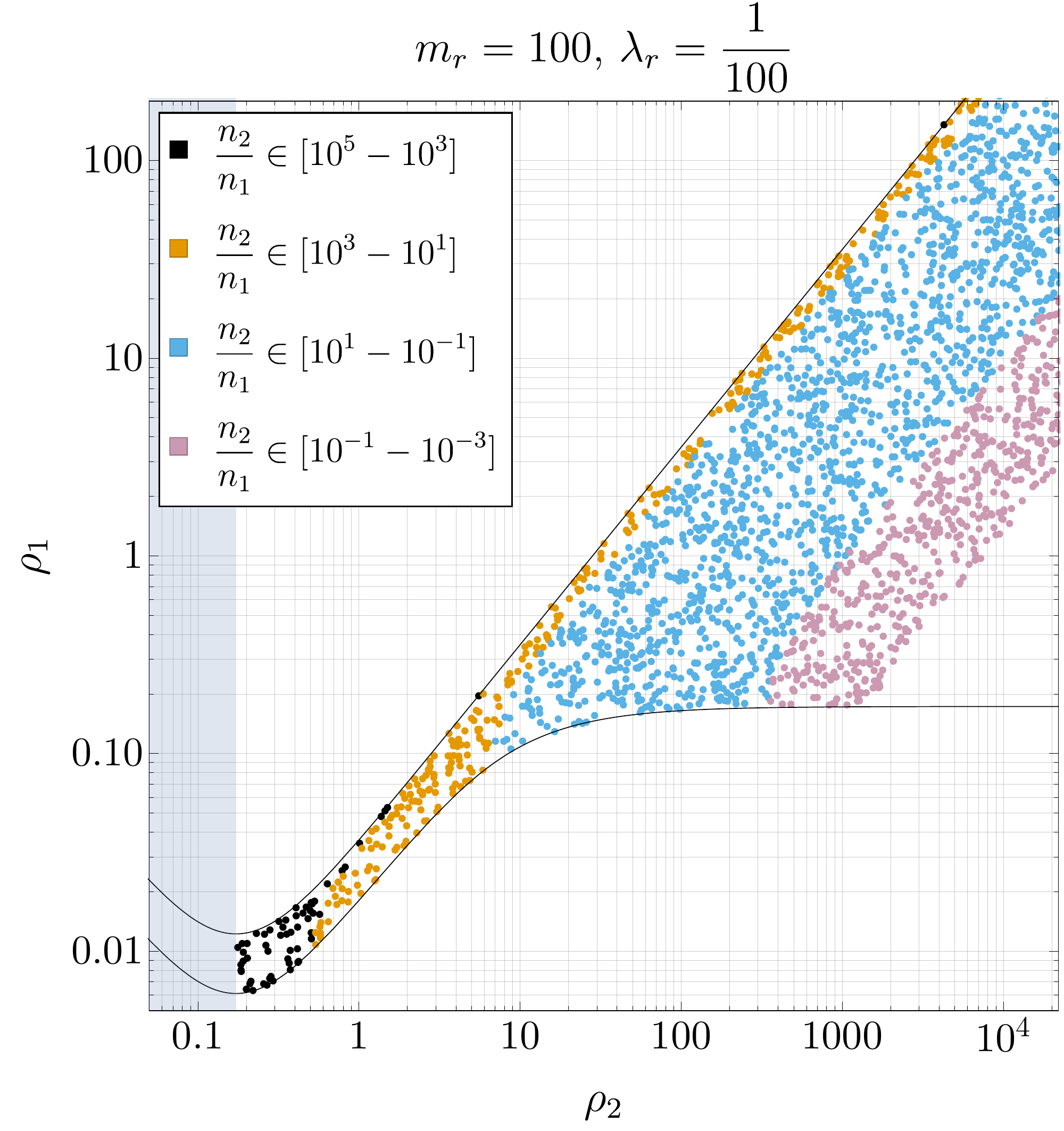} \,
 \includegraphics[scale=0.24]{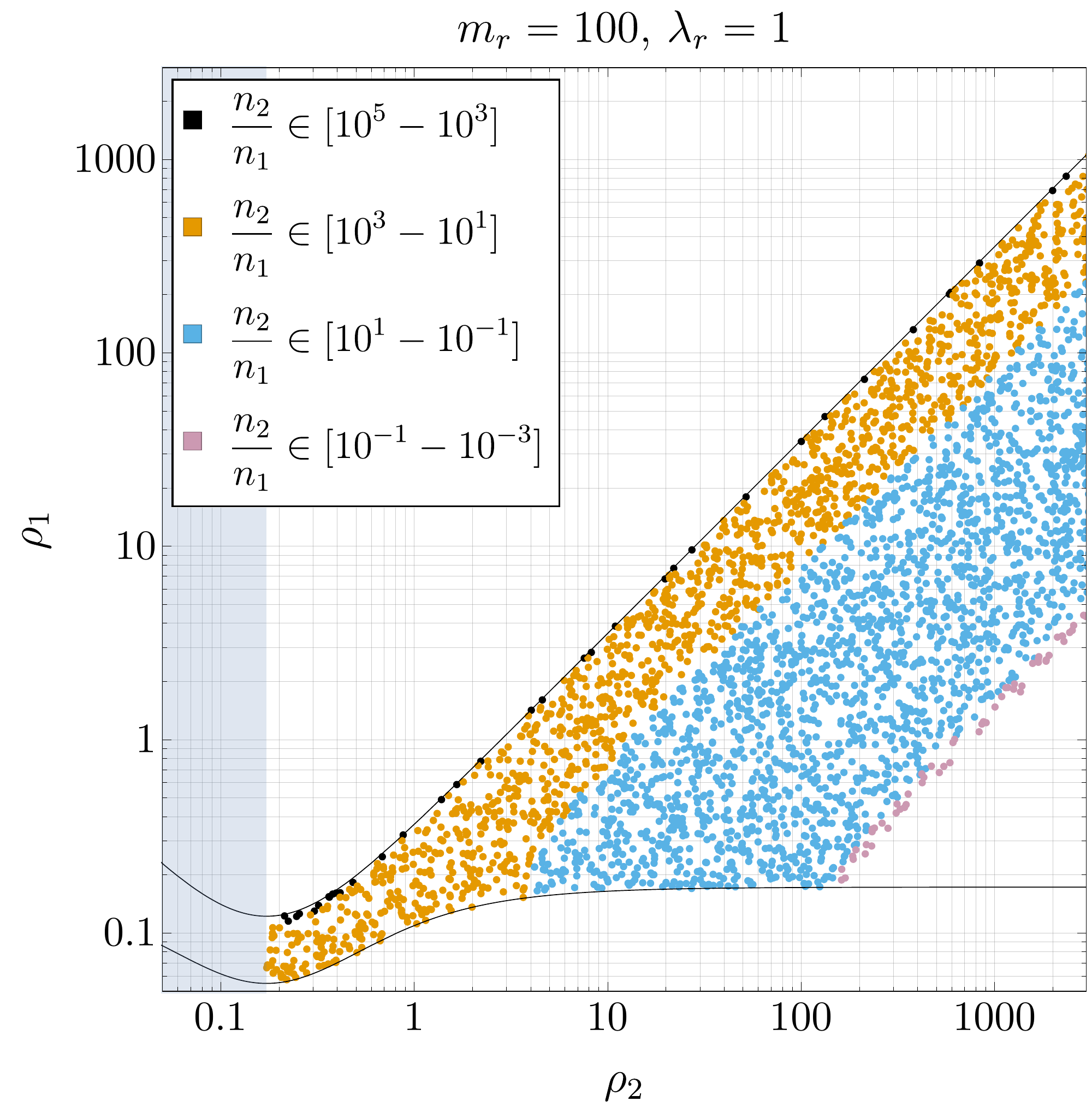} \,
 \includegraphics[scale=0.24]{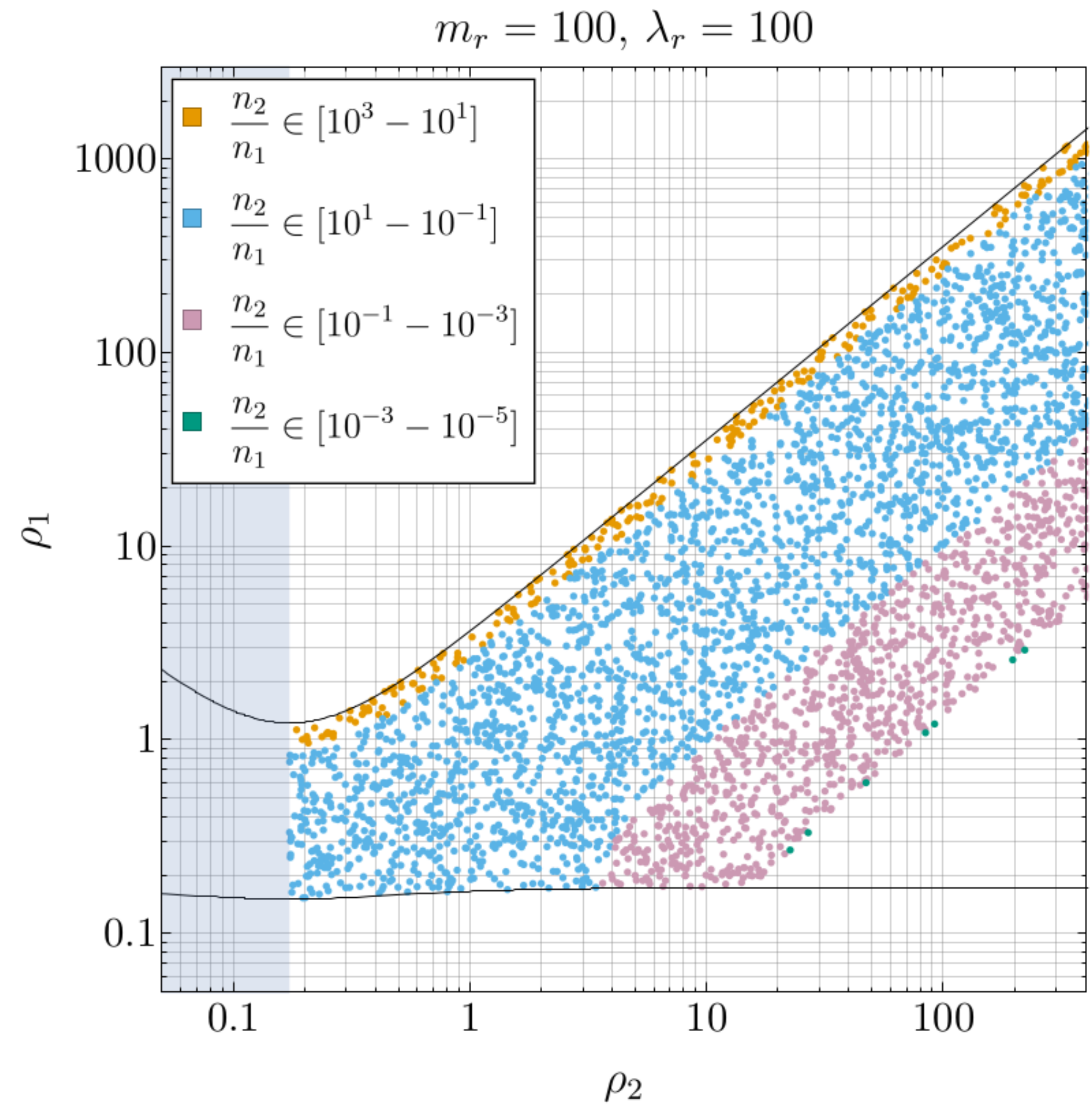}
\caption{The physical region $\mathcal B$, as defined by a random sampling of rescaled radial parameters $\r_1$ and $\r_2$. The black solid lines are the boundaries of $\Bcal$ in the $m_r\to\infty$ limit, as discussed in Section \ref{sec:mtoinf}, and the shaded region corresponds to $\r_2<\tilde\r$. 
The color of the points are determined by the rescaled number ratios $n_2/n_1$, as described in the text.
The top (bottom) row corresponds to the choice $m_r = 10$ ($m_r = 100$), whereas the left, center, and right panels correpond to $\l_r=1/100,1,100$ (respectively). Physical values can be obtained by using $R_i = M_P\,\sqrt{\l_i}\,\r_i/m_i^2$ with $i=1,2$.}
\label{Binf_rho}
\end{figure}

In this section we give the generalized results of Section \ref{sec:random} in terms of the rescaled quantities $\r_1$ and $\r_2$ rather than $R_1$ and $R_2$, the latter requiring evaluation of particular choices of parameters $m_1$ and $\l_1$. The results here can be evaluated for two-component condensates in models we have not considered in the main text.

First, we illustrate the rescaled radii $\r_1$ and $\r_2$ in Figure \ref{Binf_rho}, for $m_r = 10,100$ (top and bottom rows) and for $\l_r = 1/100,1,100$ (left, center, and right columns). To translate these results to physical quantities for a given $m_1$ and $\l_1$, one need only compute $R_i = M_P\,\sqrt{\l_i}\,\r_i/m_i^2$. For Benchmark 1, where $m_1=10^{-22}$ eV and $\l_1=10^{-94}$, one recovers Figure \ref{Binf}.

\begin{figure}[t]
\centering 
 \includegraphics[scale=0.242]{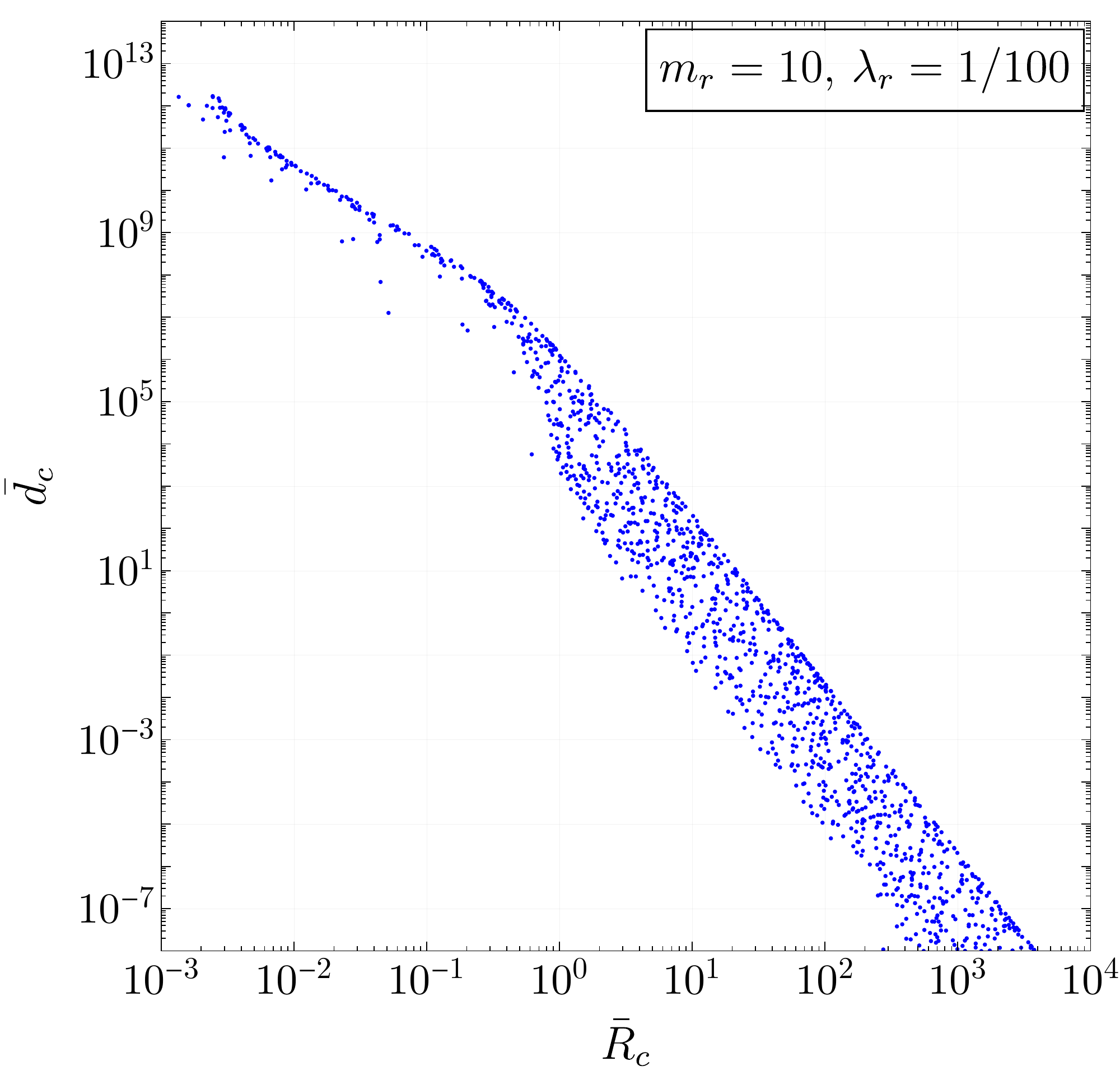} \,
 \includegraphics[scale=0.242]{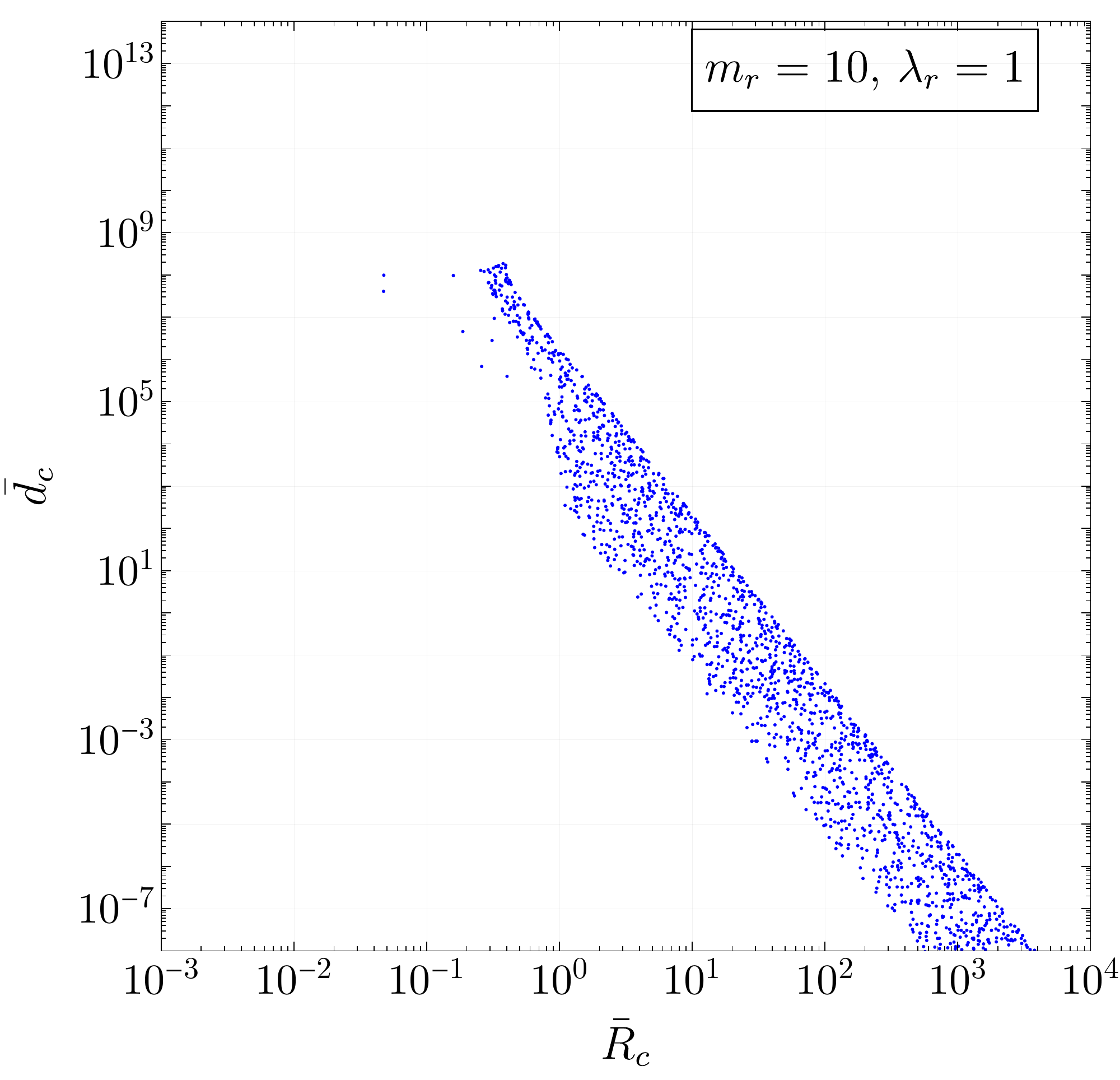} \,
 \includegraphics[scale=0.242]{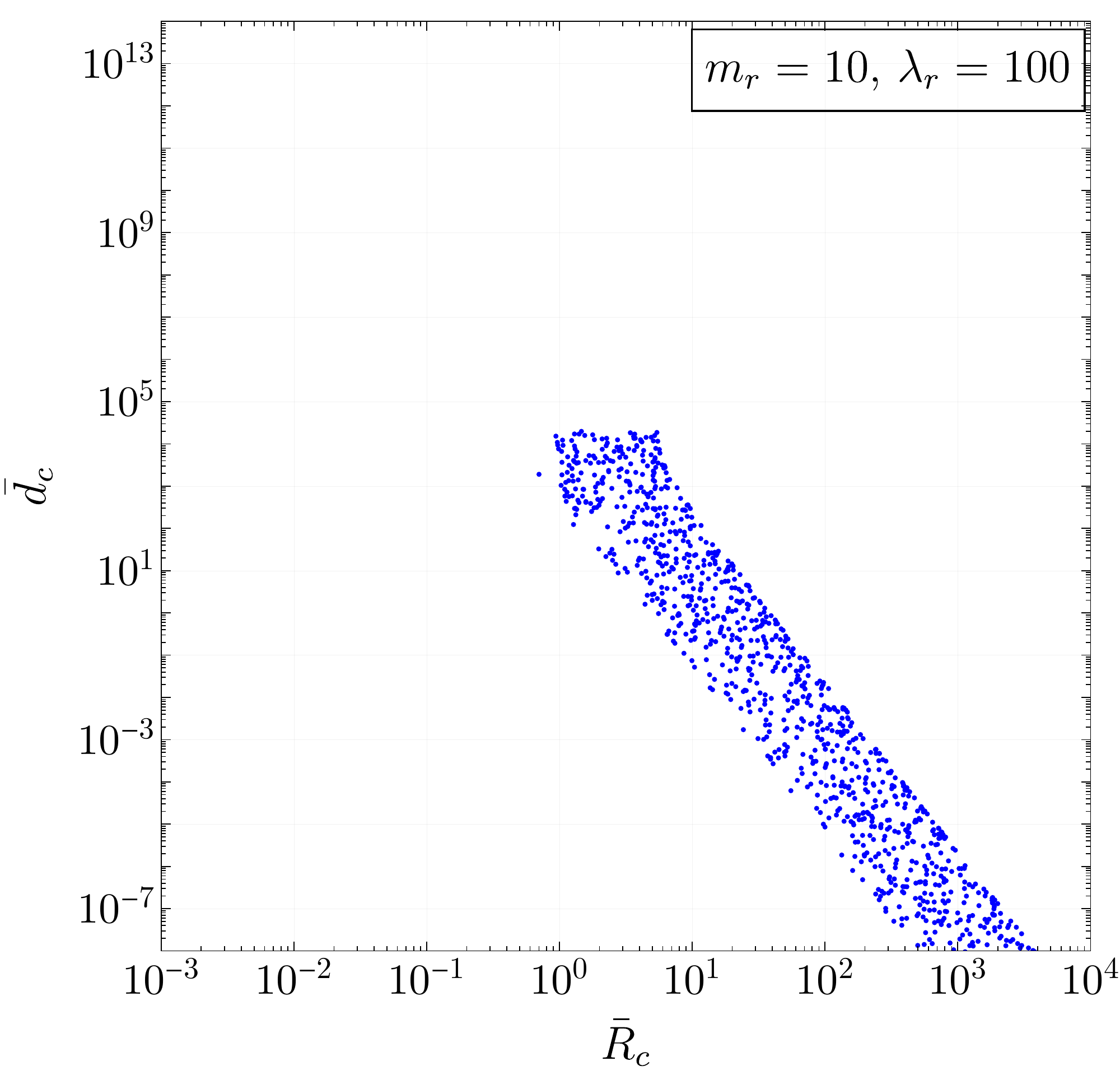}
 \\
  \includegraphics[scale=0.242]{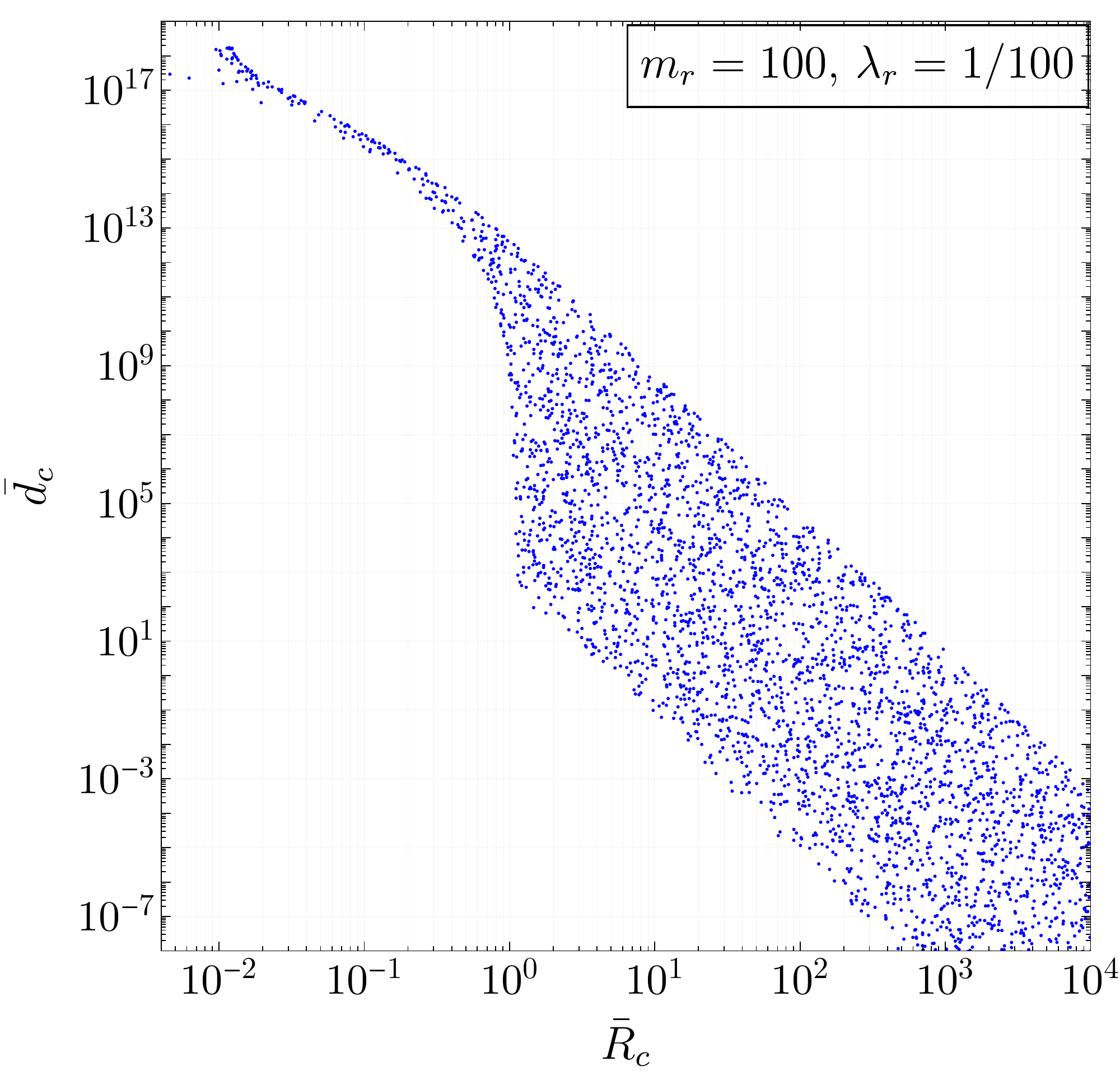} \,
 \includegraphics[scale=0.242]{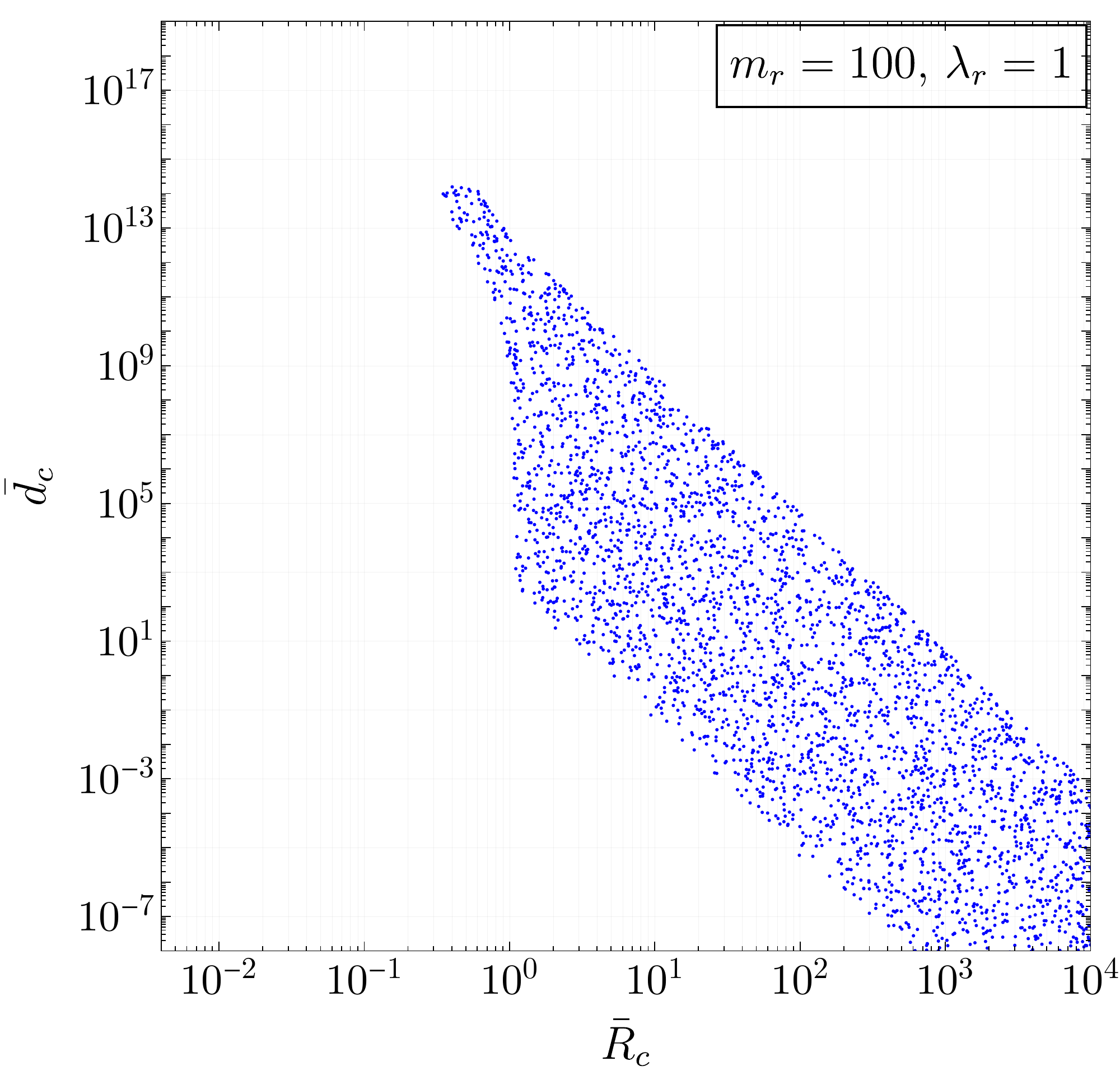} \,
 \includegraphics[scale=0.242]{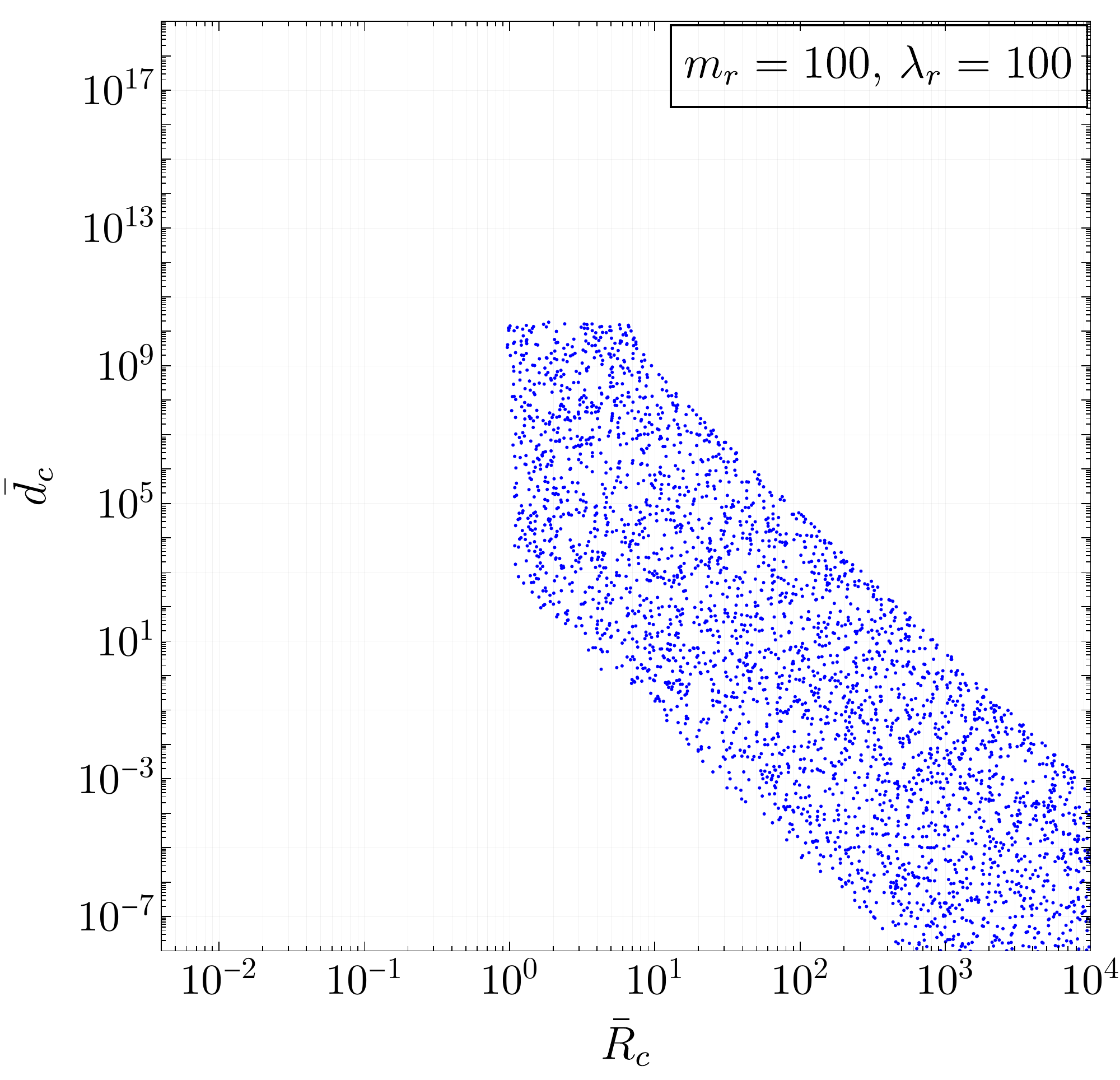}
\caption{The dimensionless density $\bar{d}_c$ and radius $\bar{R}_c$ for stable two-component condensates. The blue points represent stable configurations. 
The top (bottom) row corresponds to the choice $m_r = 10$ ($m_r = 100$), whereas the left, center, and right panels correpond to $\l_r=1/100,1,100$ (respectively). To obtain physical units, one multiplies the result on the vertical axis by $m_1^6/M_P^2\,\l_1^2$ and that of the the horizontal axis by $M_P\,\sqrt{\l_1}/m_1^2$.}
\label{fig:dcRc_DIM}
\end{figure}

To obtain model-independent results for $d_c$ and $R_c$, we extract the dependence of these quantities on the inputs $m_1$ and $\l_1$. To this end, we refer to Eq.~(\ref{rhocR}), defining $d_c = (m_1^6/M_P^2\,\l_1^2)\, \bar{d}_c$ and $R_c = (M_P\,\sqrt{\l_1}/m_1^2)\,\bar{R}_c$ with
\begin{align}
\bar{d}_c &= \frac{1}{C_2} \left[\frac{n_1}{\rho_1^3}+\frac{m_r^6}{\lambda_r^2}\frac{n_2}{\rho_2^3}\right] \\
\bar{R}_c &= \frac{A_2\,\sqrt{\l_r}}{m_r^2}\left[ \frac{m_r^3\,n_1\,\rho_1+n_2\,\rho_2}{m_r\,\sqrt{\lambda_r}\,n_1+n_2} \right].
\end{align}
The `barred' quantities $\bar{d}_c$ and $\bar{R}_c$ depend only on the sampled quantities $\r_1$ and $\r_2$, the derived values of $n_1$ and $n_2$, and the ratios $m_r$ and $\l_r$ (up to $\Ocal(1)$ numbers $A_2$ and $C_2$, resulting from the choice of profile). These dimensionless quantities are depicted in Figure \ref{fig:dcRc_DIM}, for $m_r = 10,100$ (top and bottom rows) and for $\l_r = 1/100,1,100$ (left, center, and right columns).

\clearpage

\acknowledgments
We thank H. Deng for bringing this discussion to our attention during the 2018 IPA conference and to the organizers of this conference. We are also grateful to N. Bar and C. Sun for useful discussions. L.C.R.W. thanks the Aspen Center for Physics, which is supported by National Science Foundation grant PHY-1607611, where some of the research was conducted. The work of J.E. was supported by the Zuckerman STEM Leadership Fellowship. M.L. was supported by the Ford Foundation Fellowship and NSF Graduate Research Program Fellowship.  L.S. and L.C.R.W. thank the University of Cincinnati Office of Research Faculty Bridge Program for funding through the Faculty Bridge Grant. L.S. also thanks the Department of Physics at the University of Cincinnati for financial support in the form of the Violet M. Diller Fellowship.

\bibliography{ref}

\providecommand{\href}[2]{#2}\begingroup\raggedright\begin{thebibliography}{10}

\bibitem{Arvanitaki:2009fg}
A.~Arvanitaki, S.~Dimopoulos, S.~Dubovsky, N.~Kaloper and J.~March-Russell,
  \emph{{String Axiverse}},
  \href{https://doi.org/10.1103/PhysRevD.81.123530}{\emph{Phys. Rev.}
  {\bfseries D81} (2010) 123530}
  [\href{https://arxiv.org/abs/0905.4720}{{\ttfamily 0905.4720}}].

\bibitem{Kaplan:2015fuy}
D.~E. Kaplan and R.~Rattazzi, \emph{{Large field excursions and approximate
  discrete symmetries from a clockwork axion}},
  \href{https://doi.org/10.1103/PhysRevD.93.085007}{\emph{Phys. Rev.}
  {\bfseries D93} (2016) 085007}
  [\href{https://arxiv.org/abs/1511.01827}{{\ttfamily 1511.01827}}].

\bibitem{Kaup:1968zz}
D.~J. Kaup, \emph{{Klein-Gordon Geon}},
  \href{https://doi.org/10.1103/PhysRev.172.1331}{\emph{Phys. Rev.} {\bfseries
  172} (1968) 1331}.

\bibitem{Ruffini:1969qy}
R.~Ruffini and S.~Bonazzola, \emph{{Systems of selfgravitating particles in
  general relativity and the concept of an equation of state}},
  \href{https://doi.org/10.1103/PhysRev.187.1767}{\emph{Phys. Rev.} {\bfseries
  187} (1969) 1767}.

\bibitem{Colpi:1986ye}
M.~Colpi, S.~L. Shapiro and I.~Wasserman, \emph{{Boson Stars: Gravitational
  Equilibria of Selfinteracting Scalar Fields}},
  \href{https://doi.org/10.1103/PhysRevLett.57.2485}{\emph{Phys. Rev. Lett.}
  {\bfseries 57} (1986) 2485}.

\bibitem{Turner:1983he}
M.~S. Turner, \emph{{Coherent Scalar Field Oscillations in an Expanding
  Universe}}, \href{https://doi.org/10.1103/PhysRevD.28.1243}{\emph{Phys. Rev.}
  {\bfseries D28} (1983) 1243}.

\bibitem{Press:1989id}
W.~H. Press, B.~S. Ryden and D.~N. Spergel, \emph{{Single Mechanism for
  Generating Large Scale Structure and Providing Dark Missing Matter}},
  \href{https://doi.org/10.1103/PhysRevLett.64.1084}{\emph{Phys. Rev. Lett.}
  {\bfseries 64} (1990) 1084}.

\bibitem{Sin:1992bg}
S.-J. Sin, \emph{{Late time cosmological phase transition and galactic halo as
  Bose liquid}}, \href{https://doi.org/10.1103/PhysRevD.50.3650}{\emph{Phys.
  Rev.} {\bfseries D50} (1994) 3650}
  [\href{https://arxiv.org/abs/hep-ph/9205208}{{\ttfamily hep-ph/9205208}}].

\bibitem{Hu:2000ke}
W.~Hu, R.~Barkana and A.~Gruzinov, \emph{{Cold and fuzzy dark matter}},
  \href{https://doi.org/10.1103/PhysRevLett.85.1158}{\emph{Phys. Rev. Lett.}
  {\bfseries 85} (2000) 1158}
  [\href{https://arxiv.org/abs/astro-ph/0003365}{{\ttfamily
  astro-ph/0003365}}].

\bibitem{Hui:2016ltb}
L.~Hui, J.~P. Ostriker, S.~Tremaine and E.~Witten, \emph{{Ultralight scalars as
  cosmological dark matter}},
  \href{https://doi.org/10.1103/PhysRevD.95.043541}{\emph{Phys. Rev.}
  {\bfseries D95} (2017) 043541}
  [\href{https://arxiv.org/abs/1610.08297}{{\ttfamily 1610.08297}}].

\bibitem{Lee:2017qve}
J.-W. Lee, \emph{{Brief History of Ultra-light Scalar Dark Matter Models}},
  \href{https://doi.org/10.1051/epjconf/201816806005}{\emph{EPJ Web Conf.}
  {\bfseries 168} (2018) 06005}
  [\href{https://arxiv.org/abs/1704.05057}{{\ttfamily 1704.05057}}].

\bibitem{Schive:2014dra}
H.-Y. Schive, T.~Chiueh and T.~Broadhurst, \emph{{Cosmic Structure as the
  Quantum Interference of a Coherent Dark Wave}},
  \href{https://doi.org/10.1038/nphys2996}{\emph{Nature Phys.} {\bfseries 10}
  (2014) 496} [\href{https://arxiv.org/abs/1406.6586}{{\ttfamily 1406.6586}}].

\bibitem{Schive:2014hza}
H.-Y. Schive, M.-H. Liao, T.-P. Woo, S.-K. Wong, T.~Chiueh, T.~Broadhurst
  et~al., \emph{{Understanding the Core-Halo Relation of Quantum Wave Dark
  Matter from 3D Simulations}},
  \href{https://doi.org/10.1103/PhysRevLett.113.261302}{\emph{Phys. Rev. Lett.}
  {\bfseries 113} (2014) 261302}
  [\href{https://arxiv.org/abs/1407.7762}{{\ttfamily 1407.7762}}].

\bibitem{Schwabe:2016rze}
B.~Schwabe, J.~C. Niemeyer and J.~F. Engels, \emph{{Simulations of solitonic
  core mergers in ultralight axion dark matter cosmologies}},
  \href{https://doi.org/10.1103/PhysRevD.94.043513}{\emph{Phys. Rev.}
  {\bfseries D94} (2016) 043513}
  [\href{https://arxiv.org/abs/1606.05151}{{\ttfamily 1606.05151}}].

\bibitem{Veltmaat:2016rxo}
J.~Veltmaat and J.~C. Niemeyer, \emph{{Cosmological particle-in-cell
  simulations with ultralight axion dark matter}},
  \href{https://doi.org/10.1103/PhysRevD.94.123523}{\emph{Phys. Rev.}
  {\bfseries D94} (2016) 123523}
  [\href{https://arxiv.org/abs/1608.00802}{{\ttfamily 1608.00802}}].

\bibitem{Mocz:2017wlg}
P.~Mocz, M.~Vogelsberger, V.~H. Robles, J.~Zavala, M.~Boylan-Kolchin,
  A.~Fialkov et~al., \emph{{Galaxy formation with BECDM: I. Turbulence and
  relaxation of idealized haloes}},
  \href{https://doi.org/10.1093/mnras/stx1887}{\emph{Mon. Not. Roy. Astron.
  Soc.} {\bfseries 471} (2017) 4559}
  [\href{https://arxiv.org/abs/1705.05845}{{\ttfamily 1705.05845}}].

\bibitem{Weinberg:2013aya}
D.~H. Weinberg, J.~S. Bullock, F.~Governato, R.~Kuzio~de Naray and A.~H.~G.
  Peter, \emph{{Cold dark matter: controversies on small scales}},
  \href{https://doi.org/10.1073/pnas.1308716112}{\emph{Proc. Nat. Acad. Sci.}
  {\bfseries 112} (2015) 12249}
  [\href{https://arxiv.org/abs/1306.0913}{{\ttfamily 1306.0913}}].

\bibitem{Marsh:2015wka}
D.~J.~E. Marsh and A.-R. Pop, \emph{{Axion dark matter, solitons and the
  cusp–core problem}},
  \href{https://doi.org/10.1093/mnras/stv1050}{\emph{Mon. Not. Roy. Astron.
  Soc.} {\bfseries 451} (2015) 2479}
  [\href{https://arxiv.org/abs/1502.03456}{{\ttfamily 1502.03456}}].

\bibitem{Irsic:2017yje}
V.~Ir\v{s}i\v{c}, M.~Viel, M.~G. Haehnelt, J.~S. Bolton and G.~D. Becker,
  \emph{{First constraints on fuzzy dark matter from Lyman-$\alpha$ forest data
  and hydrodynamical simulations}},
  \href{https://doi.org/10.1103/PhysRevLett.119.031302}{\emph{Phys. Rev. Lett.}
  {\bfseries 119} (2017) 031302}
  [\href{https://arxiv.org/abs/1703.04683}{{\ttfamily 1703.04683}}].

\bibitem{Leong:2018opi}
K.-H. Leong, H.-Y. Schive, U.-H. Zhang and T.~Chiueh, \emph{{Testing
  extreme-axion wave dark matter using the BOSS Lyman-Alpha forest data}},
  \href{https://doi.org/10.1093/mnras/stz271}{\emph{Mon. Not. Roy. Astron.
  Soc.} {\bfseries 484} (2019) 4273}
  [\href{https://arxiv.org/abs/1810.05930}{{\ttfamily 1810.05930}}].

\bibitem{Schutz:2020jox}
K.~Schutz, \emph{{Subhalo mass function and ultralight bosonic dark matter}},
  \href{https://doi.org/10.1103/PhysRevD.101.123026}{\emph{Phys. Rev. D}
  {\bfseries 101} (2020) 123026}
  [\href{https://arxiv.org/abs/2001.05503}{{\ttfamily 2001.05503}}].

\bibitem{Benito:2020avv}
M.~Benito, J.~C. Criado, G.~Hütsi, M.~Raidal and H.~Veermäe,
  \emph{{Implications of Milky Way substructures for the nature of dark
  matter}}, \href{https://doi.org/10.1103/PhysRevD.101.103023}{\emph{Phys. Rev.
  D} {\bfseries 101} (2020) 103023}
  [\href{https://arxiv.org/abs/2001.11013}{{\ttfamily 2001.11013}}].

\bibitem{Bar:2018acw}
N.~Bar, D.~Blas, K.~Blum and S.~Sibiryakov, \emph{{Galactic rotation curves
  versus ultralight dark matter: Implications of the soliton-host halo
  relation}}, \href{https://doi.org/10.1103/PhysRevD.98.083027}{\emph{Phys.
  Rev.} {\bfseries D98} (2018) 083027}
  [\href{https://arxiv.org/abs/1805.00122}{{\ttfamily 1805.00122}}].

\bibitem{Bar:2019bqz}
N.~Bar, K.~Blum, J.~Eby and R.~Sato, \emph{{Ultralight dark matter in disk
  galaxies}}, \href{https://doi.org/10.1103/PhysRevD.99.103020}{\emph{Phys.
  Rev.} {\bfseries D99} (2019) 103020}
  [\href{https://arxiv.org/abs/1903.03402}{{\ttfamily 1903.03402}}].

\bibitem{Safarzadeh:2019sre}
M.~Safarzadeh and D.~N. Spergel, \emph{{Ultra-light Dark Matter is Incompatible
  with the Milky Way's Dwarf Satellites}},
  \href{https://arxiv.org/abs/1906.11848}{{\ttfamily 1906.11848}}.

\bibitem{Broadhurst:2018fei}
H.~N. Luu, S.-H.~H. Tye and T.~Broadhurst, \emph{{Multiple Ultralight Axionic
  Wave Dark Matter and Astronomical Structures}},
  \href{https://doi.org/10.1016/j.dark.2020.100636}{\emph{Phys. Dark Univ.}
  {\bfseries 30} (2020) 100636}
  [\href{https://arxiv.org/abs/1811.03771}{{\ttfamily 1811.03771}}].

\bibitem{Amendola:2005ad}
L.~Amendola and R.~Barbieri, \emph{{Dark matter from an ultra-light
  pseudo-Goldsone-boson}},
  \href{https://doi.org/10.1016/j.physletb.2006.08.069}{\emph{Phys. Lett.}
  {\bfseries B642} (2006) 192}
  [\href{https://arxiv.org/abs/hep-ph/0509257}{{\ttfamily hep-ph/0509257}}].

\bibitem{Marsh:2013ywa}
D.~J.~E. Marsh and J.~Silk, \emph{{A Model For Halo Formation With Axion Mixed
  Dark Matter}}, \href{https://doi.org/10.1093/mnras/stt2079}{\emph{Mon. Not.
  Roy. Astron. Soc.} {\bfseries 437} (2014) 2652}
  [\href{https://arxiv.org/abs/1307.1705}{{\ttfamily 1307.1705}}].

\bibitem{Fan:2016rda}
J.~Fan, \emph{{Ultralight Repulsive Dark Matter and BEC}},
  \href{https://doi.org/10.1016/j.dark.2016.10.005}{\emph{Phys. Dark Univ.}
  {\bfseries 14} (2016) 84} [\href{https://arxiv.org/abs/1603.06580}{{\ttfamily
  1603.06580}}].

\bibitem{Berman:2020kko}
G.~P. Berman, V.~N. Gorshkov, V.~I. Tsifrinovich, M.~Merkli and V.~V.
  Tereshchuk, \emph{{Two-Component Axionic Dark Matter Halos}},
  \href{https://arxiv.org/abs/2001.10584}{{\ttfamily 2001.10584}}.

\bibitem{Chavanis:2011zi}
P.-H. Chavanis, \emph{{Mass-radius relation of Newtonian self-gravitating
  Bose-Einstein condensates with short-range interactions: I. Analytical
  results}}, \href{https://doi.org/10.1103/PhysRevD.84.043531}{\emph{Phys.
  Rev.} {\bfseries D84} (2011) 043531}
  [\href{https://arxiv.org/abs/1103.2050}{{\ttfamily 1103.2050}}].

\bibitem{Chavanis:2011zm}
P.~H. Chavanis and L.~Delfini, \emph{{Mass-radius relation of Newtonian
  self-gravitating Bose-Einstein condensates with short-range interactions: II.
  Numerical results}},
  \href{https://doi.org/10.1103/PhysRevD.84.043532}{\emph{Phys. Rev.}
  {\bfseries D84} (2011) 043532}
  [\href{https://arxiv.org/abs/1103.2054}{{\ttfamily 1103.2054}}].

\bibitem{Eby:2014fya}
J.~Eby, P.~Suranyi, C.~Vaz and L.~C.~R. Wijewardhana, \emph{{Axion Stars in the
  Infrared Limit}}, \href{https://doi.org/10.1007/JHEP11(2016)134,
  10.1007/JHEP03(2015)080}{\emph{JHEP} {\bfseries 03} (2015) 080}
  [\href{https://arxiv.org/abs/1412.3430}{{\ttfamily 1412.3430}}].

\bibitem{Peebles:2000yy}
P.~J.~E. Peebles, \emph{{Fluid dark matter}},
  \href{https://doi.org/10.1086/312677}{\emph{Astrophys. J.} {\bfseries 534}
  (2000) L127} [\href{https://arxiv.org/abs/astro-ph/0002495}{{\ttfamily
  astro-ph/0002495}}].

\bibitem{Goodman:2000tg}
J.~Goodman, \emph{{Repulsive dark matter}},
  \href{https://doi.org/10.1016/S1384-1076(00)00015-4}{\emph{New Astron.}
  {\bfseries 5} (2000) 103}
  [\href{https://arxiv.org/abs/astro-ph/0003018}{{\ttfamily
  astro-ph/0003018}}].

\bibitem{Li:2013nal}
B.~Li, T.~Rindler-Daller and P.~R. Shapiro, \emph{{Cosmological Constraints on
  Bose-Einstein-Condensed Scalar Field Dark Matter}},
  \href{https://doi.org/10.1103/PhysRevD.89.083536}{\emph{Phys. Rev.}
  {\bfseries D89} (2014) 083536}
  [\href{https://arxiv.org/abs/1310.6061}{{\ttfamily 1310.6061}}].

\bibitem{Dev:2016hxv}
P.~S.~B. Dev, M.~Lindner and S.~Ohmer, \emph{{Gravitational waves as a new
  probe of Bose–Einstein condensate Dark Matter}},
  \href{https://doi.org/10.1016/j.physletb.2017.08.043}{\emph{Phys. Lett.}
  {\bfseries B773} (2017) 219}
  [\href{https://arxiv.org/abs/1609.03939}{{\ttfamily 1609.03939}}].

\bibitem{Rodrigues:2017vto}
D.~C. Rodrigues, A.~del Popolo, V.~Marra and P.~L.~C. de~Oliveira,
  \emph{{Evidence against cuspy dark matter haloes in large galaxies}},
  \href{https://doi.org/10.1093/mnras/stx1384}{\emph{Mon. Not. Roy. Astron.
  Soc.} {\bfseries 470} (2017) 2410}
  [\href{https://arxiv.org/abs/1701.02698}{{\ttfamily 1701.02698}}].

\bibitem{Deng:2018jjz}
H.~Deng, M.~P. Hertzberg, M.~H. Namjoo and A.~Masoumi, \emph{{Can Light Dark
  Matter Solve the Core-Cusp Problem?}},
  \href{https://doi.org/10.1103/PhysRevD.98.023513}{\emph{Phys. Rev. D}
  {\bfseries 98} (2018) 023513}
  [\href{https://arxiv.org/abs/1804.05921}{{\ttfamily 1804.05921}}].

\bibitem{Eby:2016cnq}
J.~Eby, M.~Leembruggen, P.~Suranyi and L.~C.~R. Wijewardhana, \emph{{Collapse
  of Axion Stars}}, \href{https://doi.org/10.1007/JHEP12(2016)066}{\emph{JHEP}
  {\bfseries 12} (2016) 066}
  [\href{https://arxiv.org/abs/1608.06911}{{\ttfamily 1608.06911}}].

\bibitem{Braaten:2016kzc}
E.~Braaten, A.~Mohapatra and H.~Zhang, \emph{{Nonrelativistic Effective Field
  Theory for Axions}},
  \href{https://doi.org/10.1103/PhysRevD.94.076004}{\emph{Phys. Rev.}
  {\bfseries D94} (2016) 076004}
  [\href{https://arxiv.org/abs/1604.00669}{{\ttfamily 1604.00669}}].

\bibitem{Eby:2017teq}
J.~Eby, P.~Suranyi and L.~C.~R. Wijewardhana, \emph{{Expansion in Higher
  Harmonics of Boson Stars using a Generalized Ruffini-Bonazzola Approach, Part
  1: Bound States}},
  \href{https://doi.org/10.1088/1475-7516/2018/04/038}{\emph{JCAP} {\bfseries
  1804} (2018) 038} [\href{https://arxiv.org/abs/1712.04941}{{\ttfamily
  1712.04941}}].

\bibitem{Namjoo:2017nia}
M.~H. Namjoo, A.~H. Guth and D.~I. Kaiser, \emph{{Relativistic Corrections to
  Nonrelativistic Effective Field Theories}},
  \href{https://doi.org/10.1103/PhysRevD.98.016011}{\emph{Phys. Rev.}
  {\bfseries D98} (2018) 016011}
  [\href{https://arxiv.org/abs/1712.00445}{{\ttfamily 1712.00445}}].

\bibitem{Braaten:2018lmj}
E.~Braaten, A.~Mohapatra and H.~Zhang, \emph{{Classical Nonrelativistic
  Effective Field Theories for a Real Scalar Field}},
  \href{https://doi.org/10.1103/PhysRevD.98.096012}{\emph{Phys. Rev. D}
  {\bfseries 98} (2018) 096012}
  [\href{https://arxiv.org/abs/1806.01898}{{\ttfamily 1806.01898}}].

\bibitem{Eby:2018ufi}
J.~Eby, K.~Mukaida, M.~Takimoto, L.~C.~R. Wijewardhana and M.~Yamada,
  \emph{{Classical nonrelativistic effective field theory and the role of
  gravitational interactions}},
  \href{https://doi.org/10.1103/PhysRevD.99.123503}{\emph{Phys. Rev.}
  {\bfseries D99} (2019) 123503}
  [\href{https://arxiv.org/abs/1807.09795}{{\ttfamily 1807.09795}}].

\bibitem{Eby:2018dat}
J.~Eby, M.~Leembruggen, L.~Street, P.~Suranyi and L.~C.~R. Wijewardhana,
  \emph{{Approximation methods in the study of boson stars}},
  \href{https://doi.org/10.1103/PhysRevD.98.123013}{\emph{Phys. Rev.}
  {\bfseries D98} (2018) 123013}
  [\href{https://arxiv.org/abs/1809.08598}{{\ttfamily 1809.08598}}].

\bibitem{Chavanis:2016dab}
P.-H. Chavanis, \emph{{Collapse of a self-gravitating Bose-Einstein condensate
  with attractive self-interaction}},
  \href{https://doi.org/10.1103/PhysRevD.94.083007}{\emph{Phys. Rev.}
  {\bfseries D94} (2016) 083007}
  [\href{https://arxiv.org/abs/1604.05904}{{\ttfamily 1604.05904}}].

\bibitem{Boehmer:2007um}
C.~G. Boehmer and T.~Harko, \emph{{Can dark matter be a Bose-Einstein
  condensate?}},
  \href{https://doi.org/10.1088/1475-7516/2007/06/025}{\emph{JCAP} {\bfseries
  0706} (2007) 025} [\href{https://arxiv.org/abs/0705.4158}{{\ttfamily
  0705.4158}}].

\bibitem{Kaplinghat:2019dhn}
M.~Kaplinghat, T.~Ren and H.-B. Yu, \emph{{Dark Matter Cores and Cusps in
  Spiral Galaxies and their Explanations}},
  \href{https://doi.org/10.1088/1475-7516/2020/06/027}{\emph{JCAP} {\bfseries
  06} (2020) 027} [\href{https://arxiv.org/abs/1911.00544}{{\ttfamily
  1911.00544}}].

\bibitem{Lelli:2016zqa}
F.~Lelli, S.~S. McGaugh and J.~M. Schombert, \emph{{SPARC: Mass Models for 175
  Disk Galaxies with Spitzer Photometry and Accurate Rotation Curves}},
  \href{https://doi.org/10.3847/0004-6256/152/6/157}{\emph{Astron. J.}
  {\bfseries 152} (2016) 157}
  [\href{https://arxiv.org/abs/1606.09251}{{\ttfamily 1606.09251}}].

\bibitem{Veale:2017eqa}
M.~Veale, C.-P. Ma, J.~E. Greene, J.~Thomas, J.~P. Blakeslee, J.~L. Walsh
  et~al., \emph{{The MASSIVE survey – VIII. Stellar velocity dispersion
  profiles and environmental dependence of early-type galaxies}},
  \href{https://doi.org/10.1093/mnras/stx2717}{\emph{Mon. Not. Roy. Astron.
  Soc.} {\bfseries 473} (2018) 5446}
  [\href{https://arxiv.org/abs/1708.00870}{{\ttfamily 1708.00870}}].

\bibitem{Bar:2019pnz}
N.~Bar, K.~Blum, T.~Lacroix and P.~Panci, \emph{{Looking for ultralight dark
  matter near supermassive black holes}},
  \href{https://doi.org/10.1088/1475-7516/2019/07/045}{\emph{JCAP} {\bfseries
  1907} (2019) 045} [\href{https://arxiv.org/abs/1905.11745}{{\ttfamily
  1905.11745}}].

\bibitem{Banerjee:2019epw}
A.~Banerjee, D.~Budker, J.~Eby, H.~Kim and G.~Perez, \emph{{Relaxion Stars and
  their detection via Atomic Physics}},
  \href{https://doi.org/10.1038/s42005-019-0260-3}{\emph{Commun. Phys.}
  {\bfseries 3} (2020) 1} [\href{https://arxiv.org/abs/1902.08212}{{\ttfamily
  1902.08212}}].

\bibitem{Chavanis:2019bnu}
P.-H. Chavanis, \emph{{Mass-radius relation of self-gravitating Bose-Einstein
  condensates with a central black hole}},
  \href{https://doi.org/10.1140/epjp/i2019-12734-7}{\emph{Eur. Phys. J. Plus}
  {\bfseries 134} (2019) 352}
  [\href{https://arxiv.org/abs/1909.04709}{{\ttfamily 1909.04709}}].

\bibitem{Eby:2018zlv}
J.~Eby, M.~Leembruggen, P.~Suranyi and L.~C.~R. Wijewardhana, \emph{{Stability
  of Condensed Fuzzy Dark Matter Halos}},
  \href{https://doi.org/10.1088/1475-7516/2018/10/058}{\emph{JCAP} {\bfseries
  1810} (2018) 058} [\href{https://arxiv.org/abs/1805.12147}{{\ttfamily
  1805.12147}}].

\bibitem{Broadhurst:2019fsl}
T.~Broadhurst, I.~de~Martino, H.~N. Luu, G.~F. Smoot and S.-H.~H. Tye,
  \emph{{Ghostly Galaxies as Solitons of Bose-Einstein Dark Matter}},
  \href{https://doi.org/10.1103/PhysRevD.101.083012}{\emph{Phys. Rev. D}
  {\bfseries 101} (2020) 083012}
  [\href{https://arxiv.org/abs/1902.10488}{{\ttfamily 1902.10488}}].

\end{thebibliography}\endgroup


\begin{thebibliography}{100}
\bibitem{GalFits} D. C. Rodrigues, A. del Popolo, V. Marra, and P. L. C. de Oliveira, ``Evidences against cuspy dark matter halos in large galaxies."
MNRAS 470 (2): 2410-2426 (2017).
arXiv: 1701.02698
 \bibitem{Tufts} H. Deng, M. P. Hertzberg , M. H. Namjoo, and A. Masoumi, ``Can Light Dark Matter Solve the Core-Cusp Problem?"
Phys.Rev. D98 (2018) no.2, 023513.
arXiv: 1804.05921



\bibitem{Axiverse}
A. Arvanitaki, S. Dimopoulos, S. Dubovsky, N. Kaloper, J. March-Russell,
``String axiverse."
Phys.\ Rev.\ {\bf D81} (2010) 123530,
arXiv: 0905.4720


\bibitem{Clockwork}
D. E. Kaplan and R. Rattazzi, ``Large field excursions and approximate discrete symmetries from a clockwork axion."
Phys.Rev. D93 (2016) no.8, 085007
arXiv: 1511.01827

\bibitem{Huetal}
W. Hu, R. Barkana and A. Gruzinov, ``Cold and Fuzzy Dark Matter."
Phys.Rev.Lett. \textbf{85} (2000) 1158-1161,
DOI: 10.1103/PhysRevLett.85.1158
arXiv: 0003365

\bibitem{MarshPop}
D.J.E. Marsh and A. Pop, ``Axion dark matter, solitons, and the cusp-core problem."
MNRAS, 451, 2479 (2015),
DOI: 10.1093/mnras/stv1050,
arXiv: 1502.03456

 \bibitem{Sim1} H.-Y. Schive, T. Chiueh, and T. Broadhurst, ``Cosmic Structure as the Quantum Interference of a Coherent
Dark Wave.” Nature Phys. 10 (2014) 496–499. arXiv: 1406.6586
 \bibitem{Sim2} H.-Y. Schive, M.-H. Liao, T.-P. Woo, S.-K. Wong, T. Chiueh, T. Broadhurst, and W. Y. P. Hwang, ``Understanding the Core-Halo Relation of Quantum Wave Dark Matter from 3D Simulations.” Phys. Rev. Lett. {\bf113} no. 26, (2014) 261302. arXiv: 1407.7762
 \bibitem{Sim3} B. Schwabe, J. C. Niemeyer, and J. F. Engels, ``Simulations of solitonic core mergers in ultralight
axion dark matter cosmologies.” Phys. Rev. D {\bf94} no. 4, (2016) 043513. arXiv: 1606.05151
 \bibitem{Sim4} J. Veltmaat and J. C. Niemeyer, ``Cosmological particle-in-cell simulations with ultralight axion dark matter.” Phys. Rev. D {\bf94} no. 12, (2016) 123523. arXiv: 1608.00802
 \bibitem{Sim5} P. Mocz, M. Vogelsberger, V. H. Robles, J. Zavala, M. Boylan-Kolchin, A. Fialkov, and L. Hernquist, ``Galaxy formation with BECDM: I. Turbulence and relaxation of idealized haloes.” Mon. Not. Roy. Astron. Soc. 471 no. 4, (2017) 4559–4570. arXiv: 1705.05845
\bibitem{MarshSilk}
D.J.E. Marsh and J. Silk, ``A Model for Halo Formation With Axion Mixed Dark Matter."
MNRAS, Volume 437, p2652 (2014),
DOI:  10.1093/mnras/stt2079
arXiv: 1307.1705

 \bibitem{LymanAlpha} V. Irsic, M. Viel, M. G. Haehnelt, J. S. Bolton, and G. D. Becker, ``First constraints on fuzzy dark matter from Lyman-$\a$ forest data and hydrodynamical simulations.” Phys. Rev. Lett. {\bf119} (2017) 031302. arXiv: 1703.04683
 \bibitem{BarBlum}  N. Bar, D. Blas, K. Blum, and S. Sibiryakov, ``Galactic Rotation Curves vs. Ultra-Light Dark Matter: Implications of the Soliton -- Host Halo Relation." arXiv: 1805.00122
 \bibitem{BarBlum2} N. Bar, K. Blum, J. Eby, and R. Sato, ``Ultra-light dark matter solitons in disk galaxies." arXiv: 1903.03402

\bibitem{JijiFan}
Jiji Fan, "Ultralight Repulsive Dark Matter and BEC." Phys.Dark Univ. 14 (2016) 84-94
DOI:  10.1016/j.dark.2016.10.005
arXiv: 1603.06580

\bibitem{Tye}
T. Broadhurst, H. N. Luu, and S.H.H. Tye, "Multiple Ultralight Axionic Wave Dark Matter and Astronomical Structures."
arXiv: 1811.03771

 \bibitem{Kaup} D.J. Kaup, ``Klein-Gordon Geon.'' Phys. Rev 172 (1968) 1331.
\bibitem{RB} R. Ruffini and S. Bonazzola, ``Systems of Self-Gravitating Particles in General Relativity and the Concept of an Equation of State.'' Phys. Rev. 187 (1969) 1767.

  \bibitem{ChavanisMR} P.H. Chavanis, ``Mass-radius relation of Newtonian self-gravitating Bose-Einstein condensates with short-range interactions: I. Analytical results.'' Phys. Rev. D {\bf84} (2011) 043531. arXiv: 1103.2050
 \bibitem{ChavanisMR2} P.H. Chavanis and L. Delfini, ``Mass-radius relation of Newtonian self-gravitating Bose-Einstein condensates with short-range interactions: II. Numerical results.'' Phys. Rev. D {\bf84}(2011) 043532. arXiv: 1103.2054.
   \bibitem{ESVW} J. Eby, P. Suranyi, C. Vaz, and L.C.R. Wijewardhana, ``Axion Stars in the Infrared Limit.'' JHEP 1503 (2015) 080.  arXiv:1412.3430 

\bibitem{FDMCollapse} J. Eby, M. Leembruggen, P. Suranyi, and L.C.R. Wijewardhana, ``Stability of Condensed Fuzzy Dark Matter Halos.” JCAP 1810 no.10 (2018) 058. arXiv: 1805.12147

\bibitem{Smoot} T. Broadhurst, I. de Martino, H. N. Luu, G. F. Smoot, and S.H.H. Tye, ``Ghostly Galaxies as Solitons of Bose-Einstein Dark Matter." arXiv: 1902.10488

\bibitem{Ansatz} J. Eby, M. Leembruggen, L. Street, P. Suranyi, and L.C.R. Wijewardhana, "On Approximation Methods in the Study of Boson Stars."
Phys. Rev. \textbf{D98} (2018), 123013,
10.1103/PhysRevD.98.123013
arXiv: 1809.08598

 \bibitem{ChavanisCollapse} P.H. Chavanis, ``Collapse of a self-gravitating Bose-Einstein condensate with attractive self-interaction.'' Phys. Rev. D {\bf94} (2016) 083007. arXiv: 1604.05904
 \bibitem{ELSW} J. Eby, M. Leembruggen, P. Suranyi, and L.C.R. Wijewardhana, ``Collapse of Axion Stars,'' JHEP 1007 (2016) 066.  arXiv:1608.06911.
 
  \bibitem{BohmerHarko} C. G. B\"ohmer and T. Harko, ``Can dark matter be a Bose-Einstein condensate?." JCAP 0706:025 (2007). arXiv: 0705.4158

\end{thebibliography}
\bibliographystyle{JHEP}

\end{document}